\documentclass[a4paper,11pt]{article}
\usepackage{pos}

\usepackage{nicefrac}
\usepackage{scalerel}

\newcommand{\Fig}[1]{figure~\ref{#1}}
\newcommand{\Eq}[1]{Eq.~(\ref{#1})}
\newcommand{\Eqs}[2]{eqs.~(\ref{#1}) and (\ref{#2})}
\newcommand{\Sec}[1]{section~\ref{#1}}

\graphicspath{{figures/}}

\title{\center Unfolding Particle Physics Hierarchies 
 with Supersymmetry and Extra Dimensions}
\author[a]{Raman Sundrum}
\date{May 31, 2023}
\affiliation[a]{ Maryland Center for Fundamental Physics \\
Department of Physics, 
University of Maryland \\
College Park 20742, USA}
\emailAdd{raman@umd.edu}
\FullConference{
  TASI 2022 – ``Ten Years After the Higgs Discovery: Particle Physics Now and Future''\\
  June 6 - July 1 2022\\
 Boulder, CO USA
}

\tableofcontents

\abstract{This is a written version of lectures delivered at TASI 2022 ``Ten Years After the Higgs Discovery: Particle Physics Now and Future''. Mechanisms and symmetries beyond the Standard Model (BSM) are presented capable of elegantly and robustly generating the striking hierarchies  we observe in particle physics. They are shown to be among the central archetypes  of quantum effective field theory and to strongly resonate  with the tight structure and phenomenology of the Standard Model itself, allowing one to motivate, develop and test  a worthy successor. 
The (Little) Hiearchy Problem is discussed within this context. The lectures culminate in specific BSM case-studies, gaugino-mediated (dynamical) supersymmetry
breaking to generate the weak/Planck hierarchy, and (in less detail)
extra-dimensional wavefunction overlaps to generate flavor hierarchies. }

\begin{document}
\maketitle

\section{Introduction}

The Standard Model (SM) describes an orchestra of elementary particles playing tightly interwoven melodies in the symphony of Nature \cite{Georgi-book}. And yet, it is an unfinished symphony. The SM gauge and Yukawa couplings display intriguing patterns that require new mechanisms  to explain. 
The enigmas of Dark Matter and the origins of the matter-antimatter asymmetry also point beyond the SM. The incorporation of a fully realistic quantum gravity represents a significant challenge. 
Against the backdrop of plausible BSM physics at far-UV scales, 
electroweak symmetry breaking (EWSB) is very fragile, 
posing another thorny mystery, the Hierarchy Problem. 

In these lectures, I want to survey the direct and indirect evidence for the very hierarchical structure of fundamental physics and to provide powerful and overarching quantum field theory (QFT) mechanisms beyond the Standard Model (BSM) capable of elegantly generating this structure. Part of the job is to fully appreciate the beautiful themes already at work within the SM, so as to guide us in how the symphony might extend further. 
New themes should harmonize with the old.
Of course, we will need to carefully account for the state of play along different experimental frontiers and the prospects for their improvement. It is in the context of this ambitious BSM undertaking that the (in)famous Hierarchy Problem will be discussed intuitively, rather than as a philosophically dubious concern of the SM in isolation. 

The lectures will culminate in a specific  BSM structure, not because I think it is the inevitable successor to the SM but because it makes a good  ``case study'', illustrating   robust QFT principles, methodology and phenomenological detective work along the way. The key new ingredients are extensions of relativistic spacetime, supersymmetry (SUSY) and extra dimensions, which will be strongly motivated from both top-down and bottom-up considerations. The key old ingredient to be recycled is dimensional transmutation as seen in QCD, capable of generating exponential hierarchies.
The Hierarchy Problem will be solved (modulo the Little Hierarchy Problem) within the framework of ``Gaugino-Mediated SUSY Breaking'' \cite{gaugino-med-susyBreak}, where SUSY is ultimately broken ``dynamically'' (via dimensional transmutation). Along the way, I will give a low-resolution introduction to the extra-dimensional wavefunction overlap mechanism for generating flavor (Yukawa-coupling) hierarchies \cite{flav-hier-extraDim}.

Apology: The goal here is to present a coherent conceptual framework for particle physics, but of course to do that concretely requires equations, which necessarily involve factors of $2, \pi, i$ and minus signs. I have done a modest job of trying to self-consistently get the right factors of $2$ and minus signs in the time I had, but I am sure that there are still several errors. I have done a better job with factors of $\pi$ and $i$. I hope this still allows the lectures to be readily comprehensible, and the reader can go through derivations more carefully for themselves or consult the more careful references provided (accounting for slight differences of convention). Since the lectures are founded on the profound mathematical identity,
\begin{equation}
e^{\rm moderate} = {\rm Large},
\end{equation}
I have been especially careful to ensure that I have no mistakes in the exponents that appear.

\subsection{The End of Particle Physics}

Let me begin with a few things you all know, but I want to look again with fresh eyes and marvel at the enigmas of fundamental physics.
Elementary particles are categorized in terms of  two spacetime quantum numbers, Mass and Spin, as well as some internal quantum numbers. 
Both the mass and spin have maximum allowed values, in each case involving General Relativity in interesting but different ways. These are the ``ends'' of 
 particle physics in the mass and spin directions. 
 
 Since $E = mc^2 = \$$ is the central consideration in particle physics, we begin with mass. A point particle has a classical  Schwarzchild radius $\sim G_N m$ as well as effectively a quantum mechanical ``size'' given by its Compton wavelength $1/m$. When the former is larger, the particle 
 effectively is within its Schwarzchild radius and is predominantly a classical black hole rather than an elementary quantum  particle. This happens when $m \gg 1/\sqrt{G_N}$, so that the Planck scale
 $M_{Planck} \equiv 1/\sqrt{8 \pi G_N} = 2 \times 10^{18}$ GeV marks the high end of particle physics in the mass direction, and the onset of black hole physics. 
 
 Relatedly, particle physics is the exploration of the smallest distances $\ell$, which by the uncertainty principle and relativity requires concentrating energy $E \geq |\vec{p}\,|  > 1/\ell$.  And yet if this concentration of energy is within its Schwarzchild radius $\ell \ll G_N E$, it will again gravitationally form a black hole. We are therefore unable to probe distances shorter than the Planck length $\ell_{Planck} = \sqrt{G_N} \sim 1/M_{Pl}$. 

Given these considerations, we are led to the following paradigm.
A full dynamics of quantum gravity operating at the highest energy/mass scales {\it matches onto} ( reduces to) a quantum effective field theory (EFT) below $M_{Pl}$, describing matter, radiation and general relativity with pointlike quanta. This EFT unfolds as we follow its renormalization group (RG) flow to lower energies and larger distances. A roughly parallel unfolding takes place in cosmic history as the universe expands and cools from high temperatures at the Big Bang to the cold temperatures of outer space today. 

\subsection{The View from the Top}


Superstring theory offers a UV-complete formulation of quantum gravity \cite{superstring-book}. It has a tight internal consistency, great beauty and unity, and the virtue of concreteness of its perturbative structure. String constructions can have quasi-realistic features. 
And who knows, maybe some incarnation of string theory is even true. But it at least gives us a concrete means to envision the quantum gravity heights and what might descend from there into EFT. 

The fundamental objects are one-dimensional strings which approximate point-particles  on distances longer than the string length parameter $\ell_{\rm string} > \ell_{\rm Planck}$. The lightest vibrational modes of the superstring appear as pointlike gravitons, gauge bosons and charged particles at low energies, but excited vibrational modes of the string appear as high mass $\sim 
m_{\rm string} \equiv 1/\ell_{\rm string} < M_{Pl}$ and higher-spin resonances. The non-renormalizable perturbative expansion of quantum general relativity, in terms of the dimensionless coupling $G_N.Energy^2$, threatens to exit perturbative control as energies grow towards $M_{Pl}$.  But in string theory this UV growth of the effective coupling is cut off while it is still weak, by energies $\sim m_{\rm string}$, retaining perturbative control. 

Remarkably, the simplest string theory constructions require higher-dimensional spacetime and SUSY for their self-consistency. There is a stringy analog of the kind of gauge-anomaly cancellation requirement familiar in chiral gauge theories such as the SM itself, which restricts the possible field content. But in string theory, spacetime dimensions are themselves fields (on the string worldsheet), and anomaly cancellation 
determines their number to be $10$!
Furthermore, for the stringy vibrational modes to include fermionic effective particles and to ensure vacuum stability (absence of tachyons), SUSY is required. In this way, extra spacetime dimensions and SUSY  are strongly motivated from a top-down quantum gravity perspective. The only question is how low in energies these exotic ingredients and their associated higher symmetries are manifest, and then  ultimately hidden at current experimental energies.

\subsection{Fundamental Physics is Hierarchical!}

\begin{figure}[hbt!]
    \centering
    \includegraphics[width=0.92\linewidth,trim={3cm 2.8cm 8cm 3.5cm},clip]{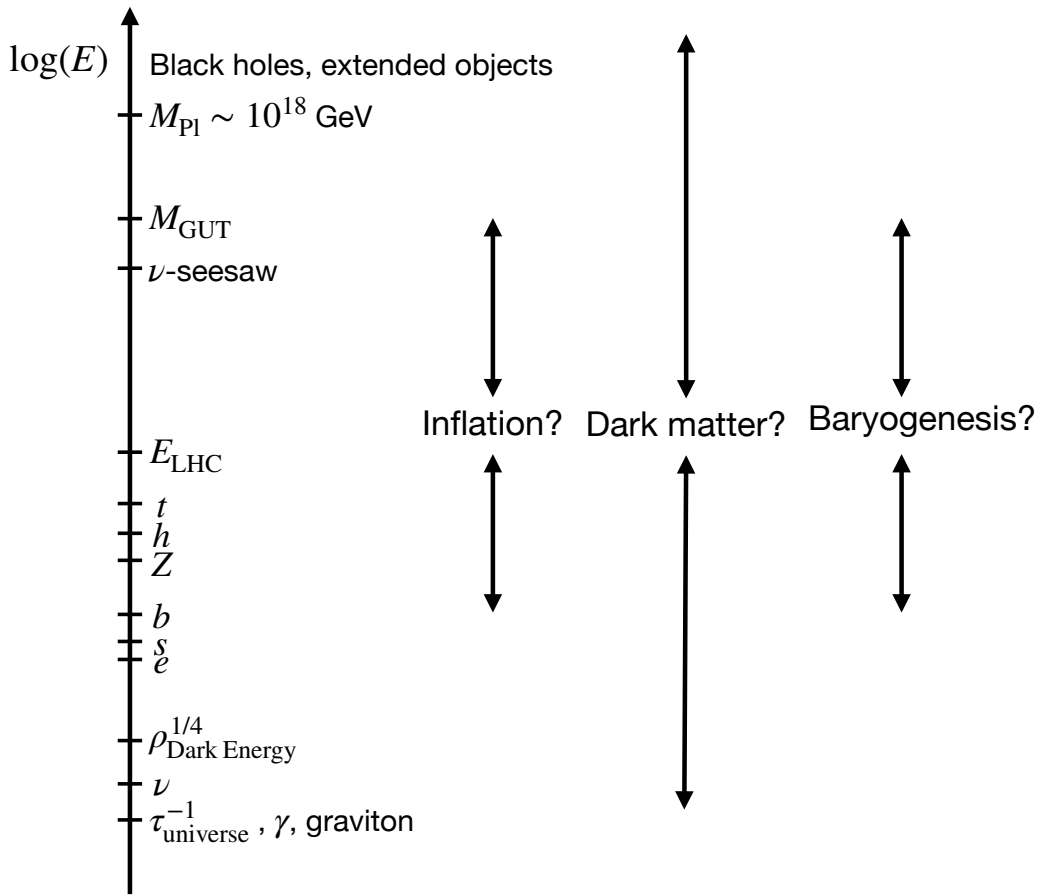}
    \caption{Cartoon of the hierarchical structure of fundamental physics}
    \label{fig:hierarchy_cartoon}
\end{figure}
\begin{figure}[hbt!]
    \centering   
    \includegraphics[width=0.7\linewidth]{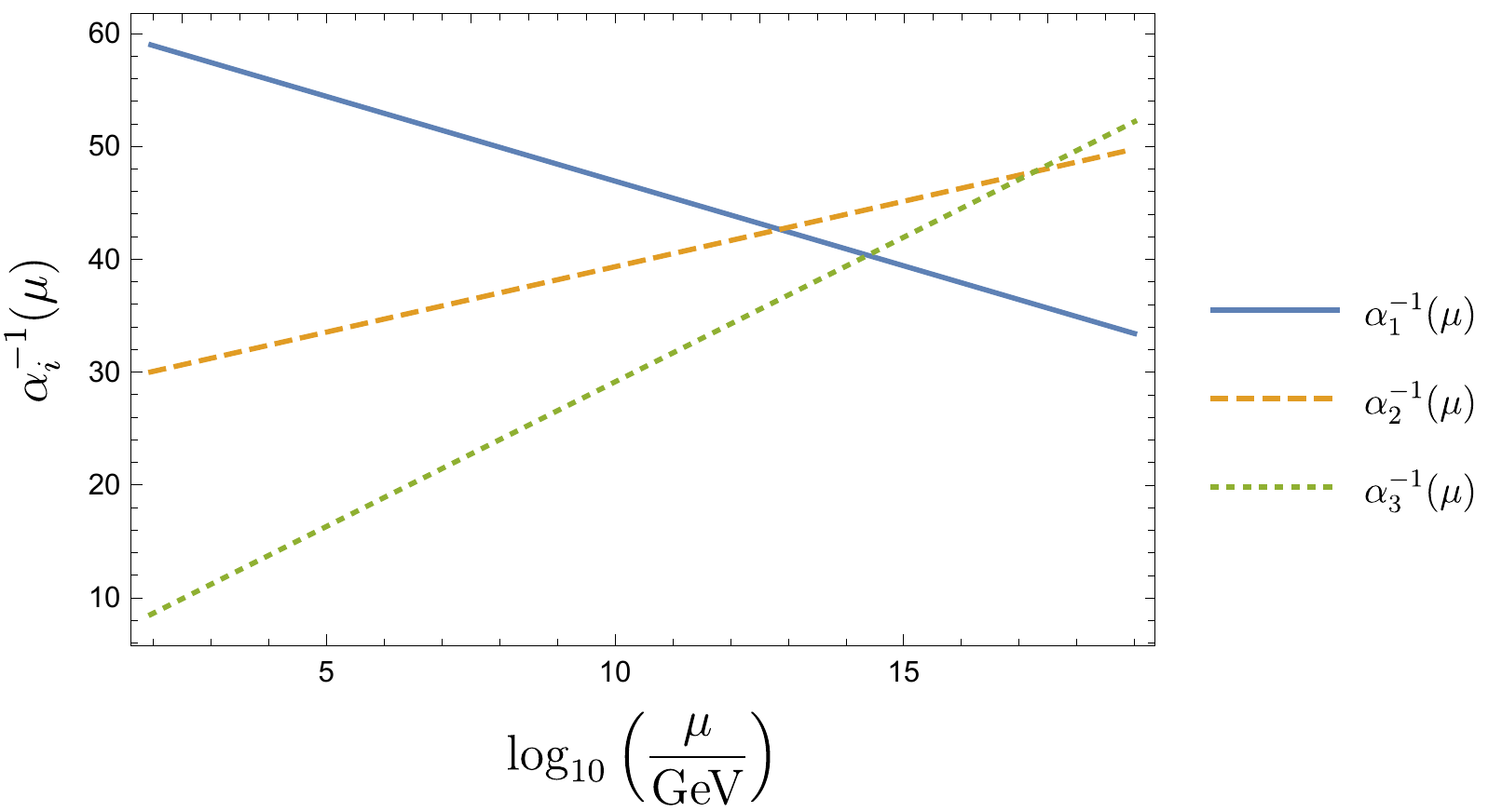}
    \caption{Running of SM gauge couplings at one loop, hinting at a grand unification at extremely high energies}
    \label{fig:unification}
\end{figure}

Consider the cartoon of fundamental physics in \Fig{fig:hierarchy_cartoon}, laid out in terms of energy/mass scales. It is meant to be like an ancient map, starting with actual experimental data but trailing off into broad theoretical prejudices and guesswork, and far from complete. It is an explorers map for those who hope to sail to its extreme reaches by every means possible.  This kind of synthesis of precision data and plausible theory is illustrated in \Fig{fig:unification}. 
Starting with the disparate measured values of Standard Model (SM) gauge couplings at the weak scale, their SM RG evolution shows a striking ``near'' coincidence 
at extremely high energies, suggesting a common origin. Indeed, on closer inspection, the different SM fields and their quantum numbers fit neatly like puzzle pieces into ``grand unified theories'' (GUTs), where some missing pieces have gotten very large masses $\sim M_{GUT}$  by a grand version of the Higgs mechanism. See \cite{gut-review} for a review. GUTs would explain why the gauge couplings run up to a nearly unified value at $\sim M_{GUT}$. As you probably know, there are attractive realizations of GUTs involving Supersymmetry (SUSY) and the extra puzzle pieces it necessitates, but here I wanted to show that just the data and SM extrapolation already suggests something interesting is going on orders of magnitude above collider energies. 

From the gargantuan size of our universe to the Planck scale, and everything in between, fundamental physics seems remarkably hierarchical. What powerful and economical mechanisms might underlie such hierarchical structure? 
One goal of these lectures is to show how supersymmetric and higher-dimensional dynamics can robustly and elegantly generate  the observed hierarchical structure in particle physics, at least in its non-gravitational aspects.

\subsection{Exponential Hierarchy from Non-Perturbative Physics}

Rather than giving some sort of formal definition of what it means to successfully ``explain'' hierarchical structure, 
I want to just remind you of a beautiful mechanism that captures its spirit, namely Dimensional Transmutation. Our goal will then be to generalize this in some way that applies it to the broader set of particle physics hierarchies. Let us specialize to the case of QCD,  for simplicity approximating the light ``up'' and ``down'' quarks as massless, and neglecting the other quark flavors altogether. We can imagine this arising as an EFT just below the Planck Scale and ask how robust it is that the observed proton mass is so many orders of magnitude lighter. 

Massless QCD is parametrized by $\alpha_{QCD}(\mu)$ where $\mu$ is the RG scale. The physical proton mass must be an RG-invariant function of 
$\alpha_{QCD}(\mu)$ and $\mu$, which uniquely determines its form,
\begin{align}\label{eq:ProtonMass}
    m_{proton} = \mu \, e^{-\int_{x_0}^{\alpha(\mu)} dx \,\, \left(\nicefrac{1}{\beta(x)}\right) }.
\end{align}
The RG flow is given by $\frac{d \alpha}{d \ln \mu} \equiv \beta(\alpha)$\footnote{It is straightforward to check that \Eq{eq:ProtonMass} is indeed RG-invariant by taking $d/d \ln \mu$ of it.}. The parameter $x_0$ is just an order one integration constant of the RG solution.  From the perspective of the Planck scale, 
\begin{align}
    m_{proton} = M_{Pl} \, e^{-\int_{x_0}^{\alpha(M_{Pl})} dx \,\, \left( \nicefrac{1}{-b x^2 + \cdots}\right)} \sim {\cal O}(M_{Pl}) \,\, e^{ \nicefrac{-1}{b\alpha(M_{Pl})}}  \,,
\end{align}
where we have chosen $\mu \sim M_{Pl}$ and used asymptotic freedom of QCD at such high scales to $1$-loop approximate $\beta(\alpha) \approx - b \alpha^2$. Here $b = 29/6\pi $ is the gauge-algebraically determined $1$-loop coeffcient for $2$-flavor QCD. 

We see that $m_{proton} \approx$ GeV, $m_{proton}/M_{Pl} \sim 10^{-18}$ emerges if $\alpha_{QCD}(M_{Pl}) \sim {\rm few} \times 10^{-2}$. If we consider any value of $\alpha_{QCD}(M_{Pl})$  between say $0$ and $\sim 1$ to have been equally likely to have emerged upon matching to some unspecified Planckia quantum gravity  a priori, then we only need to be ``lucky'' enough that $\alpha_{QCD}$ is modestly small in order to understand why the proton is orders of magnitude lighter than $M_{Pl}$.\footnote{We could more generally consider any other smooth likelihood for $\alpha_{QCD}$ in this range.} Note that this powerful mechanism is fundamentally non-perturbative in nature,  $e^{- {\rm constant}/\alpha}$ is the classic function whose perturbative Taylor expansion in $\alpha^n$ vanishes to all orders. 

The inspiration and goal that follows from this example is 
to discover BSM QFT mechanisms so that modest hierarchies ${\cal O}(10^{-1} - 10^{-2})$ in far-UV couplings and mass-parameters/$M_{Pl}$ unfold in the IR in some roughly analogous manner, being exponentially stretched out to the very hierarchical structure we observe or anticipate in particle physics. I hope you agree that this should be one of our ends for particle theory. Let us proceed to uncover the means to this end. 

\section{Spinning Tales}

Thus far, particle experiments operate at $E \ll M_{Pl}$ and therefore can only detect particles with $m \ll M_{Pl}$. That is, from a Planckian view of particle physics the particles we have seen crudely satisfy $m \approx 0$. Symmetries (and approximate symmetries) provide plausible, economic mechanisms for understanding the robustness of $m = 0$ (and $m \approx 0$). These ``protective'' symmetries vary according to the spins of the (nearly) massless particles under consideration, and together provide a powerful grammar underlying
 the story of particle physics, what we have seen and have not seen, and what we can hope to see.

\subsection{Spin-$0$}
The classic mechanism behind having massless spin-$0$ particles is their realization as Nambu-Goldstone (NG) bosons of the spontaneous symmetry breaking (SSB) of an internal global symmetry. The symmetry's Noether current can be approximated in terms of the NG field $\phi$:
\begin{equation}
J_{\mu} \sim \partial_{\mu} \phi + \text{non-linear~in~fields}.
\end{equation}
Conservation of this current,
\begin{equation}
    0 = \partial^{\mu} J_{\mu} \sim \Box \phi + 
\text{non-linear~in~fields},
\end{equation}
then implies $\phi$ is massless. 
If there is a small explicit symmetry breaking as well, current conservation is imperfect and $m_{\phi}$ is small but non-vanishing, making $\phi$  a ``pseudo-NG boson'' (PNGB). 

While (P)NGBs  can readily be kinematically light enough to discover, there is a downside. Under the associated internal symmetry transformation, 
$\phi \rightarrow \phi + {\rm constant} + ...$, so that symmetric couplings are necessarily derivatively-coupled, that is constructed from the invariant $\sim \partial_{\mu} \phi$. This implies that $\phi$ couplings rapidly drop (at least linearly) as we move into the IR, making $\phi$ difficult to detect even if it is light enough to be produced kinematically. We can compare this with the massless photon, whose coupling $\alpha_{em}$ also runs to weak coupling in the IR, but only logarithmically.

It is therefore not surprising that we easily detect photons, but are still struggling experimentally to discover very light axions, even if these generically motivated $U(1)$ PNGBs exist. We have done better with {\it composite} PNGBs, such as QCD pions. Their couplings also get weaker in the IR but only starting from their compositeness or strong-coupling scale $\sim$ GeV down to their mass $\sim {\cal O}(100)$ MeV. The spin-$0$ Higgs boson is an enigmatic case. Certainly $m_{higgs} = 125 {\rm GeV} \ll M_{Pl}$ but in the SM theory there is no protective mechanism for why 
``$m_{higgs} \approx 0$''. But in the BSM Compositeness paradigm, the Higgs boson may in fact be a kind of PNGB composite of some new strong force \cite{pgb-higgs}, analogous to the pion, although we have still not detected any corroborating compositeness effects.

We have not seen any spin-$0$ particle with sizeable couplings which we can confidently say is an elementary particle, the jury still being out on the Higgs boson. We have seen lots of interacting spin-$1/2$ and spin-$1$ elementary particles, but nature does seem to be very stingy with interacting elementary spin-$0$. If the only protective mechanism for light spin-$0$ is based on PNGBs, then we can roughly understand why this is. Needless to say, it would be very interesting to uncover any other protective mechanism for light spin-$0$ particles which are not necessarily derivatively-coupled and therefore not very weakly coupled in the IR. And we will.

\subsection{Spin-$1/2$}

For spin-$1/2$ the protective symmetry of $m=0$ is famously chiral symmetry, either gauged or global. For example, for a (single) standard Dirac fermion, $m \bar{\psi} \psi = 
m \bar{\psi}_L \psi_R + m \bar{\psi}_R \psi_L$ is forbidden if we require chiral invariance under $\psi_L \rightarrow 
e^{i \theta_L} \psi_L, ~  \psi_R \rightarrow 
e^{i \theta_R} \psi_R$, where $\theta_L \neq \theta_R$. The same is true of any other Lorentz-invariant (Majorana) mass terms. 

If chiral symmetry is broken at low scales or by small couplings, it explains $m \approx 0$ on the big stage of particle physics. Indeed this is just what happens in the SM, where the chiral electroweak gauge symmetries are Higgsed at the weak scale $\ll M_{Pl}$ and communicated to spin-$1/2$ fermions by (mostly small) Yukawa couplings. 

\subsection{Spin-$1$}

We begin with the Proca equation for massive spin-$1$ fields:
\begin{align}
    \partial^{\mu}F_{\mu\nu} +m^2 A_{\nu}=0 \, ,
\end{align}
where $F_{\mu \nu} \equiv \partial_{\mu} A_{\nu} - 
\partial_{\nu} A_{\mu}$ as in standard Maxwell theory. It is best if you pretend you know nothing about electromagnetism and Maxwell's equations. Rather, 
you can check that the Proca equation is effectively the unique relativistically covariant field equation which is local (that is, a {\it differential} equation) 
associated to a free massive spin-$1$ particle. It is particularly straightforward to interpret in $4$-momentum space in the rest frame $\vec{p} = 0$: 
\begin{align}
    (E^2 -m^2)A_{i} &=0 \nonumber\\
    m^2 A_{0} &=0 \,.
\end{align}
We see that in this rest frame it matches our non-relativistic expectation that the three polarization form a spatial vector $\vec{A}$, whose (anti-)particles satisfy $E = m (c^2)$ at rest. The $A_0$ ``polarization' formally appearing by relativistic covariance of  the equations of motion also vanishes by these same equations.

Now we take the $m \rightarrow 0$ limit of Proca. Furthermore, if the spin-$1$ field is not free, we can add a non-trivial right-hand side made out of the other fields that couple to it. In this way we arrive at Maxwell equations as effectively the only option for a massless spin-$1$ field:
\begin{align}\label{eq:emo-heavySpin1}
    \partial^{\mu}F_{\mu\nu} = J_{\nu} \,.
\end{align}
Of course, this is not the historical empirically-based path to Maxwell, but rather the only logical option for interacting massless spin-1 in a relativistic theory.

The massless limit has resulted in two remarkable properties. The first follows by noting that taking the $4$-divergence ($\partial^{\nu}$) of Maxwell's equations implies that  $J_{\nu}$ made out of other fields is a conserved ``current'':
\begin{align}
   0= \partial^{\nu}\partial^{\mu}F_{\mu\nu} = \partial^{\nu}J_{\nu} =0 \,.
\end{align}
Globally, the total charge $Q = \int d^3 x J_0$ is therefore conserved in time. 
The other property is that Maxwell's equations are gauge-invariant under
\begin{align}
    A_{\mu} \rightarrow A_{\mu} +\partial_{\mu} \xi \,.
\end{align}
This is related to the fact that massless spin-$1$ has only two propagating polarizations compared to the three of massive spin-$1$. 

When there are multiple massless spin-$1$ fields, $A_{\mu}^a$, possibly interacting among themselves so that they appear non-linearly in each other's currents, the general self-consistent structure is precisely non-abelian gauge theory.\footnote{It is interesting to check that even the standard non-abelian gauge theory field equations can be written in the ``abelian-like" forms of \Eq{eq:emo-heavySpin1} with abelian-like field strengths and conserved currents,  $F_{\mu \nu}^a \equiv \partial_{\mu} W^a_{\nu} - 
\partial_{\nu} W^a_{\mu}, ~ \partial^{\mu} J_{\mu}^a = 0$ (that is, with just ordinary partial derivatives as opposed to covariant derivatives).}
In this way, (non-abelian or abelian) gauge-invariance can be thought of as the protective symmetry of $m = 0$ for spin-$1$ particles. 

The next question is how to break gauge symmetry by a ``small amount'' so as to realize 
 $m \approx 0$. This is subtle because the $m=0$ limit of spin-$1$ is not smooth, in that it has two polarization for
 spin-$1$ while all non-zero masses have three polarizations. But we know the nuanced answer, it is precisely the Higgs mechanism which maintains the {\it total} number of physical polarizations in the massless spin-$1$ limit by including spin-$0$ fields. Clearly Nature gives us several examples of massless and Higgsed gauge particles/fields.

 Let us jump over spin-$3/2$ for the moment and survey even higher spins.

 \subsection{Spin-$2$}

We again begin with $m \neq 0$, which for relativistic spin-$2$ satisfies the Pauli-Fierz equation for a symmetric tensor field $h_{\mu\nu}$,
\begin{align}
    \Box h_{\mu\nu} - \partial_{\sigma} \partial_{\mu} h^{\sigma}_{\nu}  - \partial_{\sigma} \partial_{\nu} h^{\sigma}_{\mu} + \partial_{\mu} \partial_{\nu} h + \eta_{\mu \nu} \left( \partial_{\alpha}\partial_{\beta} h^{\alpha \beta} - \Box h\right)  - m^2 (h_{\mu\nu}-h)=0
\end{align}
where $h \equiv \eta^{\mu \nu}h_{\mu \nu}$.
In $4$-momentum space, its equations of motion in the rest frame $\vec{p} = 0$ reduce to $h_{ jj} = h_{00} = h_{0j} = 0$, while the remaining traceless symmetric spatial tensor components satisfy   $(E^2 - m^2) h_{i j} = 0 $. This again matches our non-relativistic expectation, in this case that the five spin-$2$ polarizations form a symmetric traceless spatial tensor with rest energy $m$. 

Now we take the $m \rightarrow 0$ limit. You can check that the Pauli-Fierz equation then reduces to the {\it linearized} Einstein Equations 
\begin{align}\label{eq:LinEinsteinEq}
    \Box h_{\mu\nu} - \partial_{\sigma} \partial_{\mu} h^{\sigma}_{\nu}  - \partial_{\sigma} \partial_{\nu} h^{\sigma}_{\mu} + \partial_{\mu} \partial_{\nu} h + \eta_{\mu \nu} \left( \partial_{\alpha}\partial_{\beta} h^{\alpha \beta} - \Box h\right) \propto T_{\mu \nu}.
\end{align}
These happen to be the usual Einstein Equations in terms of a dynamical spacetime metric $g_{\mu\nu} \equiv \eta_{\mu \nu} + 
h_{\mu \nu}$ but keeping only first-order terms in $h_{\mu \nu}$, that is the
{\it linearized} Einstein Equations. I have again put a non-vanishing right-hand side, $T_{\mu \nu}$, to represent any non-linear terms in the spin-$2$ or other fields that might arise beyond free field theory. Again, this is not the historical emperically-based path with Newton's Law and the Equivalence Principle as guides, but rather the only logical option for interacting massless spin-2 in a relativistic theory.

And again, there are two remarkable properties in this limit. There is a gauge invariance, 
\begin{equation}
h_{\mu \nu} \rightarrow h_{\mu \nu} + \partial_{\mu} \xi_{\nu} + \partial_{\nu} \xi_{\mu},
\end{equation}
with a $4$-vector gauge transformation $\xi_{\mu}(x)$. Secondly, 
by taking the $4$-divergence $\partial_{\mu}...$ of \Eq{eq:LinEinsteinEq}, we see that $T_{\mu \nu}$ must be a {\it conserved} tensor current, $\partial^{\mu} T_{\mu \nu} = 0$, with globally conserved charges $P_{\mu} = \int d^3 \vec{x}~  T_{\mu 0}(x)$. The only such $4$-vector charge consistent with relativistic interactions is the familiar $4$-momentum itself, one consequence of the Coleman-Mandula Theorem.  Thus we know that  $T_{\mu \nu}$ can be nothing else but the energy-momentum or stress tensor of all the fields. 

~

{\bf Exercise}: Consider the example of $2 \rightarrow 2$ scattering, involving four different species of spin-$0$ particles. 
 Of course, $4$-momentum $P_{\mu}$ is conserved by translation invariance, but let us ask if there could be another $4$-vector conserved charge $Q_{\mu} = \int d^3 \vec{x}$ (local charge density).  If so, in the far past and far future when all particles are well-separated, total $Q_{\mu}$ would be the sum of $Q_{\mu}$ contributions of each individual particle in isolation. Since the only $4$-vector attached to each isolated spinless particle is its $4$-momentum, we must have $Q_{\mu}^{particle} \propto p_{\mu}^{particle}$. But it is a priori possible that the proportionality constant is different for each  particle species.  
Show that given standard $4$-momentum conservation for a generic nontrivial scattering angle (in center-of-momentum frame) that the proportionality constant must be universal for all the particles involved. Therefore $Q_{\mu}$ must be nothing but $P_{\mu}$ up to this overall constant.

~

Because the spin-$2$ particles  must exchange $4$-momentum among themselves and with other particles in an interacting theory, in order to be conserved $T_{\mu \nu}$ must contain non-linear terms in $h_{\mu \nu}$ to represent their energy-momentum. That is, the spin-$2$ particles must be self-coupled. 
This situation is therefore analogous to the case of several self-interacting massless spin-$1$ particles, the self-consistent form resulting in non-abelian gauge theory with non-abelian gauge invariance, despite being expressable in abelian-like form. For spin-$2$, we also have an apparent abelian-like gauge transformation, but the self-coupling means that including the explicit form of $T_{\mu \nu}$ requires finding the non-abelian extension of the gauge invariance. This is precisely the gauge symmetry of general coordinate invariance, and the explicit form of \Eq{eq:LinEinsteinEq} is then the generally coordinate-invariant non-linear Einstein Equations. 

~

{\bf Exercise}: Consider a general coordinate transformation, $y^{\mu} = x^{\mu} - \xi^{\mu}(x)$. Given a proper distance function on spacetime in terms of a general metric, $ds^2 = g_{\mu \nu}(x) d x^{\mu} d x^{\nu}$, re-express it in terms of $y$,  
$ ds^2 = g'_{\mu \nu}(y) d y^{\mu} d y^{\nu}$. Writing $g_{\mu \nu} = \eta_{\mu \nu} + h_{\mu \nu},  g'_{\mu \nu} = \eta_{\mu \nu} + h'_{\mu \nu}$, and restricting $h, h'$ and $\xi$ to being infinitesimally small (that is, working strictly to first order in these)
show that 
\begin{equation}
  h_{\mu \nu}'(x) =   h_{\mu \nu}(x) + \partial_{\mu} \xi_{\nu}(x) +  \partial_{\nu} \xi_{\mu}(x),
\end{equation}
the abelian-like gauge transformation we derived for spin-$2$ above. Removing the restriction of being infinitesimal gives a fully non-abelian generalization of the gauge symmetry, namely that of general coordinate invariance.

~

The uniqueness of energy-momentum conservation as a conserved $4$-vector charge implies there could only be one massless spin-$2$ field, and it must incarnate as General Relativity. Indeed, Nature has given this to  us. 

\subsection{Spin $> 2$ (the other end of particle physics)}

In brief, in analogy to spins $1$ and $2$, in order to be massless, higher spin fields would have to couple to conserved currents of the appropriate higher Lorentz representation. For example, a spin-$3$ field would have require a conserved symmetric $3$-tensor current. But this would imply a globally conserved $2$-tensor charge $Q_{\mu \nu}$. Such a conserved charge is inconsistent with interactions, again as a consequence of the Coleman-Mandula Theorem. 

~

{\bf Exercise} Again consider $2 \rightarrow 2$ scattering, for simplicity involving a single species of spin-$0$ particle. Suppose there were a conserved traceless symmetric tensor conserved charge,
\begin{equation}
  Q_{\mu \nu} = \int d^3 \vec{x} ~ (\rm local~charge~density),
\end{equation} 
so that in the far past and far future it is the sum of  individual particle charges. Again for an isolated particle, its charge
must be constructed out of its
$4$-momentum, in this case $Q_{\mu \nu} \propto  p_{\mu} p_{\nu} - \eta_{\mu \nu} p^2/4$.  Show that conservation of total $Q_{\mu \nu}$ is inconsistent with a generic scattering angle (in center-of-momentum frame). 

~

Therefore for spin $> 2$ we cannot have $m = 0$ or even $m\approx 0$, but interacting spin $> 2$ ``particles'' can exist if their masses are comparable to the scale at which the EFT describing them breaks down in the UV, $m \sim  \Lambda_{UV}$. Operationally, such particles are ``composite'', in that the energies high enough to create them are close to the energies at which they cease to be described by point-particle EFT. Nature gives us many examples, such as the high-spin hadron composites of QCD, $m \gtrsim $ GeV. 
String Theory as a perturbative theory of quantum gravity  also contains many higher-spin excitations ``composed'' of the fundamental string, with masses $\sim m_{string} \lesssim M_{Pl}$. Their stringy structure becomes apparent when probed in their relativistic regime, $E \sim m_{string}$. 

In this sense, we see that spin-$2$ (and hence, General Relativity) is the ``end'' of point-particle physics in the spin direction.
Finally, we return to the exciting case of ...

\subsection{Spin-$3/2$}

For $m \neq 0$, the Rarita-Schwinger equation for a vector-spinor field $\psi_{\mu \alpha}$, where I am choosing $4$-vector indices from the middle or later parts of the Greek alphabet (in this case $\mu$) and spinor indices from the early part of the Greek alphabet (in this case $\alpha$), reads
\begin{align}
    \left(\epsilon^{\mu\nu\rho\kappa} (\gamma_{5}\gamma_{\nu})_{\alpha\beta}\partial_{\rho} +\frac{m}{2}\left[\gamma^{\mu},\gamma^{\kappa}\right]_{\alpha\beta} \right) \psi_{\kappa \beta} = 0.
\end{align}
The $\gamma$'s are the familiar Dirac matrices. As for Dirac spin-$1/2$, in the rest-frame of $4$-momentum space this equation describes a particle ($E = m$) and its antiparticle ($E = -m$), each with effectively two independent spinor components.  
Focusing on just the particle and its independent two-component spinor, 
the above is essentially the unique relativistically covariant equation which reduces in the rest frame 
to $\psi_{\mu =0, \alpha} = 0, ~ (\sigma_i)_{\alpha \beta} \psi_{\mu = i, \beta} = 0, (E - m) \psi_{\mu = i, \alpha} = 0$, where $\sigma_i$ are the Pauli matrices.  
Non-relativistically, the desired spin-$3/2$ representation is contained in the product representation $\psi_{\mu = i, \beta}$ of spin-$1$ and spin-$1/2$, while the second of these  equations projects out the unwanted
spin-$1/2$ representation in this product.

Taking the $m \rightarrow 0$ limit of the Rarita-Schwinger equation and introducing a non-zero right-hand side to represent non-linearities/interactions as previously, 
\begin{align}\label{eq:eom-massiveSpin3over2}
    \epsilon^{\mu\nu\rho\kappa} \sigma_{\nu}^{\alpha\beta} \partial_{\rho} \psi_{\kappa\beta} = {\cal J}^{\mu\alpha}(x) \, .
\end{align}
Here, again in analogy to the Dirac equation, 
we have taken advantage of the fact that in the massless limit the four-component spinor equations split into decoupled equations for two-component or Weyl spinors, of which we keep just the left-handed one. In this left-handed two-component spinor space, 
the $\sigma_{\nu}$ are again the Pauli matrices 
$\sigma_i$ along with  $\sigma_0$ being the identity.

One might wonder whether massless spin-$3/2$ is more like massless spin-$1$ which allows an arbitrary number of distinct but interacting fields with that spin, or whether it is like massless spin-$2$ which allows only one field with that spin, namely the graviton field of GR. It turns out that the answer is intermediate, in that the maximum number of species of distinct $\psi$ fields is $8$, but we will not do the detectivework here to show this. Instead, we will just consider a single species of $\psi_{\mu \alpha}$ as above, this  being the phenomenologically most interesting case. 

Once again we get a new gauge symmetry, under 
$\psi_{\mu \alpha} \rightarrow \psi_{\mu \alpha} + \partial_{\mu} \xi_{\alpha}$, where the gauge transformation $\xi_{\alpha}(x)$ is a spinor field. And by taking the $4$-divergence of \Eq{eq:eom-massiveSpin3over2} we see that ${\cal J}_{\mu \alpha}$ must be a conserved vector-spinor current, 
$\partial^{\mu} {\cal J}_{\mu \alpha} = 0$. This implies a globally conserved spinor charge, $Q_{\alpha} \equiv 
\int d^3 \vec{x} {\cal J}_{\mu = 0, \alpha}$. Since $2$-component spinors are necessarily complex, so are $\psi, {\cal J}$, and therefore the $Q_{\alpha}$ have distinct conjugate conserved charges $\bar{Q}_{\alpha}$. 

What are these unfamiliar spinor charges, ``halfway'' between standard scalar internal charges and the spacetime energy-momentum charges?

\section{The Supersymmetry Charge Algebra}

Before trying to answer this, let us see how we can relate charges in the most familiar case of some scalar conserved charges of internal symmetries, $Q_a = \int d^3 {\vec x } J_0^a$. Clearly if $Q_a$ and $Q_b$ are conserved in time, then so is their product $Q_a Q_b = \int d^3 {\vec x } J_0^a(x) \int d^3 {\vec x }' J_0^b(x')$. But the great practical utility of charge conservation comes from the charge being the sum of {\it local} contributions, whereas  
this product charge is clearly not, with the ${\vec x}$ and ${\vec x}'$ contributions being arbitrarily far apart. Instead, let us consider the commutator, $[Q_a, Q_b]$. This is indeed a new charge with local contributions, if it is non-vanishing. 
To see this, first note that $J_0^a(x)$ will be made of some products of boson fields (or field momenta) and bilinears in fermion fields at $x$, and their derivatives. (Since fermions are half-odd-integer spinors by the spin-statistics connection they must appear in even numbers in any $4$-vector $J_{\mu}$.) Let us pretend for just a moment that in the QFT all bosonic fields and their derivatives commute, and all fermions and their derivatives anticommute, and all fermions commute with bosons. (Only the last of these statements is true, but let us just pretend.) Therefore, given this fiction, all bosons and fermion bilinears and their derivatives would commute, in which case clearly $[Q_a, Q_b] = 0$. Therefore in truth, the only way to get a non-zero commutator is because of retaining at least one 
 non-trivial commutator between boson fields (including bosonic field momenta or time-derivatives of boson fields) or fermion bilinears at $x$ and 
bosons or fermion bilinears at $x'$. Such commutators always contain $\delta^3({\vec x} - {\vec x}')$ factors, so $[Q_a, Q_b]$ must be another charge with {\it local} contributions! In this way, the ``useful'' charges form a standard Lie algebra. 

We can apply this kind of detective work to the more enigmatic spinor charges $Q_{\alpha}, \bar{Q}_{\beta}$. To be precise, 
 let us remind ourselves of the $2$-component spinor representation of the Lorentz group and the relevant notation. It is based on the relativistic analog of the familiar isomorphism of the rotation group, $SO(3) \equiv SU(2)$\footnote{We will not be careful about the fact that globally $SU(2)$ is the double-cover of $SO(3)$, it is isomorphic in the neighbourhood of the identity element.}, namely $SO(3,1) \equiv SL(2, C)$. That is, the Lorentz group can be represented by complex $2 \times 2$ matrices $\Lambda$ with unit determinant. We can define a {\it left-handed} representation $\psi_L$ to be a $2$-component spinor transforming as $\psi_L \rightarrow \Lambda \psi_L$, where matrix-multiplication is implied on the right side. 

We will use the ``bar'' notation to be the same as conjugation $\bar{\psi}_{\alpha} \equiv \psi_{\alpha}^{\dagger}$, and without  explicitly writing indices it will represent a row vector $\bar{\psi} \equiv \psi^{\dagger}$ obtained from the hermitian conjugate of the column vector $\psi$.  
Therefore under Lorentz transformation, 
$\bar{\psi} \rightarrow \bar{\psi} \Lambda^{\dagger}$ with matrix multiplication implied. One can then form Lorentz $4$-vectors from the spinor products 
$\bar{\psi} \sigma^{\mu} \chi$, where  again $\sigma^0$ is the identity matrix and $\vec{\sigma}$ are Pauli matrices. Together the $\sigma^{\mu}$ are a Hermitian basis for $2 \times 2$ complex matrices. 

As with internal charges we can start with the product charges  $Q_{\alpha} \bar{Q}_{\beta}$. Once again, they are conserved but not in a useful way since they are not the sum of local contributions. But now we can focus on the anticommutator $\{ Q_{\alpha}, \bar{Q}_{\beta} \}$, which is a sum of local contributions. The reason that it is the anticommutator which has this property is because any local spinor-vector current such as ${\cal J}_{\mu \alpha}(x)$ must necessarily be the product of some bosons and an {\it odd} number of fermionic fields and derivatives at $x$, by the spin-statistics connection. By the analogous argument to the case of scalar charges, it is  $\{ Q_{\alpha}, \bar{Q}_{\beta} \}$ which necessarily contains a factor of
$\delta^3({\vec x} - {\vec x}')$. 

What is interesting is that by the spinor algebra reviewed above, and the fact that the $\sigma^{\mu}$ are a hermitian basis for all $2 \times 2$ matrices, it follows that $\{ Q_{\alpha}, \bar{Q}_{\beta} \}$ necessarily transforms as a Lorentz $4$-vector of conserved charges, and the only such conserved $4$-vector  that is the sum of local contributions allowed in local QFT is $4$-momentum:
\begin{align}\label{eq:chargeAlgebra1}
    \{ Q_{\alpha}, \overline{Q}_{\beta } \} = 2 P_{\mu} \sigma^{\mu}_{\alpha\beta}.
\end{align}
The factor $2$ is chosen by convention in the normalization of $Q_{\alpha}$. 

By analogous detective work, the anticommutators $\{ Q_{\alpha}, Q_{\beta} \}$ and $\{ \bar{Q}_{\alpha}, \bar{Q}_{\beta} \}$ can also be conserved charges with local contributions. But these transform as Lorentz anti-symmetric tensors, and no such conserved tensor charges are possible in interacting QFT.\footnote{The famous antisymmetric tensor charges of QFT are the angular momentum tensor $J_{\mu \nu}$, but they are not all conserved in the usual sense of commuting with the Hamiltonian $P_0$, in particular the boost generators $J_{0 i}$ do not.} Therefore, we can only have 
\begin{align}\label{eq:chargeAlgebra2}
    \{ Q_{\alpha}, Q_{\beta} \} =0, ~ \{ \bar{Q}_{\alpha}, \bar{Q}_{\beta} \} =0.
\end{align}

The fact that the spacetime $P_{\mu}$ charges appear in our anticommutation relations, mean that the new symmetry charges correspond to an extension of spacetime symmetry beyond the usual algebra of Poincare generators $P_{\mu}, J_{\mu \nu}$ and their commutation relations. In addition to all of these, there are also commutation relations between the $Q_{\alpha}, \bar{Q}_{\beta}$ and the 
$P_{\mu}, J_{\mu \nu}$. The commutation relations with the angular momentum generators $J_{\mu \nu}$ merely express the spinor properties of the $Q_{\alpha}, \bar{Q}_{\beta}$, so I will not bother to write these out. The commutation relations with $P_{\mu}$ express charge conservation, 
\begin{equation}\label{eq:QP-commutators}
    [Q_{\alpha}, P_{\mu}] = 0, ~  [\bar{Q}_{\beta}, P_{\mu}] = 0.
\end{equation}
(The $P_0 = H$ commutator is literally conservation of charge, while the $\vec{P}$ commutator vanishes because the charge is the total across space, so a spatial translation does not change this total.)

If we put together the anticommutation algebra of conserved spinor charges \Eqs{eq:chargeAlgebra1}{eq:chargeAlgebra2}, the usual commutator algebra of Poincare generators, and the conservation properties, \Eq{eq:QP-commutators}, and (spinor) Lorentz transformation properties of $Q_{\alpha}, \bar{Q}_{\beta}$, we arrive at the 
minimal or ${\cal N} = 1$ superalgebra.  The 
$Q_{\alpha}, \bar{Q}_{\beta}$ are ``supercharges''.\footnote{ ${\cal N}$ counts the number of different ``flavors'' of $Q_{\alpha}$, but only ${\cal N} = 1$ SUSY QFT is compatible with having chiral fermions, such as Nature exhibits.} 

Does supersymmetry (SUSY) exist in the space of QFTs? 
That is, are there QFTs that contain the superalgebra of charges, with charges realised  as integrals over local charge densities? In particular, is there a supersymmetric extension of the SM? What is the structure of such theories? How can they be realistic? 
Let us continue the qualitative deductive reasoning a little further. See Ref.~\cite{sugra-and-susy} for a canonical text on general SUSY field theory construction, and Ref.~\cite{mssm-review} for construction and phenomenology of SUSY extension of the Standard Model. See also Refs.~\cite{susy-review-shirman} and \cite{susy-book-Terning}.

\section{Superpartners}\label{sec:superpartners}

Broadly, SUSY charges relate fermions and bosons, remarkable given their very different behavioral properties. To see this, start with 
the conservation of $Q_{\alpha}$, expressed canonically as $[H, Q_{\alpha}] = 0$. Consider the energy eigenstate of any bosonic particle,  $|B \rangle$, with energy $E_B$. Then $|F \rangle \equiv Q_{\alpha} |B \rangle $ must be a fermion by the spin-statistics connection because $Q_{\alpha}$ changes the angular momentum of the state it is acting on by a $1/2$ unit. Thus, $[H, Q_{\alpha}] |B \rangle = 0$ implies 
that $|F \rangle$ must also have the same energy, $E_F = E_B$. Working within each momentum subspace, this means that their masses are also the same, $m_F = m_B$. 
In this way, we see that in SUSY particles come in boson-fermion degenerate pairs, or ``superpartners'', which differ in their spin by a half unit. 

The superpartner of the massless spin-$2$ graviton $h_{\mu \nu}$ must be therefore be a massless spin-$3/2$ particle/field $\psi_{\mu \alpha}$, which we can call the ``gravitino''. Note, the fermion had to have spin $1/2$ unit different from $2$, but it could not have been spin-$5/2$ because massless spins $>2$ are forbidden as discussed earlier.

The superpartner of a massless spin-$1$ gauge boson $A_{\mu}^a$ must be a massless spin-$1/2$ fermion $\lambda_{\alpha}^a$, a ``gaugino''. The `a' index is the internal (non-spacetime) index labelling gauge generators, so that the gaugino is a fermion in the adjoint representation of the gauge group (the representation furnished by the gauge group generators themselves).
 Again, one might have thought the fermion could have been massless spin-$3/2$, but that would make it a second spin-$3/2$ field, the first being the gravitino.  But this would correspondingly require two species of supercharges $Q_{\alpha}$, incompatible with the minimal ${\cal N} =1$ SUSY we are studying.

Turning to standard (non-gaugino) spin-$1/2$ fermions, it
 will be convenient from here on to exclusively use  left-handed representation. We can convert conventional right-handed fields into left-handed fields by charge conjugation. That is, given a right-handed $2$-component spinor field $\psi_R$, its charge conjugate $(\psi_R)^c \equiv \sigma_2 
\psi_R^*$ is a left-handed spinor 
field.\footnote{ Often in the literature, one distinguishes ``dotted" and ``undotted" indices, where the dotted indices label the conjugates of the left-handed spinors. I will reduce ``clutter'' by not making this distinction in these lectures, hoping that context will give away what is intended. } 
Here $\sigma_2$ is the second Pauli matrix, and matrix-multiplication is implied. The Lorentz-invariant one can then construct from two left-handed fields is $(\psi_L \chi_L) \equiv \epsilon^{\alpha \beta} \psi_{L \alpha} \chi_{L \beta} =  \psi_{L \alpha} \chi_L^{ \alpha}$. Here, $\epsilon$ is the completely antisymmetric tensor, and one can think of it as a ``metric'' on spinors allowing one to ``raise'' the indices of a standard left-handed spinor. In this way the SM spin-$1/2$ chiral fermions can be listed as 
\begin{align}
    \psi_{\alpha} = q_{\alpha}, \ell_{\alpha}, u^c_{\alpha}, d^c_{\alpha}, e^c_{\alpha}.
\end{align}

The superpartners of these standard fermions must be spin-$0$ ``sfermion'' scalars, 
\begin{align}
   \phi = \tilde{q},  \tilde{\ell}, \tilde{u}^c, \tilde{d}^c, \tilde{e}^c ,
\end{align} 
where the tilde just labels the superpartner of the related SM field. (They could not be spin-$1$ because then they would have to be gauge fields in the adjoint representation of gauge groups and the fermions would be gauginos, which does not fit SM quantum numbers.)

The only other field in the SM is the Higgs doublet scalar field $H(x)$, and this must have a spin-$1/2$ ``Higgsino'' superpartner $\tilde{H}_{\alpha}$. There is an important subtlety regarding the SUSY Higgs content, but we will get to that later. 

If we can realize SUSY in QFT it predicts a remarkable feature. Earlier, we could only identify Nambu-Goldstone bosons as the means for having robustly massless spin-$0$ particles, but they have the limitation that their (derivative) couplings rapidly become weak in the IR. In particular, it was not clear if the Higgs boson could be considered as at least approximately a NG boson in some BSM scenario given its substantial couplings. But in SUSY because of superpartner mass-degeneracy, a massless scalar is a robust possibility if it is the superpartner of a spin-$1/2$ fermion which is robustly massless because of a chiral symmetry. If the fermion has standard gauge and Yukawa couplings, so will the scalar, such couplings falling at most logarithmically in the IR. SUSY then offers an attractive mechanism for robustly having light interacting scalars in QFT, such as the Higgs boson (seen from our Planckian perspective). 

\section{Supergravity, SUSY Breaking and the $G_N \rightarrow 0$ Limit}

\begin{figure}[hbt!]
    \centering
    \includegraphics[width=0.7\linewidth,trim={3cm 3cm 3cm 3cm},clip]{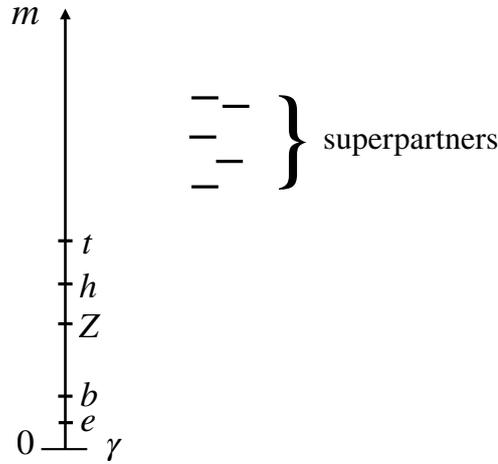}
    \caption{A plausible cartoon spectrum of a SUSY extension of the standard model. SUSY (degeneracy) is sufficiently broken to be consistent with our not having discovered superpartners yet, while being approximately valid far above the weak scale. }
    \label{fig:hierarchy-cartoon-susy}
\end{figure}

As we have seen, the graviton $h_{\mu \nu}$ is a ``gauge field'' whose associated charge is energy-momentum $P_{\mu}$, while the gravitino $\psi_{\mu \alpha}$ is a ``gauge field'' with associated charges $Q_{\alpha}, \bar{Q}_{\beta}$. But these charges obey a non-abelian type of superalgebra, \Eq{eq:chargeAlgebra1}, so together the  graviton and gravitino superpartners are ``gauge fields'' of SUSY. This gauging of the global SUSY superalgebra is sometimes called ``local SUSY''. But since it also gives a supersymmetric theory including gravity, it is also called ``supergravity'', or SUGRA for short. Like standard simple non-abelian gauge theories, local SUSY has a single ``gauge'' coupling, $G_{\rm Newton}$.


If SUSY were an exact symmetry then there would be a charged selectron with the electron's mass, and this is not seen in Nature. More generally, the absence of superpartners in experiments to date implies that SUSY must be an approximate symmetry at best, part of the robust QFT grammar of options for $m \approx 0$ from a UV perspective. 

We can guess a cartoon of a realistic particle spectrum with broken SUSY as given in \Fig{fig:hierarchy-cartoon-susy}. The superpartners are thereby heavy enough to have evaded LHC and other searches, but close enough to the weak scale that it might be playing a strong role in making a weak-scale interacting spin-$0$ Higgs a robust QFT feature. While the broken SUSY is clearly a big effect for human particle experimentalists, it represents a ``small'' breaking compared to the fundamental scale $\lesssim M_{Pl}$  at which QFT is born from quantum gravity. 

Since fundamentally, SUSY is a gauged symmetry of SUGRA, its breaking must be due to a ``super-Higgs'' effect \cite{broken-susy-and-sugra}. 
But in the global limit of this gauge theory, that is when we work in the approximation of vanishing gauge coupling $G_N \approx 0$, this Higgs-like breaking must become {\it spontaneous} SUSY breaking, analogous to what happens in the  standard Higgs mechanism in the global limit.

I have been trying to make the case to you that Nature may well be playing every trick in the book as far as $m \approx 0$ is concerned. But does it contain $m \approx 0$ for spin-$3/2$? The reason we do not know yet may not be that such a particle lies above our puny energy reach (although maybe it does) but because, like PNGBs, it has a reason to be extremely weakly coupled. And unlike the graviton, which shares its $G_N$ coupling, it cannot take advantage of Bose statistics to at least appear to us in observable classical fields. Nevertheless, its existence necessitates all of SUSY structure, and clearly searching for {\it that} experimentally is strongly motivated. 

How can we construct SUSY QFT and spontaneous SUSY breaking? To motivate the strategy, I want to digress into another important ingredient which is somewhat more intuitive (at least for theorists), namely higher-dimensional spacetime.

\section{Higher Powers and Hierarchies from Higher Dimensions}\label{sec:higherDim}

Let us consider the simplest higher-dimensional extension of 4D Minkowski spacetime  to 5D Minkowski spacetime: 
\begin{align}
    4D \,\, x^{\mu} \quad &\longrightarrow \quad 5D \,\,  X^{M} = x^{\mu}, x_5   \nonumber \\ 
    4D \,\, \text{Poincare symmetry} \quad &\longrightarrow \quad 5D \,\, \text{Poincare symmetry} \nonumber\\
    x^{\mu} \rightarrow \Lambda^{\mu}_{\nu} x^{\nu} + a^{\mu} \quad  &\longrightarrow \quad  X^{M} \rightarrow \Lambda^{M}_{N} X^{N} + A^{M} \nonumber \\
    \,\, \Lambda \in SO(3,1) \qquad  &  \qquad \quad \Lambda \in SO(4,1) \supset SO(3,1) \nonumber\\
    4D \text{ Fields} \,\, \phi(x) \quad &\longrightarrow \quad 5D \text{ Fields} \,\,  \phi(X) 
\end{align}
As presented, we posited a higher-dimensional spacetime and then noted that it could enjoy a larger 5D Poincare symmetry. But to pave the way for our later SUSY development, it is useful to think of it the other way around: if I know I have a QFT with an extension of the usual 4D Poincare spacetime symmetry, the simplest way of incorporating that is by extending the spacetime so that it is symmetric under the extended symmetry. In this way, demanding 5D Poincare symmetry leads to fields living on 5D Minkowski spacetime. 

\subsection{Compactification of the Extra Dimension}

\begin{figure}[hbt!]
    \centering    \includegraphics[width=0.9\linewidth,trim={2.cm 7cm 3cm 5.cm},clip]{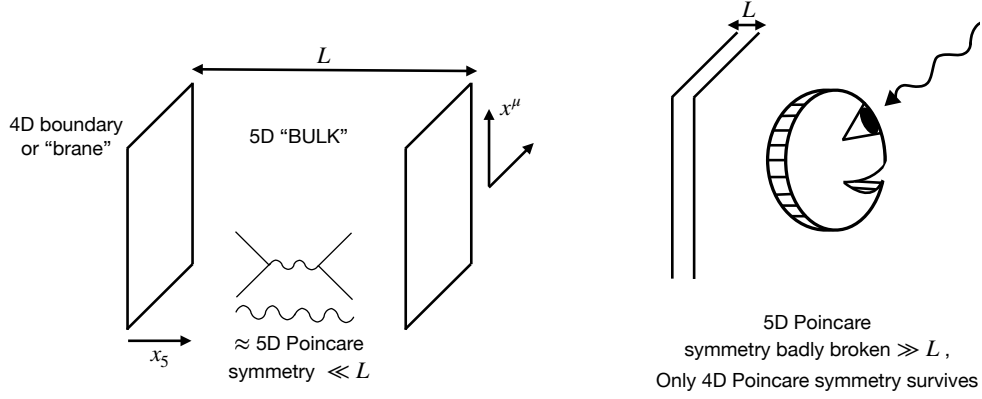}
    \caption{A tiny extra ``5th'' dimensional interval of size $L$, deep within which 5D Poincare invariance is approximately valid. 
   At much larger wavelengths, the extra dimension is invisible, 5D Poincare invariance is badly broken, and only 4D Poincare invariance survives.  
    }
    \label{fig:5Dand4Dperspective}
\end{figure}

The Poincare-invariant field equation, say Klein-Gordon, also extends straightforwardly:
\begin{align}\label{eq:5D-kleinGordon}
    (\partial_{\mu} \partial^{\mu} + m_{4}^{2}) \phi(x) =0 \longrightarrow (\partial_{\mu} \partial^{\mu} - \partial_{5}^{2} + m_{5}^{2}) \phi(X) =0.
\end{align}
Now, of course we do not live in a perfect 5D Minkowski spacetime with perfect 5D Poincare symmetry, so at best such a symmetry is badly broken at the ``low'' energies we currently probe. This allows me to introduce the notion of {\it soft} symmetry breaking, that is breaking by a dimensionful energy/mass scale such that at much higher energies (short distances) there  is a very good approximate symmetry and at lower energies (longer distances) the symmetry is not apparent. In the case of 5D, the simplest such soft breaking is depicted in \Fig{fig:5Dand4Dperspective}. The 5th dimension is a finite interval of microscopic length $L$, so that spacetime is a 5D slab-like ``bulk'',  sandwiched between two 4D boundaries or ``branes''. For short distance scattering and short wavelengths $\ll L$ in the bulk interior, as illustrated, 5D Poincare symmetry approximately holds.  For long wavelengths $\gg L$ only the 4D Poincare sub-symmetry is apparent. As depicted in \Fig{fig:5Dand4Dperspective}, at these long wavelengths the extra dimension itself cannot be resolved. 
We can say that the 5D Poincare symmetry is softly broken at the ``Kaluza-Klein'' (KK) scale $\mu_{KK} = 1/L$ down to 4D Poincare symmetry. 

Let us solve the 5D field equation on the right side of \Eq{eq:5D-kleinGordon} by separation of variables:
\begin{align}
    \phi(X) = f(x_5) \phi_{4}(x) ,
\end{align}
where the $\phi_4(x)$ factor satisfies the 4D Klein-Gordon equation
\begin{align}
    (\partial_{\mu} \partial^{\mu} + m_{4}^{2}) \phi_{4}(x) =0. 
\end{align}
Furthermore, let us assume that the fifth dimension is hidden far in the UV and that any fields the effective 4D experimentalist can probe have $m_4 \approx 0$. 
Then clearly the general 5D solution is given by 
\begin{align}
    \phi(X) &=  \left( A e^{m_5 x_5} + B e^{- m_5 x_5} \right) \phi_{4}(x) \nonumber \\ 
    & \underset{\text{typical b.c}}{\sim} e^{- m_5 x_5} \phi_{4}(x) \quad \text{or} \quad e^{+ m_5 (x_5 - L)} \phi_{4}(x).
\end{align}
The integration constants $A, B$ have to be determined by the boundary conditions at $x_5 = 0, L$.  We will not detail those, but it is easy to see that for $m_5 > L$ (the exponentials are varying signficantly across the fifth dimension) typical boundary conditions will result in the solution leaning strongly towards one boundary or the other, as indicated in the second line. These $m_4 \approx 0$ solutions, which determine (with boundary conditions) the 5D solutions, including 5D profiles, are called ``zero modes'' (or near-zero modes more accurately). In essence, at experimental ``low'' energies, the theory reduces to a 4D EFT of these zero-modes. 

See \cite{extra-dim} for reviews of extra-dimensional field theory and phenomenology. 

\subsection{Emergence of (Yukawa) Coupling Hierarchies}\label{subsec:yukawaHierarchies}

\begin{figure}[hbt!]
    \centering
    \includegraphics[width=0.5\linewidth,trim={8cm 6cm 8cm 4.5cm},clip]{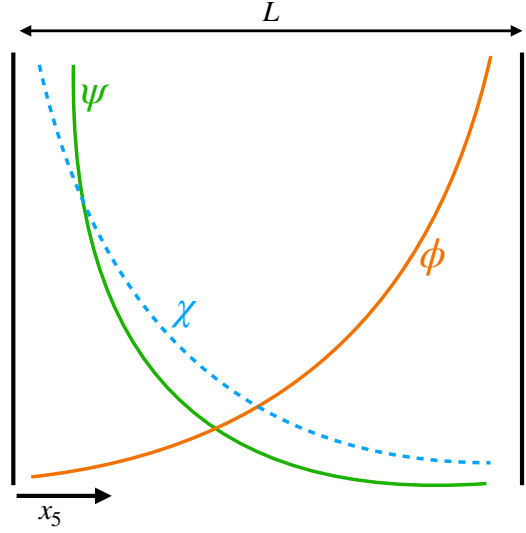}
    \caption{Some typical ``zero''-modes for three 5D fields, exhibiting exponential profiles depending on their 5D masses.  }
    \label{fig:sequestering}
\end{figure}
Now consider three species of 5D fields, $\chi(X), \psi(X), \phi(X)$, where the first two have zero-modes  leaning towards $x_5 = 0$ and the last towards $x_5 = L$, as depicted in \Fig{fig:sequestering}. We will consider a simple trilinear coupling between them,
\begin{align}
    S_{5D} &\supset \lambda_5 \int d^{4}x \int_{0}^{L} d x_{5}  \,\,\psi(X) \chi(X) \phi(X) \nonumber \\
    & \underset{\text{zero-modes}}{=} \lambda_5 \int d^{4}x \int_{0}^{L} d x_{5}  \,\,e^{-m_{5\chi} x_5} e^{- m_{5\psi} x_5} e^{m_{5\phi}(x_5 - L)} \chi_{4}(x) \psi_{4}(x) \phi_{4}(x) \nonumber \\
    &  \underset{m_{5\phi} > m_{5 \psi}+m_{5\chi}}{\sim}  \lambda_5 e^{-(m_{5\chi}+m_{5\psi})L} \int d^{4} x  \,\, \chi_{4}(x) \psi_{4}(x) \phi_{4}(x). 
\end{align}
At low energies, this reduces to plugging in the zero-modes to get the second line. Note that now the $x_5$ integral can explicitly be done, resulting in an effective coupling in the 4D EFT of the zero-modes. Our philosphy will be that the fundamental 5D theory has only modest hierarchies, that is that all mass parameters and couplings are $\sim {\cal O}(1/10) - {\cal O}(10)$ in units of the KK scale $1/L$ say. But at low energies ($m_4 \approx 0$) the 4D EFT can readily generate exponential hierarchies.
 I illustrate this on the third line with the robust possibility that $m_{5 \phi} > m_{5 \psi} + m_{5 \chi }$. We see that in this case, the effective 4D trilinear coupling can be exponentially small given that the $m_5 L$ are just modestly large,
\begin{align}
    \lambda_{4,\text{eff}} \sim  \lambda_{5} e^{-(m_{5\chi}+m_{5\psi})L}.
\end{align}

 Let us apply this result to a toy model of SM quark Yukawa couplings, where ignoring spin and chirality details, $\chi$ represents the $i$th generation of a quark electroweak doublet, $\psi$ represents the $j$th generation of a  quark electroweak singlet, and $\phi$ represents the  Higgs doublet field. These SM fields start as fundamentally 5D fields, but it is only their 4D zero-modes that have been discovered. Then the trilinear coupling $\lambda$ we just studied represents the Yukawa coupling $Y_{ij}$. We therefore see that starting from a fundamental 5D Yukawa matrix of couplings $Y_{5ij}$ which are more or less randomly distributed without large hierarchies, one predicts an exponentially hierarchical effective 4D Yukawa matrix, 
\begin{align}
    Y_{4,\text{eff}}^{ij} \sim Y_{5}^{ij} e^{-(m_{5i}+m_{5j})L}.
\end{align}

This is a very interesting structure qualitatively. If the 5D masses are not particularly degenerate, one can approximately diagonalize this Yukawa matrix to give quark mass eigenvalues and CKM mixing angles
\begin{align}
    m_{4,i} &\sim e^{-2 m_{5i}L} 
    v_{\rm electroweak}\nonumber \\
    V_{ij}^{\text{CKM}} &\underset{m_{5i}>m_{5j}}{\sim} \frac{e^{-m_{5i}L}}{e^{-m_{5j}L}} \sim \sqrt{\frac{m_{4i}}{m_{4j}}}.
\end{align}
This is not a bad qualitative fit to the hierarchical quark mass and CKM structure and the correlations between them that we observe! 

To conclude,   we have just found an attractive mechanism for understanding the hierarchical form of the Yukawa structure that would otherwise remain mysterious in the SM. 
See \cite{flav-hier-extraDim} for a rapid review of obtaining realistic flavor hierarchies from extra-dimensional wavefunction overlaps, plus original references. 

\subsection{Boundary-localized fields and Sequestering}\label{subsec:bound-loc-sequestering}
\begin{figure}[hbt!]
    \centering
    \includegraphics[width=0.57\linewidth,trim={8cm 6cm 8cm 5.8cm},clip]{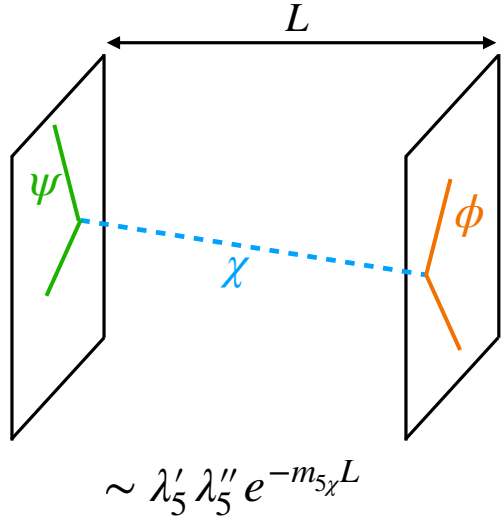}
    \caption{Two effectively boundary-localized fields, $\psi_4$ and $\phi_4$, interacting via exchange of a bulk $\chi$ field. This exchange is Yukawa-suppressed if the bulk mass is large $m_5 L \gg 1$. }
    \label{fig:sequestered-diagram}
\end{figure}
There are a couple of more  concepts to introduce in 5D, which will be important later. Let us return to $\chi, \psi, \phi$, keeping them general (without identifying them with quarks and Higgs fields). The first concept is the approximation of boundary-localization (or ``brane-localization''), which applies when $m_5 L \gg 1$. In this limit clearly the zero-modes are closely stuck close to one or other of the boundaries, and one can approximate them as propagating exclusively in 4D, either restricted to $x_5 = 0$ or $x_5 = L$. Given the exponential profiles of zero modes, this approximation kicks in quickly, so $m_5 L$ need not be too big. Having seen how we can approach this boundary-localization limit, we can simply impose boundary-localization as fundamental. For example, we can take $\psi_4$ and $\phi_4$ to be perfectly localized at $x_5= 0$ and $x_5 = L$ respectively. Consequently, they can each self-interact with themselves without suppression, but they cannot directly interact with each other by locality, since they live on the two separated boundaries. But we first take  $m_{5 \chi} L > 1$ to not be too large. This allows $\psi$ and $\phi$ to interact by exchange of $\chi$ if they have suitable trilinear couplings,
\begin{align}
    \lambda'_{5} \int d^{4} x \int_{0}^{L} d x_5 \,\psi^{2}(X) \chi(X) &\sim \lambda'_{5} \int d^{4} x  \, \psi^{2}_{4}(x) \, \chi(x, x_5 =0) \nonumber \\
    \lambda''_{5} \int d^{4} x \int_{0}^{L} d x_5 \,\phi^{2}(X) \chi(X) &\sim \lambda''_{5} \int d^{4} x   \, \phi^{2}_{4}(x) \, \chi(x, x_5 = L).
\end{align}
Therefore the $\psi-\phi$ interaction is given in terms of the 5D $\chi$ exchange,
\begin{equation}
    \propto \langle 0 | T \{ \chi(x, x_5 = 0)~ \chi(x', x_5' = L) \} | 0 \rangle \propto 
e^{- m_{5 \chi}L}.
\end{equation}
Without needing the detailed form of the 5D propagator, 
this exponential is just the Yukawa-suppression one expects whenever a massive mediator field  has to virtually traverse a distance $L$ beyond its Compton wavelength at low energies ($< m_{5 \chi}$).
The full exchange is  depicted in \Fig{fig:sequestered-diagram}.  Obviously, for $m_{5 \chi} L \gg 1$, the $\psi-\phi$ interactions effectively shut off. 
Later in the SUSY context, we will use this natural mechanism for suppressing interactions between sets of boundary-localized 4D fields. It is known as ``sequestering'' \cite{sequestering}.

\subsection{Non-renormalizability of Higher-dimensional EFT}

Higher-dimensional field theories are non-renormalizable.\footnote{Scalar field theory with only trilinear couplings is renormalizable but such a cubic potential is unbounded from below and therefore unphysical.} 
Consider Yang-Mills theory, which in 4D is the classic renormalizable QFT. In 5D, 
\begin{align}
    S_{5D} = -\frac{1}{4 g_{5}^{2}} \int d^4 x \int_{0}^{L} d x_{5} G_{MN}^{a} G_{a}^{MN} ,
\end{align}
where we have chosen for later convenience the non-canonical normalization of the gauge fields $A^a$ such that the coupling appears out the front of the action rather than in the interaction terms, 
\begin{align}
    D_{M} &= \partial_{M} + i t^{a} A_{M}({\rm x}) \nonumber \\
    G_{MN}^{a} &= \partial_{M} A_{N}^{a} - \partial_{N} A_{M}^{a} + f^{abc} A_{M}^{b} A_{N}^{c}.
\end{align}
(The field redefinition $A_M \rightarrow g_5 A_M$ restores canonical normalization.) 
We see that the 5D gauge coupling $g_5$ has mass dimension $-1/2$ and therefore the theory is non-renormalizable in the 5D regime. 

But this behavior changes below $\mu_{KK}$ for the 4D EFT.
Since $m_5 = 0$ by 5D  gauge invariance, the zero-modes have a 
$x_5$-independent profile, so that doing the $x_5$ integral for these modes yields
\begin{align}
    S_{4,\text{eff}} = -\frac{L}{4 g_{5}^{2}} \int d^4 x \, G_{\mu \nu}^{a} G^{\mu \nu}_{a} + A_{5}\text{-terms} \,.
\end{align}
We see that the effective 4D gauge coupling is now dimensionless and given by
\begin{align}\label{eq:4D-5D-gaugeCoupling}
    \frac{1}{g_{4,\text{eff}}^{2}} =  \frac{L}{g_{5}^{2}} \,.
\end{align}
This 4D Yang-Mills is asymptotically free, $g_{4, \text{eff}}$ running to become stronger in the IR.  
 
Non-renormalizability of 5D field theories is not a disaster, it however does mean that  they have to be treated by the methods of non-renormalizable EFT \cite{eft}, with a finite energy range of validity above which some more UV-complete description must take over.

\subsection{The Ultimate $m \rightarrow 0$ Limit}

SUSY has been motivated from the top down, as a possible remnant of a superstring realization of quantum gravity, as well as from the bottom up if Nature gives us every spin of elementary particle possible with $m \approx 0$, including spin $3/2$. But so far, extra spacetime dimensions have only
been motivated from the top down. Yet they too have a robust bottom-up motivation. 
 I have surveyed the different possible particle spins {\it one at a time} in terms of what mechanisms and symmetries robustly yield $m = 0$, and then the sense in which these structures might only be approximate, resulting in 
 $m \approx 0$. I could do this at weak coupling because then each particle  approximately corresponds to a free field. But what about strong coupling and $m \rightarrow 0$? We have already seen that non-perturbatively, there can be emergent mass scales via dimensional transmutation. If we really wish to go all the way to a completely massless theory non-perturbatively, both explicit mass scales and emergent mass scales must be absent. In particular, this requires a theory where even the dimensionless couplings do not run, $\beta(\alpha) = 0$. Lacking characteristic mass scales such a theory also lacks characteristic length scales, and therefore has scale-invariance as a new symmetry. There is a strong conjecture that scale-symmetry combined with Poincare symmetry  and the locality of QFT ``accidentally'' implies an even larger symmetry, namely conformal symmetry. See for example \cite{cft-conjecture}. A 4D QFT with conformal symmetry is called a conformal field theory (CFT). 

Remarkably, the famous AdS$_5$/CFT$_4$ correspondence re-expresses 4D CFT ``holographically'' as an equivalent (or ``dual'') 5D quantum gravity theory \cite{ads-cft}! The 5D theory realizes the conformal symmetry as the symmetry (isometry) of the background curved spacetime geometry, 5D Anti-de Sitter (AdS):
\begin{align}
    ds^{2}_{\text{AdS}_{5}} = R^2 \,\,\frac{\eta_{\mu\nu} dx^{\mu} dx^{\nu} - dx_{5}^{2}}{x_{5}^{2}} \, , \quad x_5 > 0 \nonumber \\
  \text{where }  R= \text{constant AdS radius of curvature} \,.
\end{align}
This is analogous to how the 5D Poincare group is the symmetry of 5D Minkowski spacetime geometry. Of course, such a CFT can only be a subsector of the real world, since we know we have 4D GR with its characteristic mass scale $M_{Pl}$, and the SM at least. If one includes these elements of realism the holographically dual description is given by the Randall-Sundrum II (RS2) scenario \cite{rs2}. 

It is also possible not to have an exact CFT (exactly massless strongly interacting QFT in the IR) but one in which conformal symmetry is approximate and broken spontaneously or softly in the IR ($\ll M_{Pl}$) as well as in the UV ($\sim M_{Pl}$). These have holographic duals called Randall-Sundrum I (RS1) models or ``warped extra dimensions'' \cite{warped-extra-dim}. They are similar to the extra-dimensional framework we have sketched in the earlier subsections, the major difference being that the bulk 5D spacetime is highly curved in a manner that however cannot be detected if one does not have the energy to resolve the extra dimension. This is the meaning of ``warped'' in this context. Much of the modeling of the Composite Higgs paradigm currently takes place in this 5D EFT framework \cite{extra-dim}. The warped variant of generating Yukawa hierarchies from extra-dimensional wavefunction overlaps is CFT/AdS dual to the mechanism of ``Partial Compositeness'' 
in purely 4D \cite{flav-at-SSC} \cite{partial-compositeness}, where the Yukawa hierarchies are created by strong-coupling non-perturbative RG effects, closely related to the physics of dimensional transmutation we have discussed.

The advantage of the 5D dual RS-level descriptions is that one does not need to completely specify the 4D CFT in detail, which would be equivalent to specifying the 5D quantum gravity in detail. Rather, one can describe the 5D theory with non-renormalizable EFT. It is easier (but still non-trivial) to check self-consistency of a 5D EFT than to check that one has a viable and UV-complete 5D quantum gravity! In this way, one can explore interesting physics that strongly coupled 4D theories might produce. For example, the possibility of traversable wormholes is explored within an RS2-like framework in \cite{traversable-wormholes}. The Composite Higgs scenario is conveniently explored within the RS1 framework.  

\section{Superspace and Superfields}

With extended spacetime now being motivated as a convenient and simple way of representing theories with extended spacetime symmetries, we return to the extended spacetime symmetry of ${\cal N} =1$ SUSY. We will try to write SUSY QFTs in terms of ``superfields'' $\Phi(X)$, fields living on an extended spacetime, ``superspace'' \cite{sugra-and-susy, mssm-review}, which has SUSY as its geometric symmetry algebra:
\begin{align}
    X &\equiv (x^{\mu},\theta_{\alpha}, \overline{\theta}_{\beta}). 
\end{align}
The extra coordinates allow us to represent the action of the supercharges in the simplest possible way, as ``supertranslations'', akin to the action of translations on Minkowski coordinates:
\begin{align}
    \theta_{\alpha} &\rightarrow \theta_{\alpha} + \xi_{\alpha} \,, \, \, \overline{\theta}_{\beta} \rightarrow \overline{\theta}_{\beta} + \overline{\xi}_{\beta}.
\end{align}
The $\xi_{\alpha}$ are the transformation parameters we associate with the action of $Q_{\alpha}$ (analogous to a translation vector $a^{\mu}$ associated to the translation generator $P_{\mu}$), and therefore must be spinorial. They are independent of $x$ because we are studying the global limit of SUSY here, not local SUSY. Furthermore, since the $Q$ are anticommuting charges (ulimately coming from the fact that they come from a current that couples to the fermionic spin-$3/2$ gravitino) the $\xi_{\alpha}$ must be Grassmann numbers. For compatibility, the $\theta_{\alpha}$ must also be spinor Grassmann coordinates, 
unlike the usual c-number $x^{\mu}$ coordinates.
This is what will separate SUSY from standard extra dimensions. Matching the conjugate relationship of $Q$ and $\bar{Q}$, $\bar{\theta}_{\alpha}$ is just the conjugate of $\theta_{\alpha}$. 

In order for superspace to exhibit the ``non-abelian'' aspect, \Eq{eq:chargeAlgebra1}, of SUSY, it is crucial that the $x^{\mu}$ also transform under the supercharges: 
\begin{align}
    x^{\mu} &\rightarrow x^{\mu} - i \theta_{\alpha} \overline{\xi}_{\beta} \sigma_{\alpha\beta}^{\mu} + i \xi_{\alpha} \overline{\theta}_{\beta} \sigma_{\alpha\beta}^{\mu}
\end{align}
Note that the two transformation terms are hermitian conjugates of each other, keeping $x$ ``real'', or, more correctly hermitian as an operator. To check these symmetry transformations, 
we first recall how ordinary (infinitesimal) translations work on  fields on ordinary spacetime, $\phi(x) \rightarrow \phi(x + a) 
= \phi(x) + a^{\mu} \partial_{\mu} \phi(x)$. We thereby 
identify the associated  $4$-momentum charge as  the translation generator
$P_{\mu} = i \partial_{\mu}$ (the ``$i$'' for hermiticity). 
 Similarly,  supercharges encoding infinitesimal supertranslations on superfields on superspace
  are represented as differential operators on superspace:
\begin{align}
    Q_{\alpha} &= \frac{\partial}{\partial \theta_{\alpha}} +i \overline{\theta}_{\gamma} \sigma_{\alpha \gamma}^{\mu} \partial_{\mu} \nonumber \\
    \overline{Q}_{\beta} &= \frac{\partial}{\partial \overline{\theta}_{\beta}} + i \theta_{\gamma} \sigma_{\gamma\beta}^{\mu} \partial_{\mu} \,. 
\end{align}
The first of these encodes the transformation of superspace by $\xi$, and the second by $\bar{\xi}$. In each case, the first term on the right encodes infinitesimal translation of the $\theta$ coordinates. The second term encodes the fact that $x^{\mu}$ also receives a $\bar{\theta}$ (or $\theta$) dependent infinitesimal translation. You can check that these supercharges indeed obey the superalgebra of eqs.~(\ref{eq:chargeAlgebra1}), (\ref{eq:chargeAlgebra2}), and (\ref{eq:QP-commutators}). 

Fortunately, we will only have explicit need of the simplest type of superfield, namely a scalar field on superspace, $\Phi(X)$, where by ``scalar'' I mean that only its spacetime argument transforms under SUSY as described above. 

\section{Chiral Superspace and Chiral Superfields}

We have seen how SUSY acts on the three types of superspace coordinates $x^{\mu}, \theta_{\alpha}, \bar{\theta}_{\beta}$. Remarkably, there is a kind of ``projection'' of full superspace down to ``chiral superspace'' which is 
closed under SUSY transformations. It has 
just two types of coordinates $Y\equiv (y^{\mu}, \theta_{\gamma})$, where 
\begin{align}
    y^{\mu} \equiv x^{\mu} - i \theta_{\alpha} \sigma_{\alpha\beta}^{\mu} \overline{\theta}_{\beta},
\end{align}
in terms of the original supercoordinates. Note that because of the ``$i$'', the $y^{\mu}$ coordinate is not real (hermitian). Given the original superspace transformations, it is straightforward to check that under SUSY, 
\begin{align}\label{eq:y-theta-transfm}
    (y^{\mu},\theta_{\gamma}) \rightarrow (y^{\mu} - 2 i \theta_{\alpha} \sigma_{\alpha\beta}^{\mu} \overline{\xi}_{\beta} , \theta_{\gamma} + \xi_{\gamma}),
\end{align}
independently of $\bar{\theta}$. 

We can therefore define a special kind of scalar superfield which only depends on chiral superspace, $\Phi(y^{\mu}, \theta_{\gamma})$. Clearly, under SUSY transformations such a ``chiral superfield'' retains its form, that is independent of $\bar{\theta}$ except implicitly within $y$. 

In this simpler context, it is time to make a central point, that superfields are really just a finite collection of ordinary component fields which are in the same supermultiplet, that is they transform into each other under SUSY. To see this we simply Taylor expand the $\theta$ dependence of the chiral supefield, 
\begin{align}\label{eq:Phi-expansion-y-theta}
    \Phi(y,\theta) = \phi(y)+ \sqrt{2} \psi_{\alpha}(y)\theta^{\alpha} +F(y) \theta^2
\end{align}
Because there are only two Grassmann coordinates  here, $\theta_{\alpha}$, and each of their squares vanishes because of their anticommuting with themselves, the Taylor expansion can not contain more than one power of each Grassmann coordinate. In particular the highest power of $\theta$s in the Taylor expansion must be $\theta_1 \theta_2 = - \theta_2 \theta_1$, or equivalently in the manifestly Lorentz invariant form $\theta^2 \equiv \epsilon^{\alpha \beta} \theta_{\alpha} \theta_{\beta}$. With a finite Taylor expansion, there are a finite number of Taylor coefficients which are fields of $y^{\mu}$ alone. To make $\Phi$ Lorentz-scalar, $\phi$ and $F$ must be complex Lorentz scalars, while $\psi_{\alpha}$ must be a {\it chiral} two-component spinor. 

For some purposes, this $y$ notation is compact and makes SUSY transformations look simpler, but in general, we will want to write the fields in terms of $x$, corresponding to real spacetime. That means each component field has to be further Taylor expanded in the $-i \theta \sigma^{\mu} \bar{\theta}$ deviation of $y$ from $x$: 
\begin{align}\label{eq:Phi-expansion-x-theta}
    \Phi(y^{\mu} = x^{\mu} -i \theta \sigma^{\mu} \overline{\theta},\theta) &= \phi(x) - i \theta \sigma^{\mu} \overline{\theta} \partial_{\mu}\phi(x) - \frac{1}{4}  \theta^{2} \overline{\theta}^{2} \square\phi(x) \nonumber \\
    & \quad + \sqrt{2} \psi_{\alpha}(x)\theta^{\alpha} + 
    \frac{i}{\sqrt{2}} \theta^{2} \overline{\theta}_{\alpha} \sigma^{\mu}_{\alpha\beta} \partial_{\mu}\psi_{\beta}(x) + \sqrt{2} F(x) \theta^{2}.
\end{align}


Obviously there is a totally analogous conjugate notion of anti-chiral superspace, $\{ \bar{Y} \equiv (\bar{y}^{\mu} = x^{\mu} + i \theta_{\alpha} \sigma^{\mu}_{\alpha \beta} \bar{\theta}_{\beta}, \bar{\theta}_{\gamma})$, and the analogous anti-chiral superfield $\bar{\Phi}(\bar{y}, \bar{\theta})$.  We need this because it houses the conjugates of the component fields of the chiral superfield. 

\section{The Wess-Zumino Model} 

Finally, we are in position to write an actual simple renormalizable SUSY QFT, with remarkable properties. 
It is built from a single chiral superfield $\Phi(Y)$ (and of course its anti-chiral conjugate $\bar{\Phi}(\bar{Y}$). 

The question is how to build a SUSY invariant action. First recall how we build Poincare invariant actions, 
\begin{align}
    S = \int d^{4} x \,\, {\rm Lorentz-scalar}(x).
\end{align}
The integration measure $d^4x$ is Poincare invariant and the Lagrangian integrand is a (composite) Lorentz-scalar field, which upon integration over all its translations (all values of $x$) becomes Poincare-invariant. We build SUSY-invariant actions the same way, find a SUSY-invariant integration measure and have the integrand be a (composite) scalar superfield: 
\begin{align}
    S = \int d^{4} x \int d^{2} \theta \int d^{2} \overline{\theta} \,\, K(\Phi(Y),\overline{\Phi}(\overline{Y})) + \int d^{4} y \int d^{2} \theta \,\, W(\Phi(Y)) + \int d^{4} \overline{y} \int d^{2} \overline{\theta }\,\, \overline{W}(\overline{\Phi}(\overline{Y})).
\end{align}
Let us begin with the first term. 
Here, the Grassmann integral measure, $d^2 \theta \equiv \theta_1 d\theta_2 = 
1/2 \epsilon^{\alpha \beta} d \theta_{\alpha} d \theta_{\beta}$ is Lorentz invariant just like $d^4 x$. This and the conjugate integral measure $\int d^2 \theta \int d^2 \bar{\theta}$ are usually abbreviated to ``$\int d^4 \theta$''.  You can check easily that the full measure 
$\int d^4x d^4 \theta$  is invariant under supertranslations, since these look like   $x$-independent translations of $x$  and simple translations of $\theta, \bar{\theta}$. Since products of scalar fields are scalar fields, including scalar superfields on superspace,  we can make the integrand by (sums of) products of the superfields $\Phi, \bar{\Phi}$. This integrand is called the Kahler potential, $K$. We require that $K$ is hermitian so that the Lagrangian and Hamiltonian are. 

The action made from the Kahler potential would be SUSY invariant if $\Phi$ were any  scalar superfield.  But the fact that $\Phi$  is a chiral superfield offers another 
possibility in the second term of the action. Here the integrand is only  made of sums and  products of the chiral superfield, so that $W$, the ``superpotential'', is also a composite chiral superfield on chiral superspace. Therefore the integration measure is only over chiral superspace coordinates. Again, it is straightforward to check the measure's SUSY invariance given \Eq{eq:y-theta-transfm}. The third term is just the conjugate of the second to keep the sum hermitian.

The above action is the most general SUSY action with the fewest explicit derivatives, namely none, which is why both $K$ and $W$ are called ``potentials''. 
But derivatives will indeed arise from the Taylor expansion of the superfield into component fields, \Eq{eq:Phi-expansion-x-theta}. 

The fact that the superpotential is a function of $y$ but not $\bar{y}$ means we can shift the integration variable from $y$ to just $x$, 
\begin{align}
    \int d^{4} y \int d^{2} \theta \,\, W(\Phi(y,\theta)) = \int d^{4} x \int d^{2} \theta \,\, W(\Phi(x,\theta)).
\end{align}
As can be seen in \Eq{eq:Phi-expansion-y-theta}, there are no derivatives hidden in the Taylor expansion of $\Phi(x, \theta)$, so the superpotential part of the action does not give rise to derivatives, these only arise from the Kahler potential which depends on both $y$ and $\bar{y}$ which cannot both be shifted to $x$ simultaneously.

It is notable that $W$ is a complex analytic, or ``holomorphic'', function of the complex chiral superfield because it does not depend on the conjugate anti-chiral superfield. This yields deep insights peculiar to SUSY, which we only touch on later.

Let us do some quick dimensional analysis to write a renormalizable theory. Given the SUSY algebra we see that the Grassmann coordinates $\theta, \bar{\theta}$ have mass dimension $-1/2$. Given that the component spinor fermion will turn out to be a canonical fermion with mass dimension $3/2$, and  
 component scalar field $\phi(x)$ will be a canonical scalar field, the superfield $\Phi$ has mass dimension $1$. 
 The component scalar field $F$ has the unusual dimension $2$, and we will see why soon. The Grassmann measure $d \theta$ has the {\it opposite} dimension to the Grassmann coordinate, namely $+1/2$, to satisfy the axiomatic $\int d \theta~\theta = 1$. This means $K$ has dimension $2$, and $W$ has dimension $3$. In order to not use couplings with negative mass dimensions, the standard indicator of non-renormaliability, we are restricted to: 
\begin{align}
    K =\overline{\Phi}\Phi , \quad W=\frac{\lambda}{3} \Phi^{3} + \frac{m}{2} \Phi^{2}.
\end{align}
 Note, $K$ needs both chiral and antichiral fields to be non-trivial, so dimension $2$ restricts it to the above, and it has coefficient $1$ by choice of the normalization of $\Phi$. $W$ could clearly be any cubic polynomial in $\Phi$, but a constant term would not survive the $\int d^2 \theta$ integration, and a linear term could be redefined away by a $\Phi \rightarrow \Phi +$ constant shift.\footnote{This is as long as there are $\Phi^2$ or higher order terms. We will study a case where this is not true soon.}  This renormalizable structure is the Wess-Zumino model.

 We can now write this out in terms of the component fields using \Eq{eq:Phi-expansion-x-theta}, and then do the Grassmann integrations: 
\begin{align}
    S &= \int d^{4}x \{ \partial_{\mu} \overline{\phi}(x) \partial^{\mu}\phi(x) + \overline{\psi} i\sigma^{\mu}\partial_{\mu} \psi(x) + \overline{F}F(x) \nonumber\\
    & \quad - \lambda \phi(x) \psi^{2}(x) - \frac{m}{2} \psi^{2}(x) + (\lambda \phi^{2} + m \phi) F(x) + h.c \}.
 \end{align}
We have done an integration by parts to put it in this form.
  Here the first line comes from $K$ and the second from $W$. We see that the dimension $2$ scalar fields $F, \bar{F}$ appear quadratically and without derivatives. If you think of this action in a path integral, this means that you can easily do the Gaussian integrals over 
$F, \bar{F}$ by ``completing the square''. It is entirely equivalent to just solving the $F, \bar{F}$ equations of motion for these ``auxiliary'' fields in terms of the other fields, and plugging the result back into the action. The result is
\begin{align}
    S = \int d^{4}x \{ \partial_{\mu} \overline{\phi} \partial^{\mu} \phi + \overline{\psi} i \sigma^{\mu} \partial_{\mu} \psi - |m \phi + \lambda \phi^{2}|^{2} - ( \frac{m}{2} +  \lambda \phi) \psi^{2}(x) +h.c. \}.
\end{align}
We have arrived at a QFT with renormalizable interactions consisting of Yukawa interactions between fermion and scalar, and scalar trilinear and quartic self-interactions. But the various couplings and masses are strongly correlated, as dictated by its manifestly SUSY construction.  For example the quartic scalar coupling is the square of the Yukawa coupling. But in particular, we see the anticipated SUSY degeneracy of fermion and boson superpartners, 
$m_{\phi} = m_{\psi} = m$. SUSY guarantees this structure is maintained radiatively, upon renormalization the counterterms must also have this special form. The Wess-Zumino model finally shows us that the abstract SUSY algebra,  which we deduced from general considerations, is actually realizable within an interacting QFT.

\subsection{Robust interacting massless scalar}

We can further consider the massless limit $m \rightarrow 0$:
\begin{align}
    S \underset{m \rightarrow 0}{\rightarrow} \int d^{4}x \{ \partial_{\mu} \overline{\phi} \partial^{\mu} \phi + \overline{\psi} i \sigma^{\mu} \partial_{\mu} \psi - \lambda|  \phi|^{4} - \lambda \phi \psi^{2}(x) +h.c. \}.
\end{align}
Note that this limit is robust, because there is a new chiral symmetry when $m =0$, 
\begin{align}
    \psi \rightarrow e^{-i \alpha/3} \psi , \quad \phi \rightarrow e^{+2 i \alpha/3} \phi.
\end{align}
Such chiral symmetries can make fermions robustly massless in Yukawa QFTs, but what is special is that the robust masslessness extends to the scalar because of SUSY degeneracy. In this way, we have an interacting massless (or light) scalar with order-one coupling (with only logarithmic running).  SUSY is the only known robust mechanism for such scalars, at least in the perturbative regime. 

\subsection{R-symmetries}

The chiral symmetry is at first sight a  bit strange in that it acts differently on the two superpartners, in particular it appears we cannot assign the entire superfield $\Phi(X)$ a charge under this symmetry. Indeed, this cannot be done if we think of it as an ``internal'' symmetry which does not act on the superspace $X$ argument of the superfield. But if we also rotate the complex Grassmann coordinates, $\theta \rightarrow e^{i \alpha} \theta, 
\bar{\theta} \rightarrow e^{-i \alpha} \bar{\theta}$, then we see that by \Eq{eq:Phi-expansion-y-theta}, we can assign $\Phi$ intrinsic charge $2/3$. Such symmetries which rephase $\theta$ are known as R-symmetries. They are not required by SUSY, for example the massive Wess-Zumino model is supersymmetric but has no R-symmetries. 
(But even when absent the nature of their violation can be useful to track.)

\subsection{Non-renormalizable SUSY EFT}

We can also take $K$ and $W$ to be more general, in which case they describe a non-renormalizable but supersymmetric EFT. We can again Taylor expand $\Phi$ in Grassmann coordinates, \Eq{eq:Phi-expansion-x-theta}, and use that to then Taylor expand $K$ and $W$.
Doing the various straightforward Grassmann coordinate integrations (and some integrations by parts) yields the action in terms of the component fields, 
\begin{align}\label{eq:susy-eft-L}
    {\cal L} &= \frac{\partial^2 K}{\partial \Phi \partial \overline{\Phi}} (\phi(x))  \lbrace |\partial_\mu \phi|^2 + \overline{\psi} i \sigma \cdot \partial \psi + |F|^2 \rbrace + \frac{1}{4} \frac{\partial^4 K}{\partial^2 \Phi \partial^2 \overline{\Phi}} \psi^2 (x) \psi^2 (x) \nonumber \\
   & \quad + \left( \frac{\partial^3 K}{\partial^2 \Phi \partial \overline{\Phi}} (\frac{i}{2} \bar{\psi} \sigma^{\mu} \psi \partial_{\mu} \phi + \psi^2 \overline{F}) + \frac{\partial^2 W}{\partial \Phi^2} (\phi(x)) \psi^2 + \frac{\partial W(\phi)}{\partial \Phi} F + h.c.\right) \nonumber\\
   & \overset{\text{solve} }{\underset{F, \overline{F} \text{ eqs}}{\longrightarrow}} \frac{\partial^2 K}{\partial \Phi \partial \overline{\Phi}}(\phi(x)) \lbrace |\partial_\mu \phi|^2 + \psi i \sigma \cdot \partial \psi\rbrace + \left( \frac{\partial^2 W}{\partial \Phi^2}(\phi) \psi^2 (x) + h.c.\right) \nonumber \\
   & \quad + \left( \frac{\partial^3 K}{\partial^2 \Phi \partial \overline{\Phi}} (\frac{i}{2} \bar{\psi} \sigma^{\mu} \psi \partial_{\mu} \phi + h.c. \right) \nonumber \\
   & + \frac{1}{4} \frac{\partial^4 K}{\partial^2 \Phi \partial^2 \overline{\Phi}} \psi^2 (x) \psi^2 (x) \quad - \left( \left| \frac{\partial W}{\partial \Phi} (\phi) + \frac{\partial^3 K}{\partial^2 \overline{\Phi} \partial \Phi} \overline{\psi}^2\right|^2 \right) \Bigg/ \left( \frac{\partial^2 K}{\partial \Phi \partial \overline{\Phi} (\phi)}\right).
\end{align}
In the second step we have again integrated out the auxiliary fields $F, \bar{F}$. 

This is the most general SUSY EFT of the chiral superfield at $2$-derivative order ($1$-derivative in fermions). We will give a simple application of this structure in the next section on SUSY breaking. Terms with more derivatives would require generalizing the SUSY Lagrangian beyond $K, W$ to integrands with explicit  superspace derivatives. But in EFT the higher derivatives would be less important in the IR and therefore can be consistently dropped.

\section{An EFT of Spontaneous SUSY Breaking}\label{sec:spontSUSYbreak-eft}

We can now give the simplest example of spontaneous breaking of SUSY, with a non-renormalizable EFT. This is somewhat analogous to the well-known non-linear $\sigma$-models that describe spontaneous breaking of ordinary (non-abelian) internal symmetries in non-renormalizable EFT, which may then be further UV-completed at higher energies. 

Before specifying the theory, it is worth seeing one general implication of spontaneous SUSY breaking. By taking the spinor trace of \Eq{eq:chargeAlgebra1}, we see that the Hamiltonian in SUSY is a sum of squares of supercharges and therefore positive semi-definite\footnote{It is important for this that we are restricting ourselves to the $G_N \rightarrow 0$ limit in which SUGRA decouples and SUSY is a global symmetry.}:
\begin{align}
    H = |Q_1|^2 + |Q_2|^2 \geq 0.
 \end{align}
If the vacuum breaks SUSY, then at least one supercharge must not annihilate it, 
\begin{align}
    Q_1 |0\rangle \text{ or } Q_2 |0\rangle \neq 0,
\end{align}
which then implies that
\begin{align}
  \langle 0 |  H |0\rangle \underset{\text{pert.}}{=} V_{\text{effective}}  >0.
\end{align}
In a perturbative theory, this means that the minimum of the effective potential energy will be positive. 

In honor of non-supersymmetric $\sigma$ models, we will call our chiral superfield $\Sigma(y, \theta) = \sigma(y) + \psi_{\Sigma} \theta + F_{\Sigma} \theta^2$, rather than the generic ``$\Phi$''. Specifically, we take
\begin{align}
    K &= \overline{\Sigma} \Sigma - \frac{\overline{\Sigma}^2 \Sigma^2}{4 M^2}, \quad M\sim {\cal O}(M_{Pl}) \nonumber \\
    W &= \Lambda^2 \Sigma.
\end{align}
We note that $K$ has a non-renormalizable term, with a negative-dimension coupling. We will take the scale of nonrenormalizability to be Planckian, $M \sim {\cal O}(M_{Pl})$, so that it only requires UV-completion in the ultimate quantum-gravity/string-theory regime. Also note that we have a linear superpotential, and it cannot be redefined away as we did in the Wess-Zumino model because that required quadratic or cubic terms to do.

\begin{figure}[hbt!]
    \centering
    \includegraphics[width=0.57\linewidth,trim={2cm 3cm 2cm 3cm},clip]{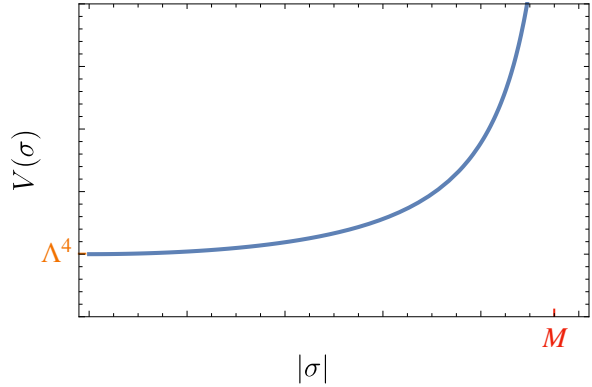}
    \caption{The scalar potential for our SUSY EFT has positive minimum, indicating spontaneous SUSY breaking.}
    \label{fig:SigmaPotLabled}
\end{figure}

From \Eq{eq:susy-eft-L}, we read off the scalar potential, 
\begin{align}
    V(\sigma) = \left| \frac{\partial W}{\partial \Sigma}\right|^2(\sigma) \Bigg/ \frac{\partial^2 K }{\partial \Sigma \partial \overline{\Sigma}}(\sigma)  = \frac{|\Lambda|^4}{1- \frac{|\sigma|^2}{M^2}},
\end{align}
illustrated in \Fig{fig:SigmaPotLabled}. Clearly, it has a local minimum  at the origin of field space, $\langle \sigma \rangle = 0$, and indeed it is positive, 
\begin{align}
    \langle \sigma \rangle =0 , \quad V_{\text{vacuum}} = |\Lambda|^4 >0.
\end{align}
 You may also notice that, taken literally, the potential can become negative for $|\sigma| > M$, suggesting that $\sigma =0$ is not the true ground state and that in fact the potential is unbounded from below, which is unphysical. However, we cannot trust such a conclusion because 
$|\sigma| > M$ lies outside strict EFT control since the EFT is breaking down at scales of order $M$. To know what really happens at large $\sigma$ would require the UV completion of this EFT to all scales. Such renormalizable SUSY breaking models  do indeed exist, but we will not need them here. In this non-renormalizable EFT, we can still conclude that $\sigma = 0$ is potentially a
  metastable (on cosmological timescales) or absolutely stable vacuuum, depending on the details of a full UV complete extension of the EFT. Either way, $\sigma =0$ has positive vacuum energy and therefore represents spontaneous SUSY breaking (at least for cosmological timescales). 

We can go back and calculate $\langle F_{\Sigma} \rangle$ from the auxiliary field equation of motion, 
\begin{align}
    F_{\Sigma} = \frac{\Lambda^{* 2}}{1-\frac{|\sigma|^2}{M^2}} ,\quad \langle F_{\Sigma} \rangle = \Lambda^{* 2} \neq 0.
\end{align}
Non-vanishing $\langle F \rangle$ VEVs are order-parameters for spontaneous SUSY breaking. This is because when we solve for the auxiliary field $F$, its square (or sum of squares when there are several chiral superfields) contributes to the effective  potential energy.

Phenomenologically, the thing that stands out is that superpartners are no longer degenerate after spontaneous SUSY breaking, 
\begin{align}
    m_{\sigma}^{2} &= \frac{|\Lambda|^4}{M^2} \nonumber \\
    m_{\psi_{\Sigma}} &= 0.
\end{align}

\subsection{The Goldstino and the Gravitino} 

The vanishing of the fermion mass is not specific to this particular model of spontaneous SUSY breaking. Rather it is a parallel of Goldstone's Theorem for ordinary internal symmetries. There, there is a massless scalar robustly arising, but for spontaneous SUSY breaking it is a massless fermion that is robustly predicted \cite{sugra-and-susy}\cite{mssm-review}\cite{susy-book-Terning}. In honor of the parallel this massless fermion is called a Goldstone fermion, or more often, ``Goldstino''.  

Again paralleling internal symmetries, where when the symmetry is gauged the gauge field ``eats'' the Goldstone particle to become massive, here too when we do include SUGRA, the gravitino ``eats'' the Goldstino to become massive \cite{broken-susy-and-sugra}, 
\begin{align}
    m_{\text{gravitino}} = ``m_{3/2}" \sim \frac{\Lambda^2}{M_{Pl}}.
\end{align}
As a result, the massless Goldstino is ultimately not part of the physical spectrum once SUGRA is included, but roughly describes the longitudinal polarizations of the massive gravitino. This is the ``super-Higgs'' mechanism.

We will refer to a sector of particle physics which spontaneously breaks SUSY as a ``hidden sector'' for phenomenological purposes, hidden in the sense that it is taken not to have SM-charged fields within it. The simplest hidden sector, which is mostly what we consider, is the non-renormalizable model we have just discussed with just the chiral gauge-singlet superfield $\Sigma$.

\section{The Renormalizable Minimal Supersymmetric SM (MSSM) \cite{mssm-review}}\label{sec:mssm}

Before we consider the gauge superfields, the SM matter fields can be elevated to chiral superfields,
\begin{align}
    Q_i &= \tilde{q}_{i}(y) + \sqrt{2} q_{i}(y)\theta + F_{q_{i}} \theta^2 \nonumber \\
    L_{i}&= \tilde{l}_{i}(y) + \sqrt{2} l_{i}(y)\theta + F_{l_{i}} \theta^2 \nonumber \\
    U_{i}^{c}&= \tilde{u}_{i}^{c}(y) + \sqrt{2} u_{i}(y)\theta + F_{u_{i}} \theta^2 \nonumber \\
    D_{i}^{c}&= \tilde{d}_{i}^{c}(y) + \sqrt{2} d_{i}(y)\theta + F_{d_{i}} \theta^2 \nonumber \\
    E_{i}^{c}&= \tilde{e}_{i}^{c}(y) + \sqrt{2} e_{i}^{c}(y)\theta + F_{e^{c}_{i}} \theta^2 \nonumber \\
    {\cal H}_{u}&= H_{u}(y) + \sqrt{2} \tilde{H}_{u}(y)\theta + F_{H_{u}} \theta^2 \nonumber \\
    {\cal H}_{d}&= H_{d}(y) + \sqrt{2} \tilde{H}_{d}(y)\theta + F_{H_{d}} \theta^2, \nonumber \\
\end{align}
(almost) as anticipated in \Sec{sec:superpartners}, and given a renormalizable SUSY action which is a straightforward generalization of the Wess-Zumino model.  
The $i$ indices refer to SM generations.
Anticipating gauging, we restrict the theory to respect global internal (non-R) symmetries $SU(3)_{QCD} \times SU(2)_{EW} \times U(1)_{Y}$ prior to their gauging. The minimal superpotential with these properties and which contains all the SM Yukawa couplings in its component expansion is\footnote{In these lectures, for simplicity I will completely neglect neutrino masses.}
\begin{align}
    W_{\text{Yukawa}} = Y_{ij}^{u} U_{i}^{c} {\cal H}_{u} Q_{j}+ Y_{ij}^{d} D_{i}^{c} {\cal H}_{d} Q_{j} + Y_{ij}^{e} E_{i}^{c} {\cal H}_{d} L_{j} \supset Y_{ij}^{e} e_{i}^{c} H_{d} l_{j} + \cdots
\end{align}

The subtlety arises because of the holomorphy of $W$ in chiral superfields. In the SM one uses the Higgs doublet to Yukawa couple to up-type fermions  and its conjugate to couple to down-type fermions, but we cannot use the conjugate anti-chiral superfield in $W$. We are therefore forced to introduce separate Higgs chiral superfields (with conjugate electroweak quantum numbers) for up-type and down-type Yukawa couplings, as indicated.  Therefore there is another electroweak-invariant renormablizable superpotential term just involving these two Higgs doublet superfields, called the ``$\mu$'' term:
\begin{align}
    W_{\mu\text{-term}} = \mu {\cal H}_{u} {\cal H}_{d}.
\end{align}

There are other possible $SU(3)_{QCD} \times SU(2)_{EW} \times U(1)_{Y}$-symmetric renormalizable superpotential terms possible, such as $D^c L Q$, but all of these can be forbidden if we impose another symmetry, ``R-parity'', in which every SM field is parity-even and every superpartner of a SM field is parity-odd.\footnote{Both Higgs scalar fields are even, and their Higgsino superpartners are odd.} R-parity is not strictly necessary, but it is definitely simplifying (my main reason here) and also makes the lightest superpartner stable and therefore potentially a dark matter candidate. The Kahler potential is so tightly constrained by renormalizablility and the SM symmetries, that (after familiar wavefunction diagonalization and normalization) it has the canonical form
\begin{align}
    K = \overline{Q}_{i} Q_{i}+ \overline{L_{i}} L_{i}+ \overline{U}_{i}^{c} U_{i}^{c} + \overline{D}_{i}^{c} D_{i}^{c} + \overline{E}_{i}^{c} E_{i}^{c} + \overline{{\cal H}_{u}} {\cal H}_{u} + \overline{{\cal H}_{d}}{\cal H}_{d}.
\end{align}

\subsection{Gauge Superfields} 

Given that ordinary gauge fields $A_{\mu}$ are real-valued, their superfields including gauginos are found in what are called ``real superfields'', but more precisely they are scalar superfields which are hermitian (and not chiral): 
\begin{align}
    V(x,\theta,\overline{\theta}) = \overline{V}(x,\theta,\overline{\theta}).
\end{align}
They are written ``$V$'' rather than ``$\Phi$'' because they are also sometimes called ``vector superfields'', presumably because they contain the $4$-vector gauge fields, but they are scalar fields on superspace in that only their superspace arguments transform under SUSY. They are also called ``gauge superfields''.

Now turn to charged matter fields. Ordinarily a gauge transformation of a matter field transforms as 
\begin{align}\label{eq:gaugetransform_matter}
    \psi(x) \longrightarrow e^{i \alpha^{a}(x)t^{a}} \psi(x),
\end{align}
where the $t^a$ are relevant matrix-valued generators of the gauge group under which $\psi$ is a charged multiplet. 
But in SUSY these fields, including the gauge transformation itself, must be elevated to superfields. Since our matter fields are chiral superfields, this must also be true of the gauge transformation. We therefore have
\begin{align}
    \Phi(y, \theta) &\longrightarrow e^{\Lambda^{a}(y,\theta)\,t^{a}} \,\,\Phi(y, \theta) \nonumber \\
    \overline{\Phi}(\overline{y}, \overline{\theta}) &\longrightarrow  \overline{\Phi}(\overline{y}, \overline{\theta}) \,\, e^{\overline{\Lambda}^{a}(\overline{y}, \overline{\theta})\,t^{a}} \,.
\end{align}
$\bar{\Phi}$ is taken to be a conjugate row vector of the charged multiplet, while $\Phi$ is a column. 
Note that the gauge transformation $\Lambda^a$ can no longer be real (hermitian), so we can absorb the usual ``i'' in \Eq{eq:gaugetransform_matter} into its definition. 

The need for gauge superfields to ensure gauge invariance of the kinetic terms ($K$) of the matter fields is easy to see. For simplicity consider the renormalizable  abelian case without gauge superfields: 
\begin{align}
    K = \overline{\Phi} \Phi \longrightarrow \overline{\Phi} \,\,e^{\overline{\Lambda}}\, e^{\Lambda} \,\,\Phi(y, \theta).
\end{align}
It is not invariant because $\Lambda \neq - \bar{\Lambda}$, they are not even fields of the same type. This is fixed by introducing the abelian gauge superfield $V$, transforming as
\begin{align}
    V(x, \theta, \overline{\theta}) \longrightarrow  V(x,\theta, \overline{\theta}) - \Lambda(y, \theta) - \overline{\Lambda}(\overline{y}, \overline{{\theta}}).
\end{align}
Straightforwardly then, 
\begin{align}
    S &\supset \int d^{4}x \int d^{4} \theta \, \,\overline{\Phi} e^{V} \Phi
\end{align}
is both gauge-invariant and SUSY-invariant. Even though we will continue to write $K = \bar{\Phi} \Phi$, think of it as short-hand for this gauge-invariant inclusion of $V$ to make the Lagrangian gauge-invariant.

For non-abelian gauge theory, we still have 
\begin{align}\label{eq:S-nonAbelian}
        S &\supset \int d^{4}x \int d^{4} \theta \,\, \overline{\Phi} e^{V^{a} \,t^{a}} \Phi, 
\end{align}
with gauge-invariance following from
\begin{align}
    e^{V^{a} \,t^{a}} \longrightarrow e^{-\overline{\Lambda}^{a} \, t^{a}} \, e^{V^{b} \,t^{b}} \, e^{-\Lambda^{c} \, t^{c}}.
\end{align}

\subsection{Wess-Zumino Gauge}

Ordinary axial gauge $A_3(x) = 0$ is a Lorentz-violating but sometimes useful partial gauge-fixing condition, 
which effectively reduces the number of gauge-field components. A general gauge field can be put in this form by a suitable gauge transformation. In analogy, Wess-Zumino 
gauge is a SUSY-violating but (Lorentz-preserving) partial gauge-fixing which reduces the number of component fields of $V$:
\begin{align}
    V^{a}(x, \theta, \overline{\theta}) = A^{a}_{\mu}(x) \theta_{\alpha} \bar{\sigma}_{\alpha\beta}^{\mu} \overline{\theta}_{\beta} + (\lambda^{a}(x) \theta) \overline{\theta}^{2} + (\overline{\lambda}^{a}(x) \overline{\theta}) \theta^{2} + \frac{1}{2} D^{a}(x) \theta^{2} \overline{\theta}^{2},
\end{align}
where $\bar{\sigma}^i = - \sigma^i, \bar{\sigma}^0 = \sigma^0$.

\subsection{Gauge field strength and gauge field action}

It is possible to construct a gauge-invariant {\it chiral} superfield
as a composite of the gauge superfield $V$, 
\begin{align}
    {\cal W}_{\alpha}^{a} {\cal W}^{a \,\alpha}(y, \theta) &= \frac{1}{2} \lambda_{\alpha}^{a} \lambda^{a\, \alpha} + \cdots + \nonumber \\
    & \quad + (\frac{1}{4} G_{\mu\nu}^{a \, 2} + \frac{i}{4} \tilde{G}_{\mu\nu}^{a} G^{a \, \mu\nu} +  \overline{\lambda}^{a} i \sigma.D \lambda^{a} + \frac{1}{2} D^{a}D^{a}) \theta^{2},
\end{align}
where $\tilde{G}_{\mu \nu} \equiv 1/2 ~\epsilon_{\mu \nu \rho \sigma}G^{\rho \sigma}$.
We will not need the terms linear in $\theta$. 
Each factor ${\cal W}^a_{\alpha}$ is a {\it spinor chiral} superfield generalizing the covariant gauge field strength,  which I have not defined, but we will only need its gauge-invariant ``square'' 
${\cal W}_{\alpha}{\cal W}^{\alpha}$ which transforms under SUSY like the elementary chiral superfields we have discussed. Clearly it has mass dimension $3$. Since it is chiral, we can write a renormalizable action, 
\begin{align}\label{eq:susy-gauge-S}
    S_{\rm gauge} &= \frac{1}{4g^2} \int d^{2}\theta \, {\cal W}_{\alpha}^{a} {\cal W}^{a\, \alpha} + h.c. \nonumber\\
    &= -\frac{1}{4 g^2} G_{\mu\nu}^{a \, 2}  + \frac{1}{g^2} \overline{\lambda}^{a} i \sigma.D \lambda^{a} + \frac{D^{a} D^{a}}{g^2}.
\end{align}
We see that there is a new set of auxiliary fields, $D^a$. 
I am again using the normalization where the gauge coupling is out the front of the action and not in interaction terms, but one can return to the canonical normalization by  field redefinition. 

It will be useful later to also allow non-renormalizable interactions at the two-derivative level (one-derivative for fermions) of EFT, between chiral matter superfields and the gauge superfields, 
\begin{align}
    \delta S = \int d^{2} \theta \,\, f(\Phi) {\cal W}_{\alpha}^{a} {\cal W}^{a \, \alpha} + h.c. 
\end{align}
The function of chiral superfields $\Phi$  is holomorphic for the same reason the superpotential is, and is known as the ``gauge coupling function'' because it is clearly a generalization of the renormalizable coupling $1/g^2$. That is, it is a field-dependent gauge coupling. We will allow an independent gauge coupling function for each gauge group.
We take it to be a gauge-invariant function of fields, as is ${\cal W}^a_{\alpha}{\cal W}^{a \alpha}$.\footnote{There is a more general option $f_{ab}(\Phi) {\cal W}^a_{\alpha}{\cal W}^{b \alpha}$ where $f_{ab}$ is not gauge-invariant but we will not need this, and discard it for simplicity.} 

\subsection{Component form of charged field gauged kinetic terms}\label{subsec:charged-gaugeFiled-kinetic}

Unpacking the $\int d^4 \theta \, \bar{\Phi} e^V \Phi$, the couplings of $V$ introduce the expected $\partial_{\mu} \rightarrow \partial_{\mu} - iA_{\mu}^a t^a$ in all the derivatives of the charged fields. In addition, they give rise to a set of Yukawa-type interactions of the gaugino, 
\begin{align}
    {\cal L}_{\lambda} = \frac{1}{g^2} \overline{\lambda}^{a} i \sigma.D \lambda^{a} + \overline{\phi} \lambda_{\alpha}^{a} t^{a} \psi_{\beta} \epsilon^{\alpha\beta} + h.c.
\end{align}
where the first term comes from the the gauge superfield action, \Eq{eq:susy-gauge-S}. Finally, there are terms linear in the auxiliary fields $D^a$. Taking into account the quadratic terms in these fields from the gauge action, \Eq{eq:susy-gauge-S}, and eliminating these fields using their equations of motion, gives rise to a scalar potential called the ``D-term potential'', 
\begin{align}
    V_{D} =  \frac{1}{2} \sum_{a} g_a^2 \left( \sum _{i} \overline{\phi}_{i} t^{a}_{(i)} \phi_{i} \right)^2.    
\end{align}
The index $a$ labels all gauge generators of all gauge groups, and $g_a$ denotes the gauge coupling appropriate to that generator (so a simple gauge group has the same $g_a$ for all its generators, $a$). The index $i$ labels the different charged species and $t^a_{(i)}$ are the gauge generator matrices in the representation $i$. The full scalar potential is then given by the sum of D- and F-term potentials, 
\begin{align}
    V_{\rm scalar} (\phi, \overline{\phi}) = V_{D} + V_{F}.
\end{align}

\subsection{Renormalizable Feynman rules}

\begin{figure}[hbt!]
    \centering
    \includegraphics[width=0.8\linewidth,trim={2cm 2.5cm 3cm 3cm},clip]{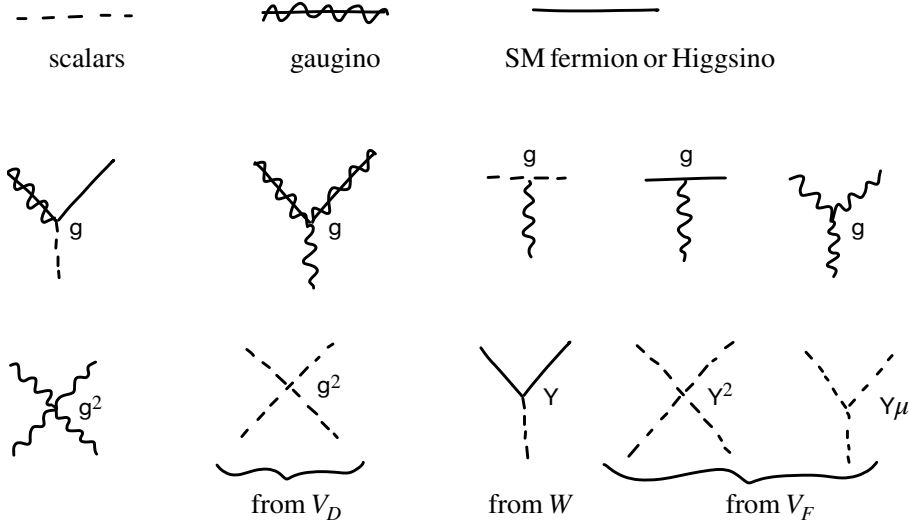}
    \caption{The schematic structure of the Feynman rules of the MSSM component fields, or indeed a generic renormalizable SUSY gauge-Higgs theory.}
    \label{fig:susy-feyn-rules}
\end{figure}

It is useful to picture the structure of the renormalizable MSSM in general terms. This is shown schematically in \Fig{fig:susy-feyn-rules}. But I have reverted to canonical normalization for the gauge fields where gauge couplings $g$ appear only in interactions not propagators, as is useful in practical work, rather than the normalization with $1/g^2$ multiplying the entire gauge kinetic terms, which is useful in some of our more theoretical and conceptual discussion. Detailed forms of these rules with all the relevant factors and representation matrices can be found in \cite{mssm-review}.  (One passes from the latter to canonical normalization by the field redefinition $A \rightarrow g A$.)

The first interaction shown is the Yukawa-type interaction of the gaugino mentioned above. The $g^2$ scalar quartic interaction comes from $V_D$. 
All the other couplings of strength $g$ or $g^2$ just follow from gauge invariance without any special SUSY considerations, just the gauge charge of the ordinary component fields. 
  The various Yukawa couplings of strength $Y$ arise from expanding the superpotential action with two fermions and one scalar field. The last two interactions arise from $V_F$. In particular, the last interaction arises from the cross-term in $V_F$ depending on both the $\mu$-term and the Yukawa coupling. 

\section{Soft SUSY breaking from Spontaneous SUSY breaking}\label{sec:soft-susyBreak}

Consider the following non-renormalizable couplings between the MSSM matter and the simple hidden sector of SUSY breaking discussed earlier:
\begin{align}
    {\cal L}_{\rm matter-hid} &= \int d^{4}\theta \, \frac{\overline{\Sigma}\Sigma}{M_{Pl}^{2}} \left( c_{ij}^{Q} \overline{Q}_{i} Q_{j} + c_{ij}^{U} \overline{U}_{i}^{c} U_{j}^{c} \right. \nonumber\\
    & \quad+ c_{ij}^{D} \overline{D}_{i}^{c} D^c_{j} + c_{ij}^{L} \overline{L}_{i} L_{j} + c_{ij}^{E} \overline{E}_{i}^{c} E_{j}^{c} \nonumber\\
    & \left. \quad + c_{u} \overline{{\cal H}}_{u} {\cal H}_{u} + c_{d} \overline{{\cal H}}_{d} {\cal H}_{d}  \right),
\end{align}
where all the $c$ coefficients are taken to be roughly ${\cal O}(1)$ in size and where $i, j = 1,2,3$ label the three SM generations.

Now, upon spontaneous SUSY breaking in the hidden sector,  we have $\langle \Sigma \rangle = 
\Lambda^2 \theta^2$. If we plug this VEV into the coupling to the MSSM matter, the $\int d^4 \theta$ is forced to integrate the Grassmann coordinates in $\langle \bar{\Sigma} \Sigma \rangle$ and not the MSSM superfields. Therefore the result can only be a potential for the MSSM scalars, 
\begin{align}
    {\cal L}_{\rm matter-hid} \underset{\Sigma \rightarrow \langle\Sigma\rangle}{\longrightarrow} \frac{\Lambda^{4}}{M_{Pl}^{2}} (c_{ij}^{Q} \overline{\tilde{q}}_{i} \tilde{q}_{j} + c_{ij}^{U} \overline{\tilde{u}}_{i}^{c} \tilde{u}_{j}^{c} +c_{ij}^{D} \overline{\tilde{d}}_{i}^{c} \tilde{d}_{j}^{c}+ c_{ij}^{L} \overline{\tilde{l}}_{i} \tilde{l}_{j} + c_{ij}^{E} \overline{\tilde{e}}_{i}^{c} \tilde{e}_{j}^{c} + c_{u} \overline{H}_{u} H_{u} + c_{d} \overline{H}_{d} H_{d}).
\end{align}
That is, we get a set of mass terms for all the scalars, roughly 
\begin{align}
    m^{2}_{\rm scalar} \sim \left( \frac{\Lambda^{2}}{M_{Pl}}\right)^{2} \gtrsim v_{weak}^2,
\end{align}
but not for their superpartner fermions. From the MSSM viewpoint this is effectively soft SUSY breaking roughly at the scale 
$\sim \Lambda^2/M_{Pl}$. As we anticipated earlier by qualitative reasoning, we will take this scale of soft SUSY breaking to be (very roughly) comparable to the weak scale $v_{weak}$. Yes, order $1-10$ factors matter hugely to experimentalists searching for such squarks and sleptons (and extra Higgs scalars) but we are not yet ready to confront experiment at this point, so let us go with this thumbnail sketch for now.

Solving for the scale of fundamental spontaneous SUSY breaking in the hidden sector, 
\begin{align}
    \Lambda \gtrsim \sqrt{M_{Pl} v_{weak}} \sim 10^{11} \, {\rm GeV}.
\end{align}
Because it is roughly the geometric mean of the Planck and weak scales it is known as the ``intermediate scale''. It is consistent to keep this big VEV of $\Sigma$ while neglecting its fluctuations in their impact on the MSSM because the fluctuation fields $\sigma, \psi_{\Sigma}$ have Planck-suppressed couplings which are tiny at accessible energies. By comparison,  the intermediate scale VEV is so large that even after Planck-suppression its impact on the MSSM is comparable to the weak scale. 

Similarly, we can also consider a non-renormalizable hidden sector coupling to the gauge superfields, 
\begin{align}\label{eq:gauge-hid-coupling}
    {\cal L}_{\rm gauge-hid} & \,\, = \int d^{2}\theta \, \frac{\Sigma}{M_{Pl}} c {\cal W}_{\alpha}^{a} {\cal W}^{a \, \alpha} \nonumber \\
    & \underset{\Sigma \rightarrow \langle \Sigma \rangle}{\longrightarrow} \,\, \frac{c \Lambda^{2}}{M_{Pl}} \lambda_{\alpha}^{a} \lambda^{a \, \alpha},
\end{align}
where there is an (implicit) independent $c$ for each MSSM gauge group. The MSSM sees this coupling as effectively a soft SUSY-breaking gaugino mass of comparable size to the scalar soft masses, 
\begin{align}
    m_{\lambda} \sim \frac{\Lambda^{2}}{M_{Pl}}.
\end{align}

Putting it all together, the MSSM is effectively softly broken by the couplings to the hidden sector, at the rough scale 
\begin{align}
    m_{\rm superpartners} \sim \frac{\Lambda^{2}}{M_{Pl}} \gtrsim v_{weak}.
\end{align}
With this simple form of soft SUSY breaking only affecting superpartner masses, the  Feynman
diagrammatics above will look schematically the same as the SUSY limit, but in detail the propagators for the different component particles will reflect the SUSY-breaking masses.

\section{The ``$\mu$-Problem'' and the Giudice-Masiero Mechanism}

There is only one explicit mass parameter (not arising from dimensional transmutation) in the MSSM prior to soft SUSY breaking, namely the $\mu$ parameter:
\begin{align}
    \int d^{2}\theta \, \mu {\cal H}_{u} {\cal H}_{d} = \mu \tilde{H}_{u} \tilde{H}_{d} + \cdots
\end{align}
One can think of it as primarily giving rise to a Dirac mass, $\mu$, for Higgsinos. In the SUSY limit it also gives the same mass to the Higgs scalars of course, but after SUSY breaking we have seen that there are other significant  corrections to the latter. There are then two parametrically distinct mass scales, $\mu$ and $\Lambda^2/M_{Pl}$, the first supersymmetric in form and the second the result of SUSY breaking. If $\mu \ll \Lambda^2/M_{Pl}  \sim v_{weak}$ then the Higgsinos would be much lighter than the weak scale and would have already been discovered given their electric and weak charges. And if $\mu \gg \Lambda^2/M_{Pl}$ then the entire ${\cal H}_{u,d}$ multiplets would have supersymmetric mass $\mu$ far above the weak scale so that they could not execute EWSB. Therefore, we need $\mu \sim \Lambda^2/M_{Pl}$. However, given the orders of magnitude hierarchy between the Planck and weak scales, this is a striking and unexplained coincidence, an unsatisfactory feature in our push to understand all large  hierarchies where seen {\it and} absences of large hierarchies where not seen. It is 
called the ``$\mu$ problem''. We turn now to its simplest and ultimately quite satisfying solution. 


First notice that $\mu = 0$ is robust in that it is protected by a new global symmetry in that limit, ``Peccei-Quinn'' (PQ) symmetry, 
\begin{align}
    U(1)_{\rm PQ} : \quad &{\cal H}_{u} \longrightarrow e^{i \alpha} {\cal H}_{u} \nonumber\\
    &{\cal H}_{d} \longrightarrow e^{i \alpha} {\cal H}_{d}.
\end{align}
Instead, we introduce new couplings to the hidden sector that are peculiar to the two Higgs supermultiplets because they are in conjugate gauge representations: 
\begin{align}
    {\cal L}_{\rm Higgs-hid} & \,\, = \int d^{4}\theta \, \left( c'  \frac{\overline{\Sigma}}{M_{Pl}} {\cal H}_{u} {\cal H}_{d} + c'' \frac{\overline{\Sigma} \Sigma}{M_{Pl}^{2}} {\cal H}_{u} {\cal H}_{d} + h.c. \right) \nonumber \\
    &\overset{\Sigma \rightarrow \langle \Sigma \rangle}{\underset{\overline{\Sigma} \rightarrow \langle \overline{\Sigma} \rangle}{\longrightarrow}} \left( \frac{c' \Lambda^{2}}{M_{Pl}} \int d^{2}\theta \, {\cal H}_{u} {\cal H}_{d} + h.c.\right) +  \left( \frac{c'' \Lambda^{4}}{M_{Pl}^{2}} \, H_{u} H_{d} + h.c.\right).
\end{align}
We see that the $c' \sim {\cal O}(1)$ term gives us an effective $\mu_{eff} = c' \Lambda^2/M_{Pl}$ term after SUSY breaking in the hidden sector, and solves the $\mu$ problem by realizing this apparent SUSY-preserving mass as fundamentally a SUSY breaking effect. This is the Giudice-Masiero mechanism \cite{Giudice:1988yz}.
It can be viewed as preserving Peccei-Quinn symmetry if we assign $\Sigma$ charge $2$ under it. The second term is called an effective soft-SUSY-breaking ``$B \mu$ term'' $= c'' \Lambda^4/M_{Pl}^2$, and can be useful in model-building EWSB as we will see later. But clearly this would violate Peccei-Quinn symmetry. The easiest way to allow both terms is to simply assume that the hidden sector and its couplings violate Peccei-Quinn explicitly, but the MSSM in  isolation does not. We will make this more plausible, and radiatively stable,
from an extra-dimensional perspective later.

\section{SUSY Phenomenology}

\subsection{Importance of the $125$ GeV Higgs Sector}\label{subsec:125GeV-higgs}

\begin{figure}[hbt!]
    \centering
    \includegraphics[width=0.95\linewidth,trim={1cm 8cm 0cm 6cm},clip]{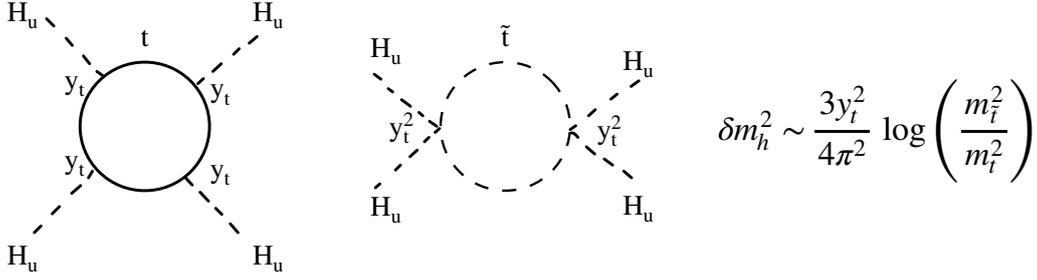}
    \caption{Top/stop loops can have a significant impact on Higgs quartic self-coupling, and hence physical Higgs scalar mass,  if the stop-top mass splitting is significantly larger than the weak scale.}
    \label{fig:susy-higgs-corr}
\end{figure}
Here, I want to give a schematic sense of the importance of the Higgs mass and SUSY-breaking radiative corrections, ignoring some of the subtleties related to the $2$-Higgs-doublet nature of the MSSM. We will do a more refined job once we have sharpened a model of SUSY-breaking $c$ parameters.

In the non-supersymmetric SM, the physical Higgs boson mass is given at tree-level by 
\begin{align}
    m_{h}^{2} = \lambda_{h} v_{weak}^{2},
\end{align}
where $\lambda_h$ is the Higgs-doublet quartic self-coupling. In the (softly-broken) MSSM, you can see from the scalar potential contributions, the only self-couplings of the (two) Higgs doublets arise in the D-term potential, and therefore $\sim g_{EW}^2$. In detail this implies a tree-level bound, $m_h \leq m_Z$, regardless of how EWSB is shared between the two Higgs doublet VEVs. In terms of the mass-squareds (more relevant for bosons) we see that the observed Higgs strongly violates this bound:
\begin{align}
    m^{2}_{h} = (125 \,{\rm GeV})^{2} \sim 2 m^{2}_{Z}!
\end{align}
This implies that if SUSY is playing a role near the weak scale, Higgs radiative corrections are important. The largest such radiative corrections are expected from the largest Higgs coupling, the Yukawa coupling to the top quark, as illustrated in \Fig{fig:susy-higgs-corr}. 

We see that the simplest fit to $m_h = 125$ GeV is to have a rather heavy stop, $m_{\tilde{t}} \sim 10$ TeV, ``just'' out of LHC reach!

\subsection{Direct LHC Searches}

\begin{figure}[hbt!]
    \centering
    \includegraphics[width=0.65\linewidth,trim={5cm 6cm 7cm 5cm},clip]{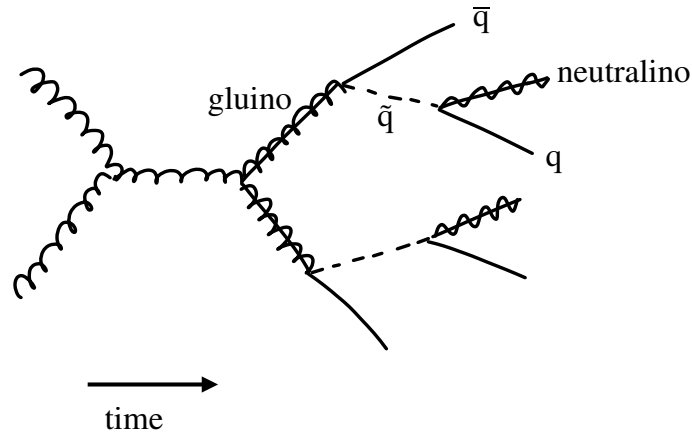}
    \caption{Gluino pair production at a hadron collider, initiated by gluons, decaying into quarks and neutralinos. If the neutralinos are stable they would be a good candidate for dark matter particles. }
    \label{fig:superpartner-prod-lhc}
\end{figure}
As you know, all searches for superpartners to date have come up empty, so these translate into bounds on superpartner masses/couplings \cite{superpartner-bounds}.
Very roughly, a variety of LHC searches have resulted in bounds,
\begin{align}
    & m_{\rm gluino} \gtrsim 2 \, {\rm TeV}, \,\, m_{\rm Wino} \gtrsim {\rm TeV} \nonumber \\
    & m_{\rm squarks} \gtrsim  {\rm TeV}, \,\, m_{\rm sleptons} \gtrsim 100(s)\,{\rm GeV}.
\end{align}
A typical process that the LHC searches for is given in \Fig{fig:superpartner-prod-lhc}. Here, two gluonic partons from within the colliding protons are pair-creating a pair of gluinos. Each gluino then promptly decays into a (anti-)quark (jet) and squark, and the squark in turn decays into a quark (jet) and ``neutralino''. We are assuming that the lightest superpartner, stable by $R$-parity conservation, is electrically and color-neutral, hence ``neutralino''. I have drawn it as one of the neutral electroweak gauginos (though it may be a more complicated superposition of superpartner gauge-eigenstates). Such a stable neutralino is an example of a WIMP (Weakly Interacting Massive Particle, implicitly stable).
Because of its neutrality (and color-neutrality), a WIMP behaves roughly like a heavy neutrino, escaping the LHC detectors, only ``detected'' as  missing (or imbalance in the) energy-momentum of the event. 

\subsection{WIMP Dark Matter Direct Detection}

A stable WIMP is, broadly speaking, a good candidate for the particles making up the Dark Matter of the universe \cite{dm-review}. If that is the case, then beyond  trying to pair-produce WIMPs at  colliders as sketched above, we can also try to detect WIMPs in our galaxy sweeping through the Earth. For example, the WIMPs can scatter off nuclei in underground detectors, the details depending on the specifics of their weak interactions. The likelihood of detection is amplified by the potentially large flux of WIMPs and the large volume of the detector target, as compared to pair-production and detection at particle colliders. Again, as you know, despite heroic efforts WIMP dark matter particles have not been directly detected to date. But there still remains room for discovery.

\section{Flavor and CP Problems of SUSY (and BSM more generally)}

In the experimental search for new BSM physics there are two modes. There is the mode of simply producing and detecting the new particles ``on-shell'' (or having them be created by the early universe and hit our detectors). But in this section we consider the other mode of processes involving SM particles mediated by virtual ``off-shell'' exchanges of the new particles, typically because the energy of the process is well below the BSM masses as the price for higher statistics and hence precision. Our greatest ability to discriminate BSM effects is within processes in which the competing SM effects are relatively small for SM-structural reasons. Flavor-violating and CP-violating processes at low energies (by high-energy collider standards) are two of the most powerful categories in this regard, because in the SM they are strongly constrained by the CKM flavor/weak-interaction structure and the resulting GIM mechanism.

\subsection{SM FCNCs and CP-Violation}

\begin{figure}[hbt!]
    \centering   \includegraphics[width=0.5\linewidth,trim={3cm 5cm 3cm 5cm},clip]{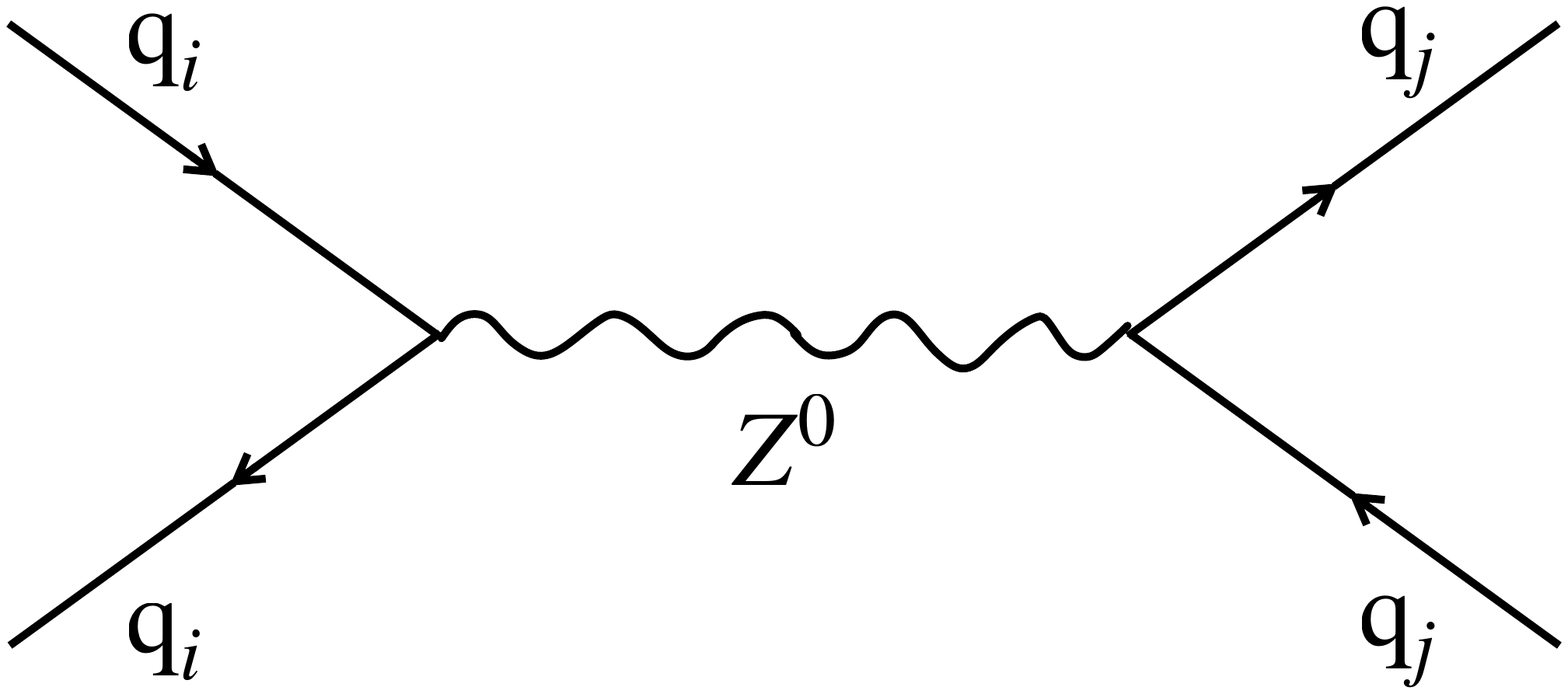}
    \caption{}
    \label{fig:qScattering}
\end{figure}

\begin{figure}[hbt!]
    \centering    \includegraphics[width=0.95\linewidth,trim={0cm 5.5cm 0cm 5.5cm},clip]{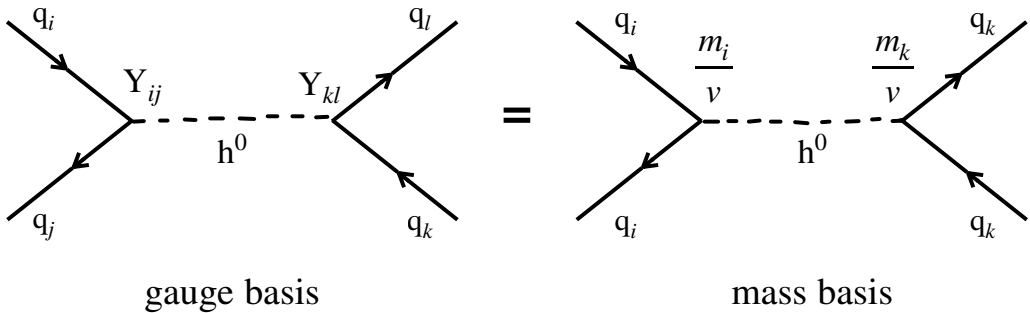}
    \caption{Tree-level neutral current processes in the standard model are flavor-preserving, a striking consequence of its minimal structure.}    \label{fig:qScatteringGaugevsMassBasis}
\end{figure}

The SM GIM mechanism strongly constrains Flavor-Changing Neutral Currents (FCNCs). The classic ``neutral current'' is virtual $Z^0$-exchange, illustrated in \Fig{fig:qScattering} between quarks, the neutral version of the ``charged current'' exchange of $W^{\pm}$ in the standard Fermi approximation to weak interactions. For charged currents the $W$ has coupling $g V^{CKM}_{ij}$ to any pair of quark and antiquark flavors $i, j$ with net charge $1$, where $V^{CKM}$ is the CKM matrix emerging from the Yukawa couplings. By contrast,
 the $Z^0$ coupling is necessarily diagonal in the species of the pair of quarks it couples to. This is a consequence of the minimal gauge coupling from covariant derivatives in the quark kinetic terms. Therefore  flavor $i$ is both entering and leaving the effective neutral current exchange, and the same is true of flavor $j$. (Or equivalently, quark flavors $j$ and antiquark flavor $j$ are being produced.) So this is a flavor-conserving neutral current, not a flavor-violating one. A less trivial ``neutral current'' arises from Higgs $h^0$ exchange, illustrated between quarks in \Fig{fig:qScatteringGaugevsMassBasis}. Even though the Yukawa coupling matrix is not diagonal in the gauge basis, the quark mass matrix is proportional to it, $m_{ij} = Y_{ij} v_{weak}$, so that the Yukawa couplings are flavor-diagonal in the mass eigenbasis. So once again, this neutral current is flavor-conserving. 

\begin{figure}[hbt!]
    \centering    \includegraphics[width=1\linewidth,trim={0cm 6.5cm 0cm 6.5cm},clip]{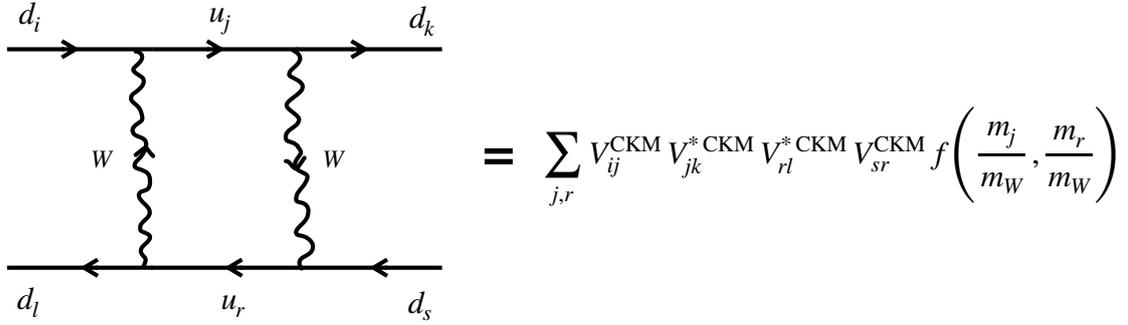}
    \caption{FCNCs appear at one loop within the standard model by sewing together a pair of  tree-level flavor-changing charged currents. Even here, there are significant suppressions due to the unitarity of the CKM matrix.}
    \label{fig:SMfcnc}
\end{figure}

We have to go to $1$-loop to produce FCNCs in the SM, where we can create an effective neutral current by exchanging a $W^+ W^-$ pair, and taking advantage of the flavor-changing $W$ couplings. This is illustrated in the box-diagram of \Fig{fig:SMfcnc} for (virtual) $W$ exchanges between down-type quarks of various generations, at low energies (well below $m_W$). I have just given the dependence on the CKM matrix elements from the four vertices, while the loop integral itself is given by $f$ at low energies. I will make my point by pretending that all the up-type quarks are light, $m_j, m_r \ll m_W$. Then the loop-level FCNC in the figure simplifies to $\approx f(0,0) \delta_{ik} \delta{\ell s} +$ small, by the unitarity of the CKM matrix in the SM. That is, even at loop level, the individual contributions to FCNCs cancel to a large extent!\footnote{Obviously the top quark violates the assumption of small quark mass and requires special treatment, which I will not go into. But suffice it to say, it enters with rather small mixing angles $V^{CKM}_{ts}, V^{CKM}_{td}$, so that its contributions to FCNCs are also relatively small.} The small deviations from this limit predict small but observable FCNCs which closely match observation. Any contributions from BSM must be small enough to hide in the current experimental and calculational error bars! 

Now let us turn to CP violation, which can appear in conjunction with flavor-changing processes or within purely flavor-conserving processes.  In QFT, CP-violation arises from complex phases in couplings that cannot be eliminated by field redefinitions. In the renormalizable SM, with its three generations (and neglecting the tiny neutrino masses for simplicity), at the perturbative level there is just one such phase in $V^{CKM}$. This gives rise to a highly restrictive form for CP-violating effects in weak-interaction processes within hadronic flavor physics, and there is a beautiful body of experimental evidence corroborating this single-source of CP-violation. Non-perturbatively and subtly, there is another complex phase that is allowed, the ``$\Theta$ angle'',  which would give rise to CP-violation in strong-interaction hadron physics (without involving weak-interactions at all), most famously predicting a neutron electric dipole moment. But no such strong-interaction CP-violation has been observed, giving rise to the stringent limit $\Theta < 10^{-10}$. Other than $\Theta$, Nature seems to have turned on every coupling allowed by the renormalizable structure of the SM at an observable level, so the nearly-vanishing $\Theta$ poses a puzzle ``The Strong CP Problem''. Famously, there is a plausible BSM resolution to this puzzle by adding an axion to the SM, subject to the brilliant Peccei-Quinn mechanism. For a review see \cite{hook-cp-review}.
We will assume that this addition to the SM has been made, so that only the CKM phase remains as the source of CP violation in the SM$+$axion theory. 

Given the SM ($+$ axion) structure that so strongly suppresses FCNCs and CP-violation, in agreement with current precision low-energy flavor and CP tests, makes these very sensitive channels for searching for BSM. Let us see how these channels are sensitive to SUSY.

\subsection{Superpartner-mediated  FCNCs}

\begin{figure}[hbt!]
    \centering    \includegraphics[width=1\linewidth,trim={0cm 5.5cm 0cm 5.5cm},clip]{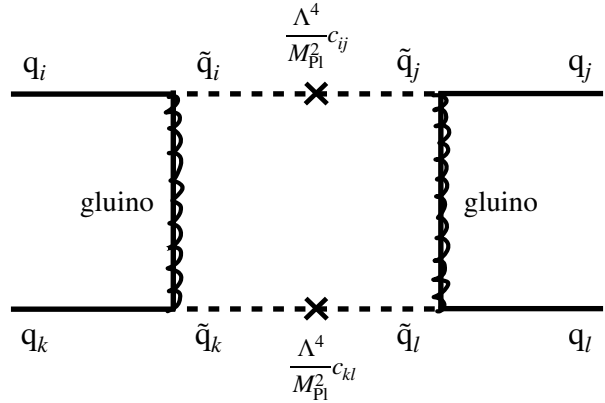}
    \caption{The difference between the mass eigenbases of quarks and squarks after generic SUSY breaking leads to FCNCs mediated by gluinos.}
    \label{fig:SUSYfcnc}
\end{figure}

To exemplify how SUSY can introduce new FCNC contributions we consider  new
box diagrams, analogous to those of the SM but with the $W$s replaced by gauginos. Gluino exchange comes with the strongest coupling, so we focus on this in \Fig{fig:SUSYfcnc}.\footnote{Note that gluon exchange in a box diagram itself would not give FCNCs because gluons always couple the same quark and antiquark flavor, unlike $W$s. The same is true for gluino couplings {\it before} SUSY breaking, but not after SUSY breaking, as \Fig{fig:SUSYfcnc} schematically illustrates.} In the figure the quark flavors are in their mass basis, and I am labelling the squark flavors in this same basis, so that gluinos connect quarks and squarks of the same flavor. However, I have indicated the soft SUSY-breaking squark mass-squareds which are not necessarily diagonal in this flavor basis because the $c_{ij}$ are a priori general matrices in flavor-space, as introduced in \Sec{sec:soft-susyBreak}. If we were to diagonalize these soft SUSY-breaking squark mass$^2$ matrices, we would find that the quark mass basis was not the same as the squark mass basis. This is in analogy to how the up-type quark mass basis is different from the down-type mass basis in the SM, so there is the nontrivial CKM generation-changing matrix for up-down couplings to $W$ after EWSB. Similarly the gluino will have generation-changing couplings between quark and squark mass eigenstates after SUSY breaking. Unlike (most of) the up-type quarks in \Fig{fig:SMfcnc}, all the squarks in \Fig{fig:SUSYfcnc} are  comparably as heavy as the gluino, and therefore we expect significant FCNCs for generic $c_{ij} \sim {\cal O}(1)$. 

The effective flavor-violating four-quark interaction arising from integrating out the superpartners is however suppressed by $\sim 1/m_{superpartner}^2$, by dimensional analysis. Therefore for generic $c_{ij}$ we can suppress BSM 
FCNCs by taking large enough superpartner masses. Given generic $c_{ij}$ and current constraints, this leads to bounds of roughly 
$m_{squarks} > 1000$(s) of TeV! \cite{bsm-flavor-constaints} 

\subsection{Superpartner-mediated CP-violation}
\begin{figure}[hbt!]
    \centering    \includegraphics[width=0.55\linewidth,trim={5cm 5.5cm 5cm 5.5cm},clip]{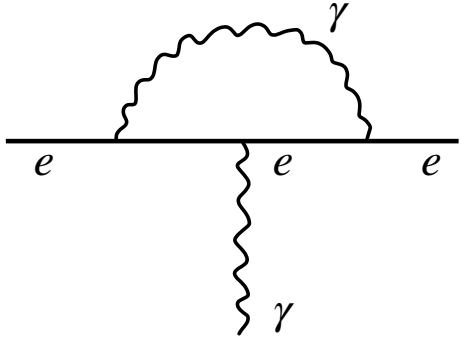}
    \caption{The classic diagram generating the anomalous magnetic moment of the electron in QED.}
    \label{fig:eMagneticMom}
\end{figure}

\begin{figure}[hbt!]
    \centering    \includegraphics[width=0.55\linewidth,trim={5cm 5.5cm 5cm 5cm},clip]{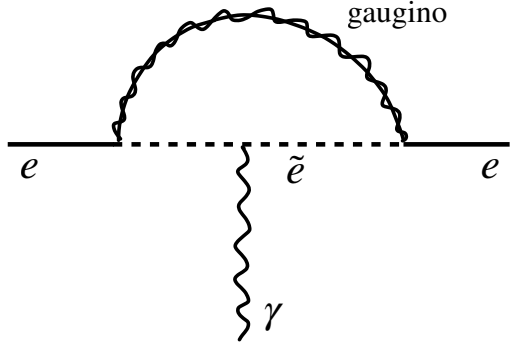}
    \caption{A supersymmetric-loop analog of the previous diagram which can generate a distinctive CP-violating electric-dipole moment if there are CP-violating phases in the superpartner masses/couplings.}
    \label{fig:SUSYeMagneticMom}
\end{figure}
We can illustrate non-flavor-violating CP-violation by the CP-violating electron electric dipole moment (EDM). We can orient ourselves by first considering the classic $1$-loop QED contribution to the electron {\it magnetic} dipole moment of \Fig{fig:eMagneticMom}. QED does not give rise to an electron EDM because it does not contain CP-violating phases in the renormalizable theory. But we can now consider a superpartner analog of this loop, \Fig{fig:SUSYeMagneticMom}. It will give a BSM contribution to the electron magnetic dipole moment. Because it is suppressed by one power of the superpartner mass, it 
can easily be much smaller than the QED contribution and therefore hard to detect. 
But if there are irreducible CP-violating phases in the gaugino or slepton masses/couplings then the diagram will also contribute to an electron EDM. For ${\cal O}(1)$ phases, the non-observation of such an EDM then requires
 $m_{superpartner} \gg$ TeV \cite{bsm-eEDM-constraints}.

\subsection{Moral for BSM from flavor and CP considerations}\label{subsec:cp-considerations} 

We have reviewed the key to how the SM ($+$  axion) strongly suppresses FCNCs and CP-violation, namely that the only sources of flavor and CP violation appear in the Yukawa coupling matrices, $Y^u_{ij}, Y^d_{ij}, Y^e_{ij}$, and that these are automatically diagonal and real in the quark/lepton mass eigenbases (neglecting neutrino masses). 
We have also seen that SUSY with generic $c_{ij}$'s (and generic BSM more generally) at collider-accessible scales is typically ruled out by the powerful flavor and CP experimental constraints we already have. This strongly suggests that:

~

\noindent {\it A viable SUSY (or BSM) theory should retain the feature of having the Yukawa couplings as the sole source of flavor- and CP-violation, and that there should be some theoretical structure that plausibly enforces this.}\footnote{ This should at least be the case to good approximation.}

~

We next describe an attractive structure of this type.\footnote{The ansatz of using copies of the Yukawa matrices in multiple places in a Lagrangian to violate flavor/CP in a manner that extends the GIM mechanism is called Minimal Flavor Violation (MFV), reviewed in \cite{min-flav-violation}, but here we are seeking a dynamical explanation of such an extension of GIM rather than an ansatz.} 

\section{Combining ``bosonic'' ($x_5$) and ``fermionic'' ($\theta, \bar{\theta}$) extra dimensions} 

\begin{figure}[hbt!]
    \centering    \includegraphics[width=1\linewidth,trim={0.cm 5cm 2cm 5.cm},clip]{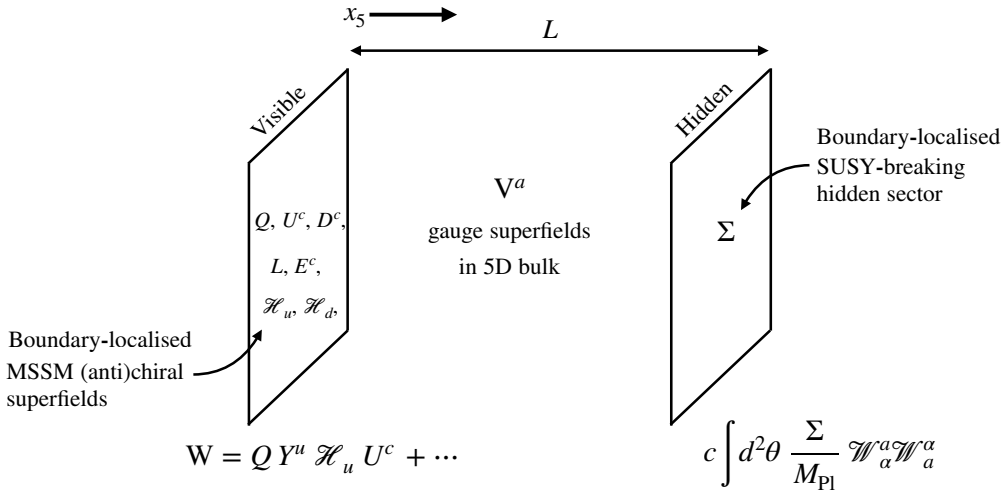}
    \caption{The higher-dimensional framework for a toy model of ``gaugino-mediated SUSY breaking'', with boundary-localized MSSM chiral superfields sequestered from a boundary-localized hidden sector.
    The gauge superfields propagate in the 5D bulk and mediate SUSY breaking in flavor-independent manner that  solves the SUSY Flavor Problem.}
    \label{fig:SUSYextraDimStructure}
\end{figure}
I want to first focus on  a toy model, but it an almost realistic one. It is a synthesis of supersymmetric and extra-dimensional dynamics, again of the type that might plausibly emerge from something like superstring quantum gravity below the Planck scale. It is depicted in \Fig{fig:SUSYextraDimStructure}. 
It is critical that the extra dimension discussed from here on is {\it not the same} as the extra dimension discussed in \Sec{sec:higherDim} as being responsible for the generation of SM Yukawa (fermion mass and CKM) hierarchies. For modularity, we can think of these as two different extra dimensions.  I will assume here that the Yukawa coupling hierarchies have been achieved by some mechanism, perhaps that of subsection~\ref{subsec:yukawaHierarchies}, but make no further reference to this. I will continue to refer to the relevant extra dimension as the ``$5$th'' (and not the ``$6$th'').
In particular, in the current context and higher-dimensional spacetime, the MSSM chiral superfields, containing quarks, leptons, Higgses and their superpartners, are all 4D fields localized on the left-hand boundary of 5D. We call this the ``visible boundary''. The hidden sector responsible for spontaneous SUSY breaking, which minimally is just the $\Sigma$ superfield model introduced earlier, is localised to 4D on the right-hand boundary. We call this the ``hidden boundary''. The MSSM gauge supermultiplets are however elevated to 5D superfields propagating in the ``bulk'' of the higher-dimensional spacetime. This model is called ``gaugino-mediated SUSY breaking'' ($\tilde{\rm g}$mSB) \cite{gaugino-med-susyBreak} for reasons we will see below.

The visible boundary dynamics in isolation is described by the 4D Kahler potential and superpotential for the MSSM chiral superfields detailed in \Sec{sec:mssm}, while the hidden boundary dynamics in isolation is described by the 4D Kahler potential and superpotential described in \Sec{sec:spontSUSYbreak-eft}. But the boundaries are not isolated because they can both interact with the gauge superfields. The leading such couplings to the charged MSSM matter at $x_5=0$ is just given by their  supersymmetric gauge couplings as described in \Eq{eq:S-nonAbelian}, and unpacked in subsection~\ref{subsec:charged-gaugeFiled-kinetic}. 
Since the hidden field $\Sigma$ is a gauge singlet, it can only have couplings at $x_5 =L$ to supersymmetric gauge field strengths, such as the $c/M_{Pl}$ of \Eq{eq:gauge-hid-coupling}. We can focus on the 4D EFT below the KK scale $1/L$. 5D gauge invariance forces the gauge fields to have $m_5 = 0$, so that their $0$-modes are $x_5-$independent. Therefore there are no extra exponentials $e^{\pm m_5 L}$ incurred in deriving the 4D EFT. It must just have the form given by \Eqs{eq:susy-gauge-S}{eq:4D-5D-gaugeCoupling}. 

But by the locality of all couplings in relativistic theories (no instantaneous action at a distance) we cannot couple the visible boundary fields (MSSM matter chiral superfields) directly to the  hidden boundary superfield $\Sigma$. Therefore the couplings $c^Q_{ij}, c^U_{ij}, c^D_{ij}, c^E_{ij}, c^L_{ij}, c^E_{ij}, c^u_{ij}, c^d_{ij}, c, c'$ must all vanish! You can check that the only remaining couplings that violate flavor symmetries (have non-trivial flavor dependence) are the supersymmetric Yukawa couplings in the visible superpotential, exactly as the moral of subsection~\ref{subsec:cp-considerations} presecribed. This 5D geographical separation of flavored superfields from the SUSY-breaking dynamics is  ``sequesetering'' \cite{sequestering} and is the key here to solving the ``supersymmetric flavor problem'' of excessive FCNCs arising via generic flavor-violating $c^{Q, U, D, L, E}_{ij}$. 

Let us now turn to CP. We will impose this as a symmetry of the bulk of 5D as well as the hidden boundary. That is, we assume that CP is only broken on the visible boundary.\footnote{If one prefers, one could impose CP as an exact symmetry of the dynamics everywhere, but with it being spontaneously broken near the Planck scale by a very heavy field which is localized to the visible boundary. Once we integrate out this heavy field, its CP-violating VEV can effectively make the visible boundary couplings CP-violating if they have Planck-suppressed couplings to it, but by locality it cannot give effective CP-violation away from the visible boundary.} This means that CP-violating phases can appear in the MSSM superpotential, in  particular in the MSSM Yukawa couplings of quark and lepton superfields. We will continue to impose Peccei-Quinn symmetry so that $\mu =0$, therefore the only CP-violation must be restricted to the irreducible single CKM phase of the quark Yukawa couplings, exactly as in the non-SUSY SM, and in accordance with the moral drawn in subsection~\ref{subsec:cp-considerations}. This solves the ``supersymmetric CP problem'' of excessive CP violation. 

But in sequestering  have we ``thrown out the baby with the bathwater''? After all, it appears the price for having $c^{ij} = 0$ for the squarks and sleptons means they will be degenerate with the quarks and leptons, and therefore we should have already have seen them experimentally. Fortunately, this is only a tree-level conclusion which is strongly radiatively corrected. The point is that quantum loops are not local and can straddle the extra dimension and connect SUSY breaking from the hidden boundary to squarks and sleptons on the visible boundary. By contrast UV divergences are always local and therefore such loop effects should be finite and calculable. Furthermore, locality of UV divergences means that sequestering itself is robust, in that all divergences and counterterms will respect this ``geographical'' structure. 

To work out the leading radiative corrections, we need to have a sense of the size of the extra dimension, $L$. We turn to this next.

\subsection{SUSY Grand Unification and the Size of the Fifth Dimension}

\begin{figure}[hbt!]
    \centering   
    \includegraphics[width=0.7\linewidth]{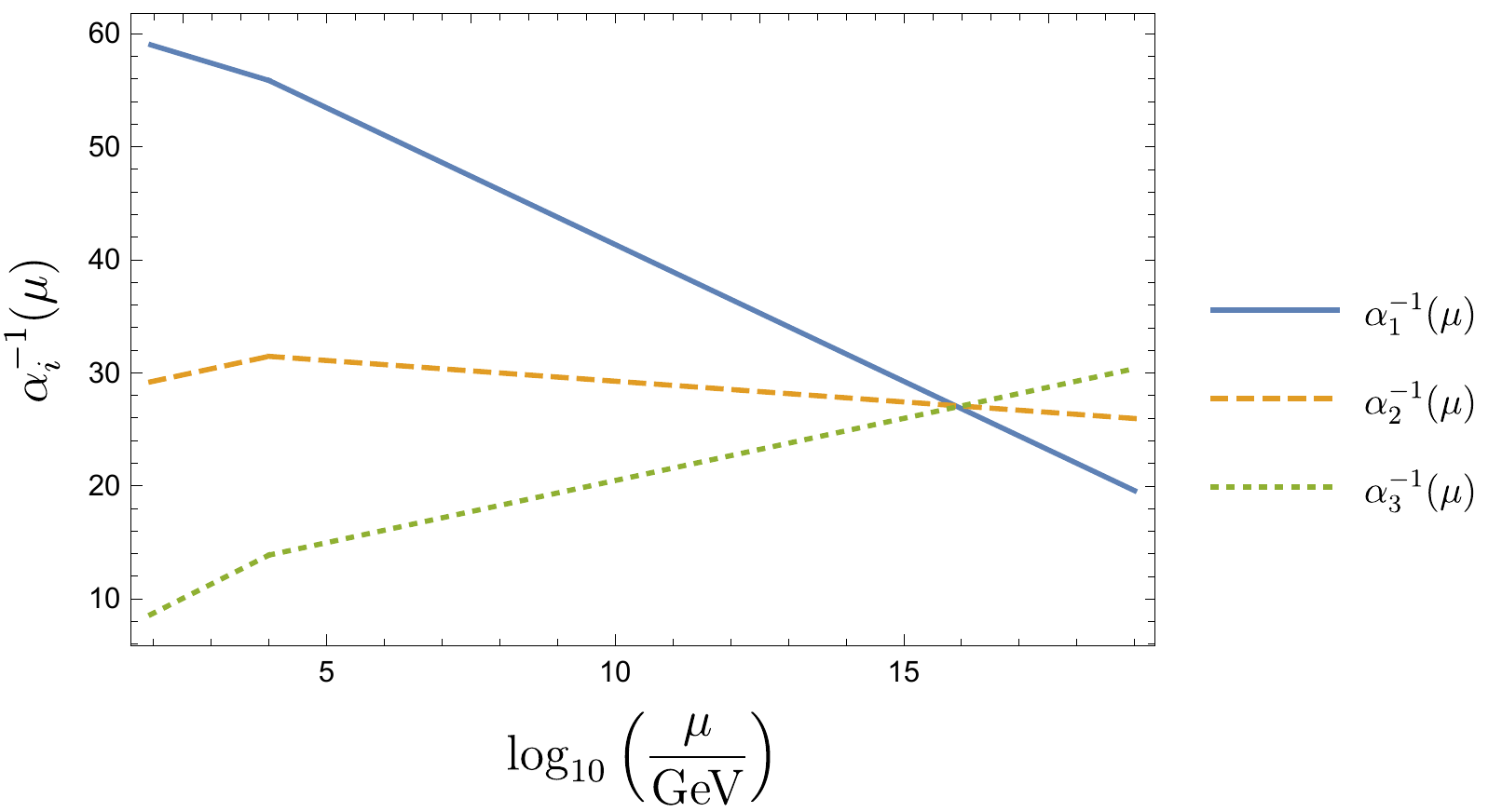}
    \caption{Running of (MS)SM gauge couplings at one loop, with superpartner masses $\sim {\cal O}(10)$ TeV, strongly suggesting grand unification not far from the Planck scale.}
    \label{fig:SUSY-unification}
\end{figure}

We can recompute the running of the SM gauge couplings in this MSSM 4D EFT. It differs from \Fig{fig:unification} in that we must include the superpartners (and two Higgs doublets versus the SM's single Higgs doublet) in the running from the $\sim {\cal O}(1-10)$ TeV scales we expect (hope) to find them. This is shown at $1$-loop approximation in \Fig{fig:SUSY-unification}.  Famously, this is a dramatic improvement over the non-supersymmetric SM in providing circumstantial evidence in favor of grand unification, given by the much closer meeting of the couplings at $10^{16}$ GeV. This is intriguingly close to the Planck scale, hinting at an even grander unification of some sort with quantum gravity.
But to retain this attractive interpretation we must trust the 4D renormalization group all the way to this high unification scale. This implies that the KK scale, above which the 4D effective description breaks down, must be even larger:
\begin{equation}
1/L \geq M_{GUT} \sim 10^{16} {\rm GeV}.
\end{equation}
For concreteness we take $1/L \sim 10^{16}$ GeV.
This is a very small extra dimension indeed, and yet with big consequences. 
 
\subsection{4D Renormalization Group (RG) evolution below the  KK scale} 

\begin{figure}[hbt!]
    \centering    \includegraphics[width=1\linewidth,trim={0.cm 4cm 0cm 5.cm},clip]{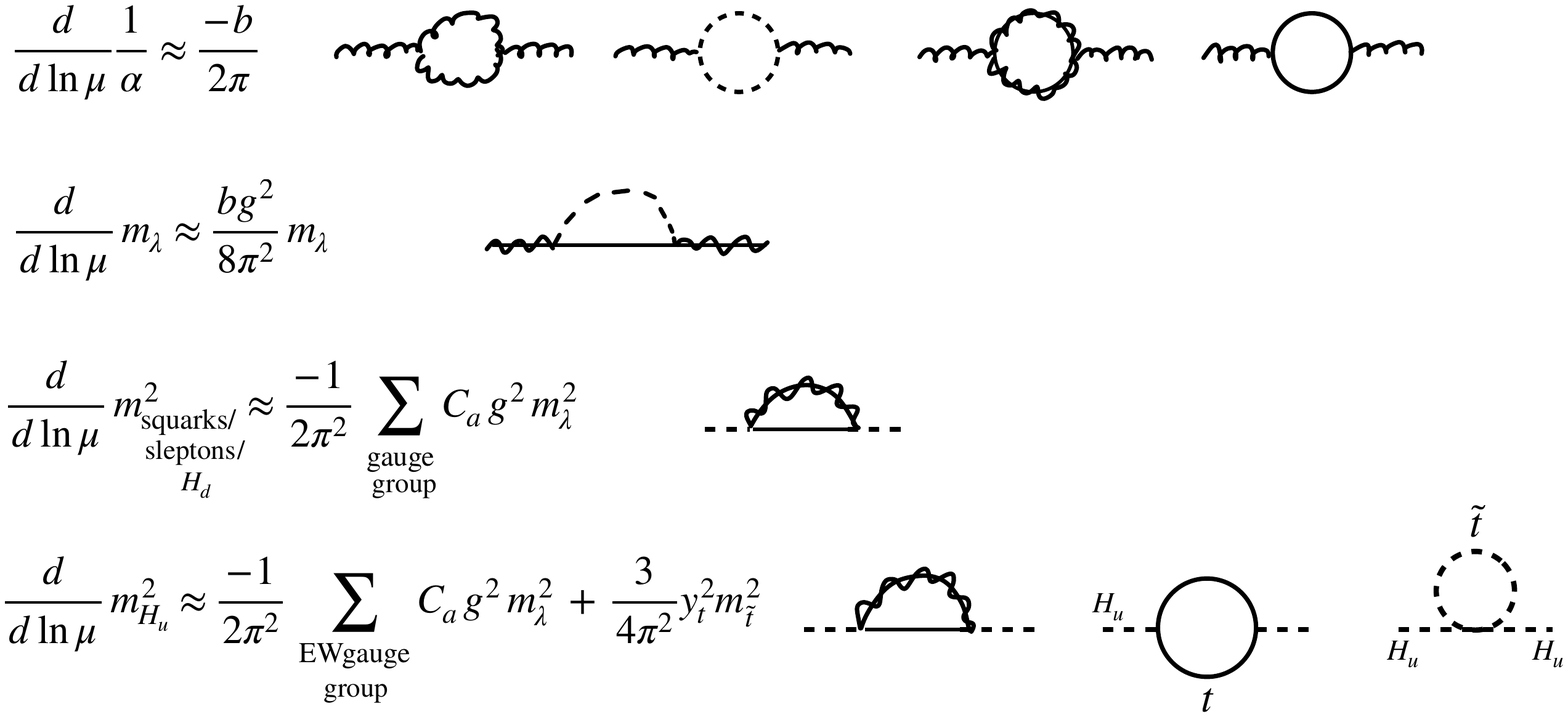}
    \caption{The one-loop running of couplings and SUSY breaking masses in the 4D EFT, including the leading mediation of SUSY breaking to the MSSM scalars.}
    \label{fig:SUSYgaugeRunning}
\end{figure}
\begin{figure}[hbt!]
    \centering    \includegraphics[width=1\linewidth,trim={0.cm 6cm 0cm 5.cm},clip]{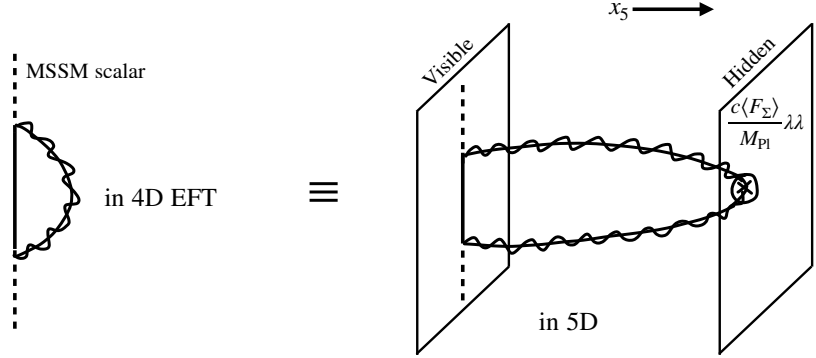}
    \caption{While gaugino-mediation of SUSY breaking to MSSM scalars is efficiently calculated in 4D EFT, fundamentally the gauginos are traversing the extra dimension to accomplish this, as depicted here.}
    \label{fig:SUSY4dEFT}
\end{figure}
The leading approximation we will use is to do tree-level matching at the KK scale $1/L$, and then do one-loop running of couplings and mass parameters below that scale. This is a consistent approach since the loop suppression is significantly compensated by the large logarithms of the very large hierarchy from $1/L$ down to the MSSM soft SUSY breaking scale $\Lambda^2/M_{Pl} \sim 10$ TeV. The good news is then that all the one-loop computation can be done within the 4D EFT, avoiding all 5D SUSY and 5D diagrammatic complexity. For readers interested in explicitly framing this model within 5D SUSY, the short-cut is to recycle our 4D superspace formalism to 5D by the methods of Ref.~\cite{5Dsusy-in-4DSuperspace}.

Now, we have already done the tree-level matching above in working out the tree 4D EFT from the 5D set-up. For the MSSM it is just given by having the $c$'s associated with visible chiral superfields all vanish. There are no renormalizable interactions in the hidden sector to resum via the RG, so it functions in the same way we reviewed in \Sec{sec:spontSUSYbreak-eft}, and as far as the MSSM is concerned we just need the hidden VEV, $\langle \Sigma \rangle = \Lambda^2 \theta^2$. We thereby find the UV boundary conditions at $1/L$ for the RG analysis of the MSSM: 
\begin{align}\label{eq:mssm-UV-BC}
    m_{\rm scalar}^{2} \left(\frac{1}{L} \sim 10^{16} \,{\rm GeV}\right) =0, \; m_{\lambda}\left(\frac{1}{L}\right) = \frac{c \Lambda^{2}}{M_{\rm Pl}} g^2.
\end{align}
That is, the gauginos have a tree-level mass via their direct coupling to the hidden sector in 5D (I have switched to canonically normalized gauginos, explaining the extra $g^2$ in \Eq{eq:mssm-UV-BC}), while all MSSM scalars have vanishing mass$^2$ parameters at this high scale. We have the  coupled $1$-loop RG equations shown in \Fig{fig:SUSYgaugeRunning}. 

The $b$'s are the one-loop gauge coupling $\beta$-function coefficients for the gauge groups, differing from the SM equivalents by adding the contributions of the scalar and gaugino loops. It turns out that the same coefficient appears in the gaugino mass running.\footnote{There is a simple enough SUSY reason for why this had to happen, but I will not go into that.} The coefficients $C_a$ are the quadratic Casimir invariants for gauge representation of the different scalars and different gauge groups. I have negelected all Yukawa couplings except for the top's, $y_t$, which is comparable in strength to the gauge couplings. It therefore appears in the $H_u$ mass RG. In principle it should also appear in the stop $\tilde{t}$ mass RG, but here I will neglect it relative to  the QCD coupling. This is not quite fair since $y_t$ and $g_{QCD}$ are not that different, but it is not a gross mischaracterization. 

Furthermore, since we only want the spirit of this RG analysis without trying to do a ``professional job'', I will make the further ``approximation'' of dropping all the running of the purely SM couplings, namely gauge and Yukawa coupling running. That is, I am dropping the gauge coupling running in \Fig{fig:SUSYgaugeRunning}, and also the gaugino mass running which shares the same RG coefficient $b$. That is, we are only keeping a crude version of the scalar mass$^2$ running because of its {\it qualitative} importance,  without which the scalar masses would be degenerate with their fermion partners. All the other running is just providing fine detail. 

The fact that as far as the MSSM is concerned, the gaugino masses are the ``seeds'' of SUSY breaking that feed radiatively into other fields, gives this basic mechanism its name, ``Gaugino-mediated SUSY Breaking" ($\tilde{g}$mSB) \cite{gaugino-med-susyBreak}. While we have cleverly avoided the need to do 5D Feynman diagrams by using 4D EFT RG equations, we can picture what these equations are fundamentally capturing at the 5D level. This is illustrated in \Fig{fig:SUSY4dEFT}. The gauginos propagate across the extra dimension to communicate SUSY breaking from the hidden boundary to the visible boundary. 

Let me confess the central sense in which this is a toy model. We imposed Peccei-Quinn symmetry which forbids the $\mu$ parameter, and there is no Giudice-Masiero mechanism where $\mu$ effectively arises from SUSY breaking because locality forbids Higgs couplings to the hidden sector. The Peccei-Quinn symmetry then guarantees that $\mu = 0$ even at loop level. Since Peccei-Quinn symmetry is the Higgsino chiral symmetry, its preservation implies the Higgsino is massless. This is, of course, completely excluded experimentally. I will describe the more realistic set-up in \Sec{sec:gaug-med-susyBreak}, but for now it is convenient to proceed with the toy model.

\subsection{Parameter Space}

Since we are ignoring the running of the gauge couplings and Yukawa couplings we can just fix these to their observed values (say evaluated for measurements at $\sim m_Z$). So these are not parameters we will want to vary, and we can drop considering them consciously as input parameters of the theory. Similarly we have tied the KK scale to the unification scale, fixed by data, so this too is not a parameter we will vary. We see then that the only  variable input parameters are the fundamental SUSY-breaking scale $\Lambda$ and the SUSY-breaking $c$ couplings to the gauge superfields. To focus on the visible sector of the MSSM, we can trade $\Lambda$  and the $c$'s for the three gauginos for the three gaugino masses 
$c g^2 \Lambda^2/M_{Pl}$, 
(which do not run in our ``approximation''). Therefore all other quantities of physical interest must be derivable from these three inputs. 

In particular, the structure of EWSB must be predictable and finite, rather than an input as in the non-supersymmetric SM.

\subsection{Radiative EWSB} 

 At tree-level, the MSSM scalars do not feel SUSY breaking, and their potential is just that given by exact SUSY, $V = V_D + V_F$. 
 Since we have no $\mu$ term in this toy model, $V_F$ arises from Yukawa couplings and connects Higgs scalars to sleptons or squarks. 
  Let us explore the non-zero Higgs directions in scalar field space, with slepton and squark scalars kept at zero, so that $V_F = 0$ along these directions. The Higgs scalar potential is then given entirely by $V_D$. Since $H_u$ and $H_d$ have conjugate electroweak quantum numbers it is easy to see that $V_D$ vanishes along the ``flat directions''\footnote{Such inequivalent degenerate vacua flat directions in complex scalar field space are common in SUSY QFT, typically as ``real-imaginary'' partners of Goldstone symmetry-breaking directions (that is, equivalent degenerate vacua related by internal symmetries).}  $H_u = H_d$.  
Therefore a predictive EWSB vacuum, with specific $\langle H_u \rangle, \langle H_d \rangle$ can only be 
determined by an effective potential
``sculpted out'' at loop level. This feature is known as ``radiative EWSB'', as our toy model will now illustrate.

We can easily solve the scalar RG equations for all but  $H_u$ (which has the extra top Yukawa corrections): 
\begin{align}
    m_{\phi}^{2}({\rm IR}) \approx \frac{C}{2 \pi^2} g^2 m_{\lambda}^{2} \ln \left(\frac{M_{\rm KK}}{m_{\rm soft}}\right).
\end{align}
Here we are just keeping the strongest gauge coupling under which the particular scalar $\phi$ is charged, and the associated gaugino mass. We are taking $m_{soft} \sim 10$ TeV, $M_{KK} = 1/L \sim 10^{16}$ GeV, and we are not very sensitive to modest deviations since it is mostly the  logarithm of the large hierarchy that matters.

For the $H_u$ RG, we need the stop mass$^2$ parameter for general RG scale $\mu$, which is simply given by 
\begin{align}\label{eq:rg-stop}
    m_{\tilde{t}}^{2} (\mu) \approx \frac{C_3}{2 \pi^2} \, g_{3}^{2}\, m_{\rm gluino}^{2} \ln \left(\frac{M_{\rm KK}}{\mu}\right).
\end{align}
Plugging this into the $H_u$ RG, we can solve to find
\begin{align}\label{eq:rg-Hu}
    m_{H_{u}}^{2}({\rm IR}) &\approx \frac{C_2}{2 \pi^2}\,  g_{2}^{2} \, m_{\rm wino}^{2} \,\ln\frac{M_{\rm KK}}{m_{\rm soft}} - \frac{3}{16 \pi^4} \, y_{t}^{2} \, C_3 \, g_{3}^{2} \, m_{\rm gluino}^{2} \left(\ln \frac{M_{\rm KK}}{m_{\rm soft}}\right)^{2} \nonumber\\
    &\approx \frac{C_2}{2 \pi^2}\,  g_{2}^{2} \, m_{\rm wino}^{2}\, \ln \frac{M_{\rm KK}}{m_{\rm soft}} - \frac{3}{8 \pi^2} \, y_{t}^{2} \, m_{\tilde{t}}^{2}({\rm IR}) \,\ln \frac{M_{\rm KK}}{m_{\rm soft}}.
\end{align}

Let us make one more switch of input parameters, trading the input $m_{gluino}$ for $m_{\tilde{t}}$ using \Eq{eq:rg-stop}. The $H_u$ mass$^2$ is then given by the last line in terms of the inputs $m_{wino}$ and 
$m_{\tilde{t}}$. We see that other than $H_u$ all other scalars are predicted to be non-tachyonic, that is they will not condense. In particular, in our toy model $m_{H_d}^2(IR) > 0$, so $\langle H_d \rangle = 0$, and therefore none of the down-type quarks and charged leptons will get EWSB masses! This is the other main unrealistic feature of the toy model. Again it will be corrected in subsection~\ref{subsec:gaug-med-susyBreak-higgsSuperfields}. But $H_u$ has radiative mass$^2$ corrections of both signs, positive from EW gaugino loops and negative from the top Yukawa coupling. Therefore it will be tachyonic in a large part of parameter space, condensing and breaking EW symmetry, and giving masses to the $W, Z$  and up-type quarks. Therefore in our toy model, $H_d$ is an entirely massive EW multiplet of scalars, while $H_u$ is the closest thing to the SM Higgs doublet in that it is responsible for EWSB. So, three of its components are eaten by the $W$ and $Z$ and the remaining one can be identified with the observed Higgs boson at $125$ GeV. Given the usual wine-bottle potential for a single Higgs doublet, this means that 
\begin{equation}\label{eq:susy-mass-Hu}
    m_{H_u}^2({\rm IR}) = - 2 (125 {\rm GeV})^2.  
\end{equation}

\section{The (Little) Hierarchy Problem}

\begin{figure}[hbt!]
    \centering    \includegraphics[width=0.9\linewidth,trim={4.cm 5cm 4cm 5.cm},clip]{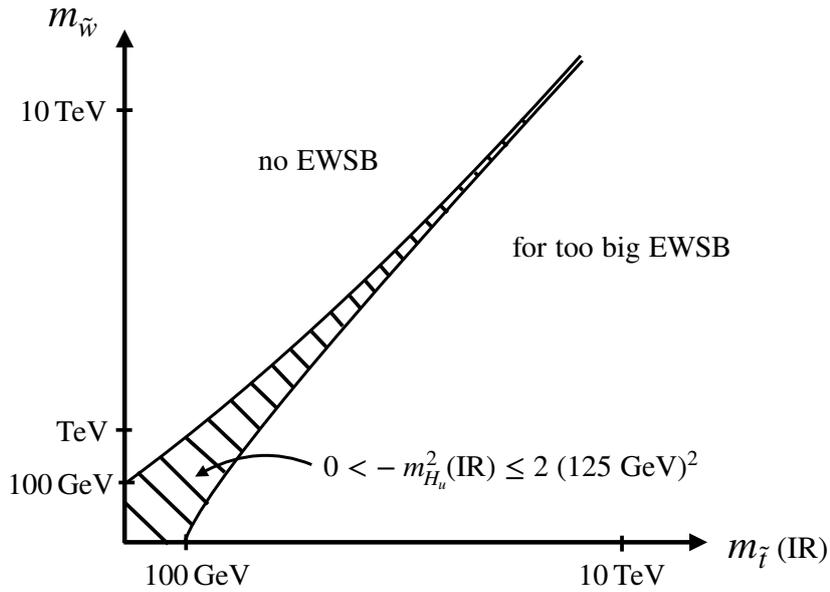}
    \caption{As illustrated here for gaugino-mediation, the larger the superpartner masses are the smaller is the sliver of parameter space in which the emergent weak scale is ${\cal O}(100)$ GeV. This suggests that  the superpartners should not lie too far above the weak scale. We are still waiting for them to show up.    }
    \label{fig:EWSBparaSpace}
\end{figure}

We are finally able to understand in this case study the technical face of the (in)famous Hierarchy Problem. 
Given the lower bounds on colored superpartners in the multi-TeV range and the large logarithms in \Eq{eq:rg-Hu}, obtaining the modest $m_{H_u}^2$(IR) of \Eq{eq:susy-mass-Hu} requires relatively fine cancellation of the two larger terms in \Eq{eq:rg-Hu}. The heavier the superpartners, the finer the cancellation needed to achieve a Higgs mass parameter as small as \Eq{eq:susy-mass-Hu}. We can picture the situation as in \Fig{fig:EWSBparaSpace}, where we are studying the theory in the (effective) input parameter space of $m_{\tilde{W}}$ and $m_{\tilde{t}}(IR)$, where we shade in the region in which EWSB takes place with
$|m_{H_u}^2({\rm IR})| \leq 2 (125 {\rm GeV})^2$. If the superpartner masses were ${\cal O}(100)$ GeV you can see that this would not require very fine cancelations. But if superpartner masses are more like (multi-)TeV, then one must live in a very narrow strip of this parameter space in order to have EWSB $\sim {\cal O}(100)$ GeV. If we had not turned on the LHC (or Tevatron for that matter) yet, given \Eq{eq:rg-Hu} and 
\Fig{fig:EWSBparaSpace}, we would gamble that superpartners were ${\cal O}(100)$ GeV. 
But so far, LHC searches up to $\sim $ TeV have found no superpartners, and furthermore the simplest way to understand the Higgs quartic coupling, discussed in subsection~\ref{subsec:125GeV-higgs}, is to have heavy stops, $m_{\tilde{t}} \sim {\cal O}(10)$ TeV. These direct and indirect arguments then suggest that if SUSY(-breaking) is relevant to EWSB we must live in the sliver of parameter space with multi-TeV superpartners. It just seems puzzling why nature would choose that over superpartners $\sim 100$ GeV, with a finely tuned cancellation in \Eq{eq:rg-Hu} at the level of one part in $10^{3 - 4}$. 

This is the SUSY version of the Little Hierarchy Problem which afflicts any BSM mechanisms in which EWSB becomes a calculable phenomenon rather than an input. Now, you might take an even more extreme view that SUSY is some vestige of superstring theory or quantum gravity near the Planck scale, but it also was broken not far below that scale. In this case, superpartners would be extremely heavy, perhaps not much lighter than the Planck scale. But extrapolating \Fig{fig:EWSBparaSpace} to such massive superpartners means that the parameter space in which EWSB $\sim 100$ GeV is extremely squeezed, the fine-cancellation is at the level of one part in $\sim 10^{30}$! This situation in which the non-SUSY SM is the  valid EFT all the way close to the Planck scale would be incredibly puzzling, and it poses the {\it Big} Hierarchy Problem. By comparison multi-TeV SUSY would seem more plausible, although we would like to know if there is some other small mechanism missing that would help resolve the Little Hierarchy Problem, or if that is agonizing unnecessarily over a relatively modest perhaps random cancellation.

In the (toy) model under study we have the gaugino masses as ultimately setting the scales of SUSY breaking in general and thereby radiative EWSB. One might therefore worry that the Big and Little Hierarchy Problems are specific to this model. But in any fundamental theory (known) {\it some} scale plays the role of the gaugino masses here. For examples: in Gauge-Higgs Unification where the Higgs incarnates as an extra-dimensional component of a gauge field the fundamental scale is given by the KK scale $m_{KK}$, in Composite Higgs theories the fundamental scale is given by the compositeness scale 
$\Lambda_{comp}$, in string theories in which SUSY is maximally broken the fundamental scale is played by the string scale $m_{string}$, and it is reasonable to expect that in a theory of quantum gravity from which EFT emerges in the IR the Planck scale $M_{Pl}$ is a kind of maximal fundamental scale  (corresponding to the minimal distance $\ell_{Pl}$).  In any of these settings it appears that there would be a fine-tuned cancellation of the sort depicted in \Fig{fig:EWSBparaSpace} unless one or more of these fundamental scales is  relatively close to the weak scale. That is, as far as we understand so far, the hierarchy problems illustrated here are in fact quite general: either new physics appears not far above the weak scale or EWSB arises as a fine cancellation of huge competing and uncorrelated contributions to the Higgs effective potential. 

\section{The UnSequestered}

\begin{figure}[hbt!]
    \centering    \includegraphics[width=0.9\linewidth,trim={1.cm 5cm 1cm 5.cm},clip]{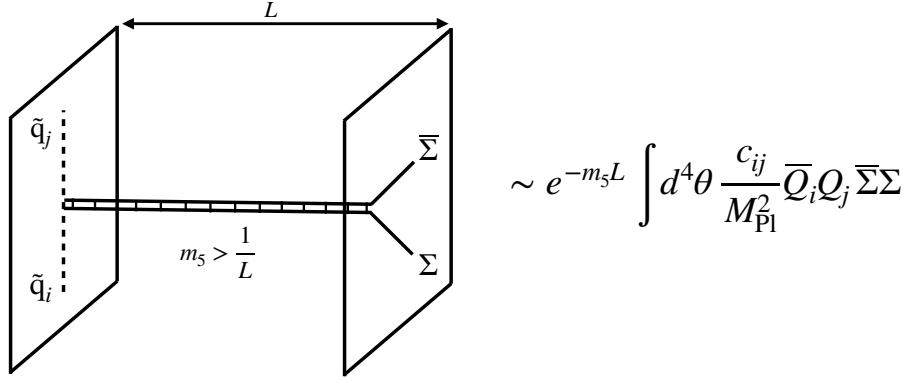}
    \caption{Flavor-violating contributions to the MSSM SUSY-breaking scalar masses can be mediated by heavy 5D $\sim$ Planckian bulk fields, whose couplings are not as constrained  as those of the gauge superfields. But these heavy exchanges across the extra dimension are Yukawa-suppressed. }
    \label{fig:SUSYunsequestered}
\end{figure}

In \Fig{fig:SUSY4dEFT} I showed how gauge superfields in the bulk (in particular the gauginos) straddle the extra dimension in communicating SUSY-breaking to the visible MSSM sector (scalars, particularly)  at loop level. But what about other possible bulk fields that might also mediate SUSY breaking across the extra dimension? Of course, the one other mandatory inhabitant of the bulk is 5D supergravity. This can and does mediate a particular type of SUSY breaking to the MSSM \cite{sequestering} but it can readily be subdominant to gaugino-mediation. There can be other fields needed to stabilize the size $L$ of the extra dimension, given that spacetime is dynamical with 5D General Relativity. See for example, Ref.~\cite{rad-stabilization}.
Such fields can also readily give only subdominant SUSY breaking contributions. 

But as (5D) effective field theorists we are always at the mercy of very heavy fields that lie above the UV cutoff of our  (5D) EFT control, minimally the KK scale $1/L$. So let us consider the generic effect of such a massive bulk field, $m_5 > 1/L$, as depicted in \Fig{fig:SUSYunsequestered}. I have estimated the size of the effect by the same arguments of subsection~\ref{subsec:bound-loc-sequestering} for massive bulk exchange between boundary localized fields. In particular there is the Yukawa suppression $e^{- m_5 L}$ for a massive field to connect distant boundaries. But without knowing the details of this massive field, it might well couple to MSSM matter with flavor-violating couplings $\propto c_{ij}$. So flavor violation and CP violation in $c_{ij}$ is not completely eliminated by sequestering, but only exponentially suppressed. Indeed, given that the maximal scale $M_{Pl}$ is not far above $1/L$, the possible $m_5 L$ of heavy states associated to string theory or quantum gravity cannot be too large. While it is possible that the exponential suppressions are sufficient to make BSM FCNC or CP-violation smaller than current bounds, it is interesting that upcoming experimental improvements  may detect the effects of the heavy bulk fields close to the Planck scale! It may be these precision low energy tests that are the first to go beyond the SM, but if the sequestering is very good then it will take high energy colliders to explicitly see the BSM physics. It is exciting to see how a BSM breakthrough may happen along very different experimental fronts, whether flavor, CP, collider resonances, or dark matter detection.

\section{Realistic Gaugino(-Higgs) Mediated SUSY Breaking}\label{sec:gaug-med-susyBreak}

\subsection{Higgs superfields in the Bulk}\label{subsec:gaug-med-susyBreak-higgsSuperfields}

\begin{figure}[hbt!]
    \centering    \includegraphics[width=1\linewidth,trim={0.cm 3cm 1cm 5.cm},clip]{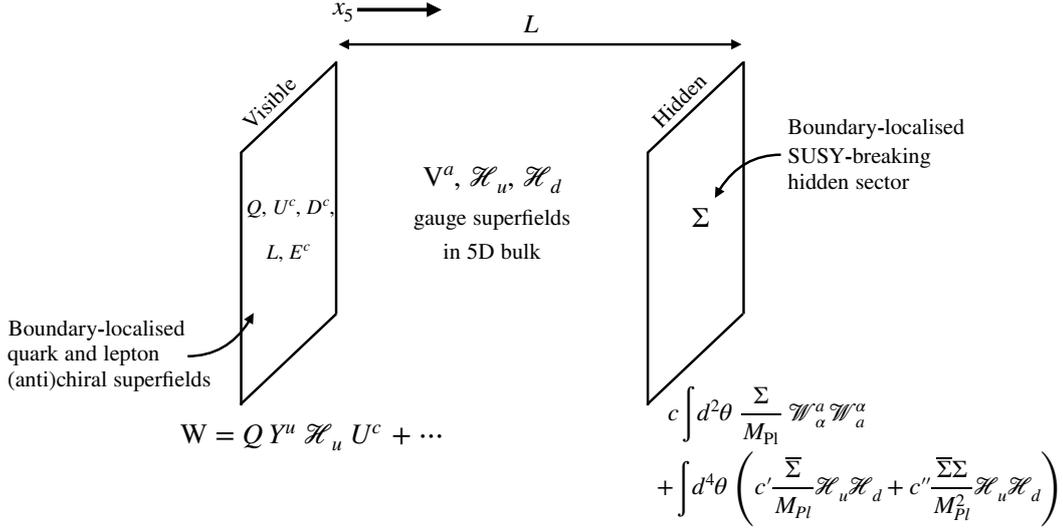}
    \caption{Improving on our toy model, the Higgs superfields now propagate in the 5D bulk, permitting the Giudice-Masiero mechanism and thereby a realistic theory.
    }
    \label{fig:SUSYextraDimStructure-higgsInBulk}
\end{figure}
 
So far, I have reviewed a toy model of SUSY breaking in the MSSM in which the down-type fermions and Higgsinos remain massless, but in which the up-type fermions and EWSB Higgs field are treated fairly realistically but with rather crude approximations in the RG analysis. Here, I briefly want to leave you with the set-up that allows a fully realistic incarnation of the MSSM and SUSY-breaking, while still solving the SUSY flavor- and CP-problems, and now the $\mu$-problem as well. It is depicted in \Fig{fig:SUSYextraDimStructure-higgsInBulk}. 

The central difference from before is that the MSSM Higgs superfields now propagate in the bulk rather than being boundary-localized. Therefore, like the MSSM gauge superfields, they can directly couple to the hidden sector on the hidden boundary and acquire SUSY-breaking masses. 
They can still have superpotential couplings with the flavored matter localized on the visible boundary.  Just as we are taking CP-violation to be localized to the visible boundary, we can take Peccei-Quinn symmetry violation to be localized to the hidden boundary, allowing arbitrary Higgs superfield couplings to $\Sigma$. This allows us to realize a $\mu$ term  naturally comparable to the other MSSM soft masses, via the Giudice-Masiero mechanism, as well as a $B \mu$ term also naturally comparable to the other soft mass-squareds. 

These two terms fix the lack of realism in our toy model. The $\mu$ term straightforwardly incorporates a Dirac mass for the Higgsinos, so they are no longer massless. $B \mu$ is the coefficient of the Higgs scalar mass-mixing, $H_u H_d$. To see its effect, we treat it as modestly small with respect to the Higgs spectrum of the toy model, where the $H_u$ was solely responsible for EWSB and the entire $H_d$ doublet was massive with zero VEV. Perturbatively, 
$B \mu H_u H_d$ will induce a tadpole for $H_d$ after $H_u$ EWSB, $B \mu \langle H_u \rangle H_d$. Given that $H_d$ is massive, such a tadpole will induce a small VEV for $H_d$,
\begin{equation}
\langle H_d \rangle \approx \frac{B \mu }{m_{H_d}^2} \langle H_u \rangle.
\end{equation}
This subdominant EWSB VEV is sufficient to give the small (compared to the weak scale) down-type fermion masses we observe without overly disturbing our discussion of the $125$ GeV observed Higgs phenomenology. But it does mean that the down-type Yukawa couplings are different from their non-SUSY SM equivalents to get the same observed fermion masses:
\begin{equation}
Y_{d, ij}^{MSSM} \approx Y_{d, ij}^{SM} \tan \beta, ~  
Y_{e, ij}^{MSSM} \approx Y_{e, ij}^{SM} \tan \beta,
\end{equation}
where 
\begin{equation}
\tan \beta \equiv \frac{\langle H_u \rangle}{\langle H_d \rangle} \gg  1.
\end{equation}

With tree-level matching at the KK scale we get our softly broken 4D MSSM but with just the flavor-dependent $c_{ij}^{Q, L, U, D, E} = 0$, while other $c$'s are non-zero and roughly order one, and containing no new CP-violating phases. It is just this feature that ensures that the SUSY flavor- and CP-problems are solved. The RG analysis is more complicated than our toy analysis because the Higgs soft masses are additional seeds of SUSY-breaking beyond the gaugino-masses, but qualitatively it is similar, and can readily yield fully realistic super-spectra. This attractive BSM framework is still called ``$\tilde{g}mSB$'', though it is perhaps more accurate to call it ``gaugino-Higgs'' mediation.

\subsection{A note on the different UV scales}

Just as for gauge theory, where the 4D and 5D couplings are different but related as in \Eq{eq:4D-5D-gaugeCoupling}, the same is true for general relativity, 
\begin{equation}
\frac{1}{G_{4D, eff}} = \frac{L}{G_{5D}}.
\end{equation}
In terms of Planck scales, 
\begin{equation}
    M_{Pl, 4D} \sim  10^{18} {\rm GeV} = M_{Pl, 5D}^{3/2} L^{1/2}.
\end{equation}
Since we are taking $1/L \sim 10^{16} {\rm GeV}$, it follows that the 5D  Planck scale is 
\begin{equation}
M_{Pl, 5D} \sim 10^{17} {\rm GeV}. 
\end{equation}

Since $M_{Pl, 5D}$ is the fundamental scale of 5D gravity, one might expect it to set the scale of non-renormalizable couplings in our theory rather than the $M_{Pl, 4D}$  we have used so far, in units of which our non-renormalizable couplings were expected to be  dimensionless $c$'s of order one. More generally, if one carefully does the 5D to 4D matching there will be various (fractional) powers of $L M_{Pl, 5D} \sim 10$ that will arise here and there in 5D Planck units. But since our philosophy is that we are allowing modestly hierarchical input couplings while aiming to generate all other hierarchies via physical exponentials, we can afford to be sloppy about such 
$L M_{Pl, 5D}$ factors. We simply absorb them into redefinitions of the $c$'s of the 4D EFT. 

If we are being sloppy about these UV distinctions between 
$1/L, M_{Pl, 5D}, M_{Pl, 4D}$, then what is the point of having invoked them in the first place?! The central point is that extremely heavy 5D particles outside our 5D EFT are expected to  have masses at most  $\sim M_{Pl, 5D}$, so the non-sequestered effects they mediate are suppressed by $e^{- M_{Pl, 5D} L} \sim 10^{-4}$. This level of suppression is very important in understanding how the SUSY flavor- and CP-problems are adequately solved. So we have invoked the modest distinctions between $L, M_{Pl, 5D}, M_{Pl, 4D}$ because they are important when exponentiated, but not otherwise. Of course, one can more carefully track these distinctions even outside exponentials, if desired. 

\section{Dynamical SuperSymmetry Breaking (DSSB)}\label{sec:dssb}

Far below the highest scales in our theory, $1/L, M_{Pl, 5D}, M_{Pl, 4D}$, lies the intermediate scale of SUSY-breaking, $\Lambda \sim 10^{11}$ GeV. The soft SUSY-breaking scale $\sim \Lambda^2/M_{Pl} \sim$ (several) TeV, and subsequent  EWSB are derived from it. Until now, we have put $\Lambda$ in by hand into the hidden sector Lagrangian, but ultimately we would like to have $\Lambda \ll M_{Pl}$ be emergent from a Planckian theory by something like dimensional transmutation. Remarkably, there are indeed elegant QFT mechanisms that accomplish this task if we generalize the hidden sector to include a suitable hidden SUSY gauge theory. Spontaneous SUSY breaking via dimensional transmutation is called ``dynamical super-symmetry breaking'' (DSSB) \cite{susyBreak-lectures}.

Here I will give the simplest example of this, with the simplicity again coming at the cost of being a non-renormalizable EFT. The core mechanism in DSSB is ``gaugino-condensation'', not to be confused with the gaugino-mediation we have discussed above. In particular, gaugino-mediation involved gauginos of the MSSM, whereas we are now focusing on the hidden sector, and the gauginos that are condensing belong to a {\it hidden sector gauge theory} under which no MSSM fields are charged.
Consider a rather minimal hidden sector containing SUSY non-abelian Yang-Mills theory (SYM), consisting of just  gauge bosons and gauginos, and our $\Sigma$ chiral superfield which remains a complete gauge singlet. The hidden Lagrangian is 
\begin{align}\label{eq:hiddenL}
    {\cal L}_{\rm hidden} = \int d^{4}\theta \, |\Sigma|^{4} - \frac{|\Sigma|^{4}}{4 M_{1}^{2}} + \left\lbrace \int d^{2} \theta \left( \frac{1}{4 g^{2}} - \frac{ \pi \Sigma}{M_{2}}\right) {\cal W}^{2}_{\alpha} + h.c. \right\rbrace
\end{align}
The two scales of non-renormalizable couplings, $M_1, M_2$ are taken to roughly Planckian in size.
We have not included any superpotential for $\Sigma$, because unlike in \Sec{sec:spontSUSYbreak-eft} where we put in a superpotential with $\Lambda$ by hand, here we want it to be generated via dimensional transmutation.  

In components, SYM is a non-abelian gauge theory with the gaugino being a single  species of ``quark'' in the adjoint representation of the gauge group (to match that of the gauge boson). We can justify the vanishing superpotential by imposing $U(1)_R$ symmetry, under which the bosons $A_{\mu}$ and $ \sigma$ (and hence superfield $\Sigma$) have vanishing charge but  their fermionic superpartners are charged. Since this means the Grassmannn coordinates $\theta$ are also charged it follows that $\int d^2 \theta~W(\Sigma)$ is forbidden. Now, this may seem like overkill, since we do want to generate a non-trivial $\Sigma$ superpotential by dimensional transmutation. Fortunately, 
$U(1))_R$ is non-perturbatively anomalous, its action on the gaugino ``quark'' being analogous to the famously anomalous $U(1)_{\rm axial}$ chiral symmetry of QCD. So, $U(1)_R$
is at best a perturbative symmetry, and non-perturbatively we can expect that it is broken. This is the perfect situation, perturbatively $W(\Sigma)$ vanishes but then it can be generated non-perturbatively. 

In QCD with massless quarks, there are various axial chiral symmetries which are spontaneously broken and result in massless composite Nambu-Goldstone bosons, but the overall $U(1)_{axial}$ is non-perturbatively anomalous and the anomalous breaking does not give rise to a NG boson. Similarly, in SYM the anomalous breaking of $U(1)_R$ does not give rise to a NG composite. Instead the spectrum of composites of the confined gauge bosons and gauginos is completely massive, with masses given by dimensional transmutation of order 
\begin{equation}
\Lambda_{SYM} \sim M_{\rm Pl} \, e^{ - \nicefrac{1}{ b_{\scaleto{SYM}{2.5pt}} \alpha_{\scaleto{SYM}{2.5pt}}(M_{\rm Pl})}}.   
\end{equation}
 Here,  $b_{SYM} = 3N/(2\pi)$ is the one-loop $\beta$-function coefficient appropriate to SYM for gauge group $SU(N)$. 
 We can therefore imagine integrating out the entire massive SYM sector and asking what the EFT for $\Sigma$ is below $\Lambda_{SYM}$. In particular, we would like to know what effective non-perturbative superpotential $W_{non-pert}(\Sigma)$ is generated. There are two big clues as to its form: (i) $W_{non-pert}$ must be holomorphic by SUSY, depending on $\Sigma$ but not $\bar{\Sigma}$, (ii) as far as SYM is concerned $\Sigma$ only appears in the combination $1/\alpha - \Sigma/M_2$ (see \Eq{eq:hiddenL}), so 
 $W_{non-pert}$ only depends on this combination (but not its conjugate). 

 Putting these clues together with general features of non-perturbative gauge theory will fix the form of $W$. Given that the dimension-$3$ superpotential must be generated by dimensional transmutation, we must have 
\begin{align}
    W_{\rm non-pert} \propto M_{\rm Pl}^{3} \, e^{- \nicefrac{3}{b_{\scaleto{SYM}{2.5pt}}\,\alpha_{\scaleto{SYM}{2.5pt}}(M_{\rm Pl})}}.
\end{align}
Therefore, given (ii), 
\begin{align}
    W_{\rm non-pert} = M_{\rm Pl}^{3} \, e^{- \nicefrac{3}{b_{\scaleto{SYM}{2.5pt}} } \left( \nicefrac{1}{\alpha_{\scaleto{SYM}{2.5pt}}(M_{\rm Pl})} - \nicefrac{\Sigma}{M_2} \right)}.
\end{align}
Equivalently, 
\begin{align}
    W_{\rm non-pert} = {\cal O}(1) \, \Lambda^{3}_{\rm SYM} \, e^{\nicefrac{3 \Sigma}{(b_{\scaleto{SYM}{2.5pt}}M_{2}})}.
\end{align}

\begin{figure}[hbt!]
    \centering    \includegraphics[width=0.8\linewidth,trim={1.cm 2.5cm 1cm 2.cm},clip]{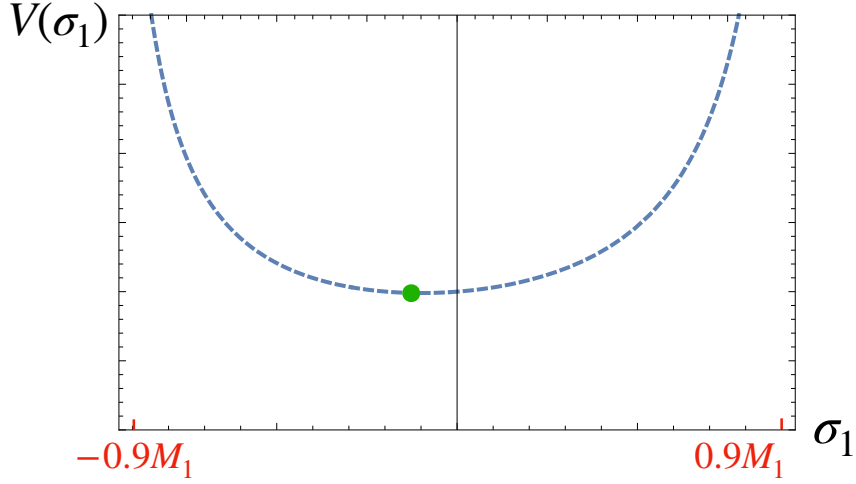}
    \caption{The scalar effective potential arising from coupling to the non-perturbative SYM dynamics. Its positive (local) minimum signals spontaneous SUSY breaking.}
    \label{fig:Sigma1PotLabled}
\end{figure}

Note, we are computing the non-perturbative correction to the superpotential even though it is very ``small'' (since we can readily have $\Lambda_{SYM} \ll M_{Pl}$ for modestly small $\alpha_{SYM}(M_{Pl})$) because otherwise it would simply vanish. We do not need to compute corrections to the Kahler potential since that is already non-trivial. Putting together $K(\Sigma, \bar{\Sigma})$ and $W_{non-pert}(\Sigma)$ in the $\Sigma$ EFT below $\Lambda_{SYM}$, we find a scalar $\sigma$ potential,
\begin{align}
    V &\sim  \frac{\Lambda_{SYM}^6}{M_{Pl}^2} \frac{ e^{\nicefrac{3(\sigma +\overline{\sigma})}{(b_{\scaleto{SYM}{2.5pt}} M_{2})}}}{1- \nicefrac{|\sigma|^{2}}{M_{1}^{2}}} \nonumber \\
    & = \frac{\Lambda_{SYM}^6}{M_{Pl}^2} \frac{e^{\nicefrac{6\sigma_{1}}{(b_{\scaleto{SYM}{2.5pt}} M_{2})}}}{1- \nicefrac{(\sigma_{1}^{2}+\sigma_{2}^{2})}{M_{1}^{2}}},
\end{align}
where we decompose into real and imaginary components, $\sigma \equiv \sigma_1 + i \sigma_2$. We easily see that for sub-Planckian VEVs, $\langle \sigma_2 \rangle = 0$ is a minimum. Plugging this in, $V(\sigma_1)$ is sketched in \Fig{fig:Sigma1PotLabled}.  Clearly, non-renormalizable EFT is breaking down at $\sigma \sim M_1$, so we should only trust it for $\langle \sigma_1 \rangle \ll M_1$.  This happens for $b_{SYM} M_2 \gg M_1$, which is the case that is plotted. In fact, we can analytically solve for the VEV to leading order in $M_1/(b_{SYM} M_2) \ll 1$ straightforwardly (which does not have to be very small for this to be a reasonable approximation), 
\begin{equation}
\langle \sigma_1 \rangle \approx - \frac{M_1^2}{3 b_{SYM} M_2}, ~
\langle \sigma_2 \rangle = 0.
\end{equation}
Since the vacuum energy density is positive,
this shows that SUSY has been spontaneously broken at the scale $\sim \Lambda_{SYM}^{3/2}/M_{Pl}^{1/2}$. 

We can think of this new hidden sector model as realizing our old model of \Sec{sec:spontSUSYbreak-eft}, in the approximation where $M_2$ is the largest scale so that we can Taylor expand the non-perturbative superpotential to leading non-trivial order in $1/M_2$, 
\begin{align}
    \int d^{2} \theta \, W_{\rm non-pert} (\Sigma) &\sim \int d^{2} \theta \, \Lambda^{3}_{\rm SYM} \left( 1- \frac{3 \Sigma}{b_{SYM} M_{2}} \right) \nonumber \\
    & \sim \int d^{2} \theta \, \frac{3 \Lambda^{3}_{\rm SYM}}{ b_{SYM} M_{2}} \, \Sigma.
\end{align}
That is, we are simply realizing our old SUSY-breaking scale via dimensional transmutation,
\begin{align}
    \Lambda \sim \frac{\Lambda_{\rm SYM}^{3/2}} {M_{Pl}^{1/2}} \sim {\cal O}(M_{\rm Pl}) \, e^{-\nicefrac{3}{2 b_{\scaleto{SYM}{2.5pt}} \alpha_{\scaleto{SYM}{2.5pt}}(M_{\rm Pl})}}.
\end{align}

Given we want $\Lambda \sim 10^{11}$ GeV,  we see that $\Lambda_{SYM} \sim 10^{14-15}$ GeV. 
In essence then, all the earlier sections utilizing the SUSY breaking module of \Sec{sec:spontSUSYbreak-eft} continue unaffected after we integrate out the hidden SYM ``hadrons'' with masses $\sim \Lambda_{SYM} \sim 10^{14-15}$ GeV, with just the understanding that $\Lambda \ll M_{Pl}$ is now elegantly arising via dimensional transmutation.

\section{Conclusions}

I have covered a variety of big-picture mechanisms and considerations in these lectures, culminating in a prime example of a  {\it comprehensive} BSM framework. Most research projects these days do not involve developing comprehensive BSM frameworks, but rather pushing on particular directions related to new experimental developments or anomalies, or interesting new theoretical mechanisms involving a modest set of particles.  Nevertheless, I hope the comprehensive big picture provides a useful backdrop, context and guidance for your own particular research activities. It certainly does for me in my current research. 

The focal point of the lectures was the grand ambition of generating {\it all} significant particle hierarchies, in the spirit of the non-perturbative mechanism of dimensional transmutation arising in QCD. To make this practicable, modest hierarchies of an order of magnitude were accepted without question in the input parameters of the final model if the mechanisms were present to exponentiate these into the larger hierarchies we see. The actual Hierarchy Problem and Little Hierarchy Problem were (re-)introduced in the context of this ambitious goal. 

I should qualify that we only attempted to understand the {\it non-gravitational} particle hierarchies, the quantum gravity Planck scale itself only featuring as the UV ``end'' of particle physics, the ultimate cutoff of particle physics EFT. But we did not attempt to address the greatest hierarchy problem, the Cosmological Constant Problem, based on the experimental observation
\begin{equation}
\rho_{dark-energy} \sim {\rm meV}^4 \ll v_{weak}^4 \ll M_{Pl}^4,
\end{equation}
which brings in gravity in an essential way. There is no known (well-accepted) mechanism for realistically solving this problem other than an appeal to the Anthropic Principle operating within a vast  multiverse of universes with differing laws and couplings \cite{polchinski-cosmoConstant-stringLandscape}. And yet, it is intriguing that global SUSY gives a symmetry reason why the vacuum energy vanishes, and that when coupled to gravity to make SUGRA theories this at least gives a very natural means of having extremely small dark energy (cosmological constant). The problem is that this powerful mechanism for small dark energy does not survive realistic SUSY-breaking. Nevertheless, (to me) it is one of the great discoveries of theoretical physics that a (non-realistic) interacting theory of particles and gravity, with diverse mass scales, can have very small dark energy if the theory is supersymmetric. 

There is a second qualification. Physics is based on the laws of nature that allow us to time-evolve an initial state {\it and} a suitable initial state as input. In these lectures, I have not discussed any of the puzzles and hierarchical structure that relate to the initial conditions of the universe, or at least very early universe conditions. See Ref.~\cite{ClineTasi-earlyCosmo} for a review. For example, the state of the universe is clearly asymmetric between ordinary matter and antimatter and the laws we discussed simply maintain this asymmetry in time. So the observed asymmetry must be due to some other mechanism operating in the early universe, broadly known as Baryogenesis.  The very high degree of homogeneity of the universe on the largest length scales poses another puzzle, most often mitigated within the framework for initial conditions known as Cosmic Inflation. Questions of cosmic initial conditions and the place of our universe within a larger multiverse are highly ambitious and yet plausibly within our ability to make progress on, by a combination of cosmological experiments and BSM  theory. I hope that these lectures and their focus on the current laws of nature will provide a useful base for the ongoing pursuit of these grand questions.

\section*{Acknowledgements} 
I am grateful to the organizers, Jiji Fan, Stefania Gori and Liantao Wang, for their invitation to lecture at TASI 2022. 
I am grateful to Arushi Bodas for assistance in  preparing the figures, equations and bibliography for this article.
This work was supported by NSF grant PHY-2210361 and by the Maryland Center for Fundamental Physics.


\end{document}